%% file: higher_twist.tex
\documentclass[11pt,a4paper]{article}
\usepackage[colorlinks=true, linkcolor=black!50!blue, urlcolor=blue, citecolor=blue, anchorcolor=blue]{hyperref}
\usepackage[font=small,labelfont=bf,margin=0mm,labelsep=period,tableposition=top]{caption}
\usepackage[a4paper,top=3cm,bottom=2.5cm,left=2.5cm,right=2.5cm,bindingoffset=0mm]{geometry}
\usepackage{amsmath}
\usepackage{multirow}
\usepackage{xcolor}
\usepackage{xspace}
\usepackage{booktabs}
\usepackage{bbold}
\usepackage{graphicx}
\usepackage{tabularx}
\usepackage{amssymb}
\usepackage{cite}
\usepackage{cleveref}
\crefformat{equation}{Eq.~(#2#1#3)}
\crefformat{figure}{{Fig.~#2#1#3}}
\crefformat{section}{{Sec.~#2#1#3}}
\crefmultiformat{figure}{Figs.~#2#1#3}{ and~#2#1#3}{, #2#1#3}{ and~#2#1#3}
\crefmultiformat{equation}{Eqs.~(#2#1#3)}{ and~(#2#1#3)}{, (#2#1#3)}{ and~(#2#1#3)}
\crefrangeformat{equation}{Eqs.~(#3#1#4) to~(#5#2#6)}
\newcolumntype{C}[1]{>{\centering\arraybackslash}p{#1}}

\numberwithin{equation}{section}
\numberwithin{figure}{section}
\numberwithin{table}{section}

\newcommand{\annnlo}{aN$^3$LO\xspace}
\newcommand{\tev}{~\mathrm{TeV}}

\newcommand{\T}[1]{\textrm{#1}}

\newcommand{\chisq}{$\chi^2$}

\begin{document}
\newgeometry{top=1.5cm,bottom=1.5cm,left=1.5cm,right=1.5cm,bindingoffset=0mm}

\vspace{-2.0cm}
\begin{flushright}
Edinburgh 2025/29
\end{flushright}
\vspace{0.3cm}

\begin{center}
  {\Large \bf Parton distributions with higher twist and jet power corrections}
  \vspace{1.1cm}

  Richard D. Ball$^{1}$, Amedeo Chiefa$^{1}$ and Roy Stegeman$^{1,2}$

  \vspace{0.2cm}

  {\it \small
    $^{1}$ The Higgs Centre for Theoretical Physics, University of Edinburgh,\\
    JCMB, KB, Mayfield Rd, Edinburgh EH9 3JZ, Scotland\\[0.1cm]
    $^2$  Quantum Research Centre, Technology Innovation Institute, Abu Dhabi, UAE\\[0.1cm]
  }

  \vspace{0.7cm}

  {\bf \large Abstract}
\end{center}

We present a global determination of parton distribution functions (PDFs) that accounts for higher twist corrections in deep-inelastic scattering (DIS) and linear power corrections for single inclusive jet and dijet production data from the LHC.
We determine these corrections and their associated correlated uncertainties using a methodology based on the theory covariance formalism, previously
used to account for nuclear uncertainties and missing higher order uncertainties (MHOUs) in global PDF determinations.
We then study the impact of the power corrections on the extracted PDFs, and
demonstrate an improved description of the data due to a reduced sensitivity
to DIS data in the low-$x$ region where higher twist uncertainties are
relatively large, and a reduced sensitivity to single inclusive jet data at
relatively low $p_T$, where hadronization corrections can be large.
Finally, we assess the impact of power corrections on observables relevant
to LHC phenomenology, including Higgs production via gluon fusion, and the
determination of $\alpha_s$. We find that these effects, while small, can
be significant, improving perturbative convergence.

\tableofcontents

\clearpage

\input{sec-introduction.tex}

\input{sec-methodology.tex}

\input{sec-pctcm.tex}
\input{sec-results.tex}

\input{sec-conclusions.tex}

\clearpage
\bibliography{higher_twist}

\end{document}

%% file: sec-introduction.tex
\section{Introduction}
\label{sec:intro}

The advent of the high energy frontier, now at a center-of-mass
energy scale of 13.6 TeV at the LHC, offers an unprecedented opportunity to
test the predictive power of the Standard Model. Remarkable progress has
been made in calculations at next-to-next-to-leading order (NNLO) and even
next-to-next-to-next-to-leading order (N$^3$LO) accuracy in perturbative QCD,
together with the inclusion of mass corrections, QED and Electroweak
corrections. Nevertheless, some significant sources of theoretical
uncertainty remain, both perturbative and non-perturbative, and in both the hard
cross-sections themselves and in the parton
distribution functions (PDFs). If our theoretical predictions are to match the
accuracy and precision of LHC data, we need to control all theoretical
uncertainties at the one percent level~\cite{Amoroso:2022eow}.

A systematic estimation of all possible sources of significant theoretical
uncertainty is important for two main reasons. First, it provides a more
reliable assessment of the uncertainties associated with theoretical
predictions, important for interpreting experimental data and establishing
a robust framework for probing potential new physics. Second, and more
subtly, since theoretical predictions at hadron colliders rely on
accurate and precise PDFs, it is necessary to also account systematically
for all the various theoretical uncertainties in the calculations used in
global PDF fits.

Theoretical uncertainties arising from missing higher-orders (MHOUs)
have been included in PDF fits using
a theory covariance matrix~\cite{Ball:2018lag} constructed from scale
variations~\cite{NNPDF:2019vjt,NNPDF:2019ubu,NNPDF:2024dpb}, or by
fitting nuisance parameters~\cite{McGowan:2022nag}. The covariance
matrix approach has also been used to incorporate the non-perturbative
uncertainties due to nuclear
effects in global fits with deep-inelastic scattering (DIS) and Drell-Yan (DY)
data on heavy nuclei and deuteron targets~\cite{Ball:2018twp,Ball:2020xqw}.

Besides theoretical uncertainties due to perturbative corrections and nuclear
effects, there are further uncertainties due to effects which are suppressed by
a power of the hard scale. In DIS, the familiar
expression for the cross-section,
factorised into a convolution of a perturbative coefficient function and
universal PDFs, is only
the leading term in the operator-product expansion. Additional terms,
known as higher twist (HT) corrections, are always present, but suppressed by
powers of the hard scale $Q^2$. These HT corrections are in essence
non-perturbative, and over the years there have been many attempts to determine
them empirically by fitting $\mathcal{O}(\Lambda^2/Q^2)$ corrections (here $\Lambda$ is the QCD scale) in the context of a PDF
fit~\cite{Virchaux:1991jc,Alekhin:1998df,Alekhin:1999af,Martin:1998np,Martin:2003sk,Blumlein:2008kz,Alekhin:2012ig,Ball:2013gsa,Thorne:2014toa,Alekhin:2017kpj,Alekhin:2022tip,Cerutti:2025yji,Harland-Lang:2025wvm,Alekhin:2025qdj}.
This is challenging, as the HT corrections can easily be confused with
the unknown terms behaving as powers of $\mathcal{O}\left(1/\log (Q^2/\Lambda^2)\right)$ arising due to MHOUs. Consequently most global
PDF fits impose kinematic cuts on DIS data to ensure that the HT corrections are
small.

For hadronic processes, such as DY or inclusive jets, there is no operator product expansion, but
factorization still holds in the limit of large momentum
transfer~\cite{Collins:1989gx}. Again there are non-perturbative contributions
suppressed by powers of the hard scale $Q_T$. For sufficiently inclusive
processes with a color singlet final state, such as DY, the power corrections
(PCs) are expected to be
$\mathcal{O}(\Lambda^2/Q_T^2)$~\cite{Beneke:1995pq,Caola:2021kzt}, and thus small except
close to threshold. However for many other hadronic observables there are
generally linear power corrections of $\mathcal{O}(\Lambda/Q_T)$. For example, linear PCs
to jet processes have been studied extensively in $e^+ e^-$
annihilation~\cite{Dasgupta:2003iq,Dasgupta:2007wa,Dasgupta:2008di,
Cacciari:2007aq}, with analytic estimates of the shift in the jet energy due to
hadronization and underlying event. Studies of jets in hadronic
collisions~\cite{Olness:2009qd} have extended these analyses to estimate the
size of theoretical uncertainties related to various non-perturbative effects in
inclusive single jet cross-sections. These estimates suggest that the linear PCs
to jet processes might be significant even up to scales as large as $100$ GeV.
Linear PCs are also expected in top production~\cite{Makarov:2023uet}, unless
one uses a short distance scheme for the top quark mass. Since linear PCs fall
off much more slowly with hard scale than HT, it is even more difficult to
disentangle them from MHOUs. To the best of our knowledge, an empirical
determination of linear PCs to jets in the context of a global PDF fit has not
yet been attempted.

In this paper, we use the theory covariance matrix formalism developed
in~\cite{Ball:2018lag,Ball:2021icz}, recently validated in
Ref.~\cite{Ball:2025xgq}, to provide reliable estimates of the corrections and
theoretical uncertainties due to HT in DIS, and to PCs in single jet inclusive
cross-sections and dijet cross-sections. The analysis is performed entirely
within the open-source NNDPF
framework~\cite{NNPDF:2021uiq}. The paper is organised as follows. In
\cref{sec:methodology} the methodology is discussed in detail. In
\cref{sec:pc-tcm} we will present results, firstly for the HT, and then for PCs
to inclusive jets and dijets.  In \cref{sec:pheno}, we will include these
corrections and uncertainties in a global PDF fit, assess its impact on the PDFs
and relative luminosities, on the theoretical predictions for fundamental LHC
processes such as the Higgs boson production, and on the determination of the
strong coupling $\alpha_s$ from hadronic data. Finally, in
\cref{sec:conclusions} we summarise our main conclusions and provide information
on how to access the PDF fits with HT and PCs in LHAPDF
format~\cite{Buckley:2014ana}.

%% file: sec-methodology.tex
\section{Methodology for the Determination of Power Corrections}
\label{sec:methodology}

In this section, we review the theory covariance methodology we will use to
estimate the theoretical uncertainties associated with power corrections. This
is based on the general formalism developed in Ref.~\cite{Ball:2021icz}, where
it is shown how to account for correlations between theory uncertainties in the
PDFs and in the matrix element used in a prediction. When applied to parameters
of physical interest, the methodology allows for the extraction of such
parameters, together with their correlated uncertainties, in such a way as to
account for all the correlations with the underlying PDFs. As such, the
methodology was recently validated in a determination of the strong coupling
\cite{Ball:2025xgq}, by comparing results to those obtained using the more
cumbersome correlated replica method~\cite{Ball:2018iqk}, and by closure testing
\cite{NNPDF:2017mvq}. In the following, we provide a brief overview of the
theory covariance methodology before moving to its specific application to power
corrections.

\subsection{The Theory Covariance Method}
The accurate and precise determination of PDFs requires careful consideration of
the theoretical uncertainties associated with theoretical predictions.
Theoretical uncertainties and their correlations can be incorporated in the
fitting procedure through the implementation of a theoretical covariance matrix
$C^{\textrm{th}}$. If we assume that these uncertainties are Gaussian and
independent of the experimental uncertainties, the total covariance matrix $C$
can be written as the sum of the experimental $t_0$ covariance
matrix~\cite{Ball:2009qv} $C^{\textrm{exp}}_{t_0}$ and a theory covariance
matrix~\cite{Ball:2018lag,Ball:2018twp,Ball:2020xqw,NNPDF:2019vjt,NNPDF:2019ubu,NNPDF:2024dpb}
incorporating in particular all MHOUs and nuclear uncertainties
\begin{equation}
  C_{ij} = C^{\textrm{exp}}_{t_0,ij} + C^{\textrm{th}}_{ij} \,,
  \label{eq:ctot}
\end{equation}
where the indices $i,j$ run over all the experimental data points.

In the NNPDF framework, the uncertainties encoded in this covariance matrix are
propagated to the PDFs through a Monte Carlo approach. A set of
$N_{\textrm{rep}}$ data replicas $D^{(k)}$ is generated satisfying
\begin{equation}
    \langle D_i^{(k)}\rangle = D_i\,,\qquad
 \langle (D_i^{(k)}-D_i)(D_j^{(k)}-D_j)\rangle = C_{ij} \,,
  \label{eq:replicas}
\end{equation}
where $D_i$ denotes the central value of the experimental data, and the angle
brackets denote an average over data replicas, in the limit
$N_{\textrm{rep}}\to\infty$.

The metric $\chi^2$ used to optimise the set of PDF parameters ${\theta}$ is
then
\begin{equation}
  \chi^2 \left( {\theta}^{(k)}, D^{(k)}\right)
  = \frac{1}{N_{\textrm{dat}}}\sum_{i,j}^{N_{\textrm{dat}}}
  \left( T_i[{\theta}^{(k)}] - D_i^{(k)} \right)
  (C^{-1})_{ij}
  \left( T_j[{\theta}^{(k)}] - D_j^{(k)} \right) \,,
  \label{eq:chi2_exp_th}
\end{equation}
where $T_i$ is the theory prediction for data point $i$. The figure of merit in
Eq.~\eqref{eq:chi2_exp_th} ensures that the relative weights of the data points
during the fit account for both experimental and theoretical uncertainties
\cite{Ball:2018lag}. The $\chi^2$ in Eq.~\eqref{eq:chi2_exp_th} is then
optimised for each replica $k$ in the Monte Carlo ensemble, using a
cross-validation procedure to prevent overfitting. In terms of Bayesian
reasoning, minimization of the $\chi^2$ corresponds to the maximization of the
conditional probability of the predictions given the data
\begin{equation}
  P(T | D) \propto \exp
  \left(
    -\frac{1}{2} (T - D)^T C^{-1} (T - D)
  \right)\,,
  \label{eq:inference_theory}
\end{equation}
where matrix notation has been adopted for simplicity.

Following Ref.~\cite{Ball:2021icz}, additional sources of theoretical
uncertainty can be modelled as fully correlated shifts in the predictions
induced by a set of $N_{\T{nuis}}$ sources of theoretical uncertainty
\begin{equation}
  T({\lambda}) = T + \sum_{\alpha=1}^{N_{\T{nuis}}}\lambda_{\alpha} \beta_{\alpha} \,,
\end{equation}
where $\beta_{\alpha}$ are vectors (in the space of the data) of correlated
shifts, and $\lambda_{\alpha}$ is the associated nuisance parameter that
specifies the size of the shift. If we promote the nuisance parameters to
stochastic variables, the conditional probability in
Eq.~\eqref{eq:inference_theory} becomes
\begin{equation}
  P(T | D {\lambda}) \propto \exp
  \left(
    -\frac{1}{2} (T + \sum_{\alpha=1}^{N_{\T{nuis}}}\lambda_{\alpha} \beta_{\alpha}- D)^T
    C^{-1}
    (T + \sum_{\alpha=1}^{N_{\T{nuis}}}\lambda_{\alpha} \beta_{\alpha} - D)
  \right)\,.
  \label{eq:inference_theory_nuisance}
\end{equation}
Then by marginalising over the nuisance parameters, assuming for the latter a
multivariate Gaussian distribution centred on zero with unit variance, the
original expression in Eq.~\eqref{eq:inference_theory} is recovered, but with
$C$ replaced by $C+S$, where
\begin{equation}
  S = \sum_{\alpha=1}^{N_{\T{nuis}}}\beta_{\alpha} \beta_{\alpha}^T \,,
  \label{eq:theory_cov}
\end{equation}
is the theory covariance matrix for the $N_{\T{nuis}}$ additional sources of
theoretical uncertainty. Hence, we can determine the optimal PDF with these new
uncertainties by following the usual fitting procedure, but with $C$
replaced by $C+S$ in Eq.\eqref{eq:replicas} and Eq.\eqref{eq:chi2_exp_th}.

After determining this PDF, we can update the prior estimate of the nuisance
parameters using information from the fit. Using Bayes' theorem, the posterior
distribution of the nuisance parameters is
\begin{equation}
  P({\lambda} | T D) \propto
  \exp \left(
    \frac{1}{2}\sum_{\alpha,\beta=1}^{N_{\T{nuis}}} (\lambda_{\alpha} - \bar{\lambda}_{\alpha})
    Z^{-1}_{\alpha \beta}
    (\lambda_{\beta} - \bar{\lambda}_{\beta})
  \right)\,,
\end{equation}
which is a multivariate Gaussian distribution with mean
\begin{equation}
  \bar{\lambda}_{\alpha}(T, D) =
  - \beta_{\alpha}^T (C+S)^{-1}
  ( T - D)\,,
  \label{eq:nuis_par_post}
\end{equation}
and covariance
\begin{equation}
  Z_{\alpha \beta} = \delta_{\alpha \beta}
  - \beta_{\alpha}^T (C+S)^{-1} \beta_{\beta} \,.
\end{equation}
Note that Eq.~\eqref{eq:nuis_par_post} depends on the theoretical predictions,
which in turn depend on the PDFs. In the context of PDF fits, theoretical
predictions are thus statistical variables. In NNPDF the ensemble of PDF
replicas is used as a proxy for the underlying distribution of PDFs, so the
marginalization of the theoretical predictions over the nuisance parameters is
performed by averaging the nuisance parameters over the PDF replicas:
\begin{equation}
  \bar{\lambda}_{\alpha} = \langle \bar{\lambda}_{\alpha}(T[\theta^{(k)}, D)\rangle
  = -\beta_{\alpha}^T (C+S)^{-1}
  (T^{(0)} - D )\,,
  \label{eq:nuis_par_post_cntrl}
\end{equation}
where the central replica $T^{(0)} = \langle T[\theta^{(k)}]\rangle$. Likewise,
the covariance matrix of the nuisance parameters is
\begin{equation}
   \bar{Z}_{\alpha \beta}
  = \delta_{\alpha \beta} -
    \beta_{\alpha}^T (C+S)^{-1}\beta_{\beta} + \beta_{\alpha}^T (C+S)^{-1}
      X (C+S)^{-1} \beta_{\beta} \,,
   \label{eq:nuis_par_cov}
\end{equation}
where $X$ is the covariance matrix of the PDF uncertainties of the theoretical
predictions, defined as
\begin{equation}
  X = \langle (T^{(k)} - T^{(0)}) (T^{(k)} - T^{(0)})^T\rangle\,.
  \label{eq:Xdef}
\end{equation}
This contains the experimental uncertainties in the data, all the theoretical
uncertainties of the theory covariance matrix used in the fit (including the new
sources of uncertainty), and the uncertainty due to the methodology in the PDF
fit.

From Eq.~\eqref{eq:nuis_par_post_cntrl} we see that the nuisance parameters are
informed by the fit, and in particular by the agreement between predictions and
data weighted by the total covariance matrix. This is the information that the
data provide on the nuisance parameters that is not absorbed by the PDFs.
This is also evident from the expression for the covariance in the space of nuisance
parameters, Eq.~\eqref{eq:nuis_par_cov}. This consists of three terms: the first
is the prior, and the second gives the reduction in uncertainty due to the
information from the data, assuming fixed PDFs. The third
term then gives the increase in uncertainty due to the uncertainty in the PDFs.
Note that
\begin{equation}
  0 < Z_{\alpha \beta} \leq \bar{Z}_{\alpha \beta} \,,
  \label{eq:nuis_par_cov_ineq}
\end{equation}
since $(C+S)^{-1}X (C+S)^{-1}$ is positive semi-definite. Therefore, the nuisance
parameters receive an extra uncertainty due to the PDF fit itself: there is less
information on the theoretical uncertainties from the data because part of it is
used in the fit to the PDFs. Note that Eq.~\eqref{eq:nuis_par_cov_ineq} implies
that when the PDF uncertainty $X$ is large, then the covariance of the nuisance
parameters can become larger than the prior.

We now show how the formalism above can be adapted for the present study. We
model the shifts to predictions due to power corrections through a functional
form that depends on a set of $N_h$ parameters $h_{\alpha}$. The details of the
functional form depend on the physical process under consideration, and will be
discussed in the next subsection. What is important here is that the power
correction induces a shift in the theoretical predictions which is linear in
$h_\alpha$:
\begin{equation}
  T(\{h\}) = T +
  \sum_\alpha h_{\alpha}T^\prime_\alpha
  \equiv
  T + \sum_\alpha \lambda_{\alpha} \beta_{\alpha} \,,
  \label{eq:linearised_shift}
\end{equation}
so if we choose for each $\alpha$
\begin{equation}
  \beta_{\alpha} = \delta h_{\alpha}T^\prime_\alpha \,,\qquad
  h_{\alpha} = {\lambda_{\alpha}}{\delta h_{\alpha}}\,.
  \label{eq:beta_h}
\end{equation}
and take the prior distribution of the nuisance parameters as unit Gaussians
centred on zero, as before, then $h_{\alpha}$ are also Gaussian variables
centred on zero, but with prior covariance $H^{\rm prior}_{\alpha\beta}=(\delta
h_{\alpha})^2\delta_{\alpha\beta}$.

The theory covariance matrix $S$ to be used in the fit will then be constructed
as in Eq.~\eqref{eq:theory_cov} using the shifts in Eq.~\eqref{eq:beta_h}. Note
that since the corrections in Eq.~\eqref{eq:linearised_shift} are linear in the
nuisance parameters, there is no need (unlike in
Refs.~\cite{AbdulKhalek:2019bux,AbdulKhalek:2019ihb,NNPDF:2024dpb,Ball:2025xgq})
to consider contributions from positive and negative shifts separately, nor to
consider complicated prescriptions: Eq.~\eqref{eq:theory_cov} is perfectly
sufficient. After performing the fit including the new theory covariance matrix,
we can compute the posterior for the nuisance parameters using
Eq.~\eqref{eq:nuis_par_post_cntrl}, and their covariance using
Eq.~\eqref{eq:nuis_par_cov}, which after the rescaling \eqref{eq:beta_h} gives
the posterior parameters $h^{\rm post}_{\alpha}$ and their covariance
\begin{equation}
  {h}^{\rm post}_{\alpha}  =\delta h_{\alpha}
  \bar\lambda_{\alpha}\,,\qquad \T{H}^{\rm post}_{\alpha\beta}  =\delta
  h_{\alpha}\delta h_\beta
  \bar{Z}_{\alpha\beta}\,,
  \label{eq:pc_ht_post}
\end{equation}
and thus the power correction. This technique can be applied equally to
determine HT corrections to DIS and power corrections to hadronic observables,
as we now discuss.

\subsection{Modelling Higher Twist Corrections in DIS}

Power corrections in predictions for DIS observables arise from higher
twist terms (twist four and above) in the operator product expansion, which are
suppressed by powers of $Q^2$ compared to the leading twist (twist two).
Although HT terms can be factorised into perturbative coefficients functions and
non-perturbative generalised parton distribution functions, the latter are at
present unknown. The HT terms are thus a source of systematic uncertainty that affects
all DIS theoretical predictions, becoming particularly relevant at low $Q^2$ and
large $x$ (low invariant mass of the final state $W^2$). Here we will only consider twist four terms, of
$O(\Lambda^2/Q^2)$. Where these are small, it is reasonable to assume that twist
six terms of $O(\Lambda^4/Q^4)$ will be negligible.

The DIS data we consider are structure function data taken on proton $F_2^p$ and
deuteron $F_2^d$  targets, ratio data $F_2^d/F_2^p$, reduced cross-sections on
proton targets, both neutral current (NC) and charged current (CC), and CC
reduced cross-section on nuclear targets. We do not include DIS data on charm or
bottom production in this study, nor data on DIS+jets.

There are two distinct approaches to modelling higher twist corrections to theoretical
predictions for DIS data: the additive approach
\cite{Alekhin:1998df,Alekhin:1999af,Blumlein:2008kz,Alekhin:2012ig,Alekhin:2017kpj,Alekhin:2022tip},
and the multiplicative approach
\cite{Virchaux:1991jc,Martin:1998np,Martin:2003sk,Ball:2013gsa,Thorne:2014toa}.
In the additive approach one
ignores all perturbative corrections to the HT. In the multiplicative approach,
one first factors out the leading twist contribution, so in effect one assumes
that the HT evolves perturbatively in the same way as the leading twist. Both
approaches are thus approximate, and which one adopts is largely a matter of
taste (see Ref.~\cite{Cerutti:2025yji} for a recent phenomenological
comparison).

We consider HT corrections as multiplicative shifts to DIS observables,
which comprise measurements of structure functions $F_2$ and reduced 
cross-sections. We model the shifts as functions of Bjorken-$x$, to be determined
empirically. Moreover, given that in principle HT can depend on nuclear effects,
the shifts for proton and deuteron targets are parametrised independently
\cite{Cerutti:2025yji}. Hence, we write the measured structure functions as
\begin{eqnarray}
  F^p_2(x, Q^2)
  &= F^{p, \textrm{LT}}_2(x, Q^2)
  \left(1 + \frac{H_2^p(x)}{Q^2}  \right)\,,\\
    F^d_2(x, Q^2)
  &= F^{d, \textrm{LT}}_2(x, Q^2)
  \left(1 + \frac{H_2^d(x)}{Q^2}  \right)\,.
  \label{eq:pc_shift_sf}
\end{eqnarray}
Since the corrections are multiplicative, we can correct the HERA NC
reduced cross-sections using the same factor:
\begin{equation}
  \sigma_{\rm NC}(x, Q^2)
  = \sigma^{\textrm{LT}}_{\rm NC}(x, Q^2)
  \left(1 + \frac{H_2^{p}(x)}{Q^2}  \right)\,.
  \label{eq:pc_shift_nc}
\end{equation}
Here we make the simplifying assumption that higher twist corrections to $F_L$
is the same as for $F_2$, since the data are not good enough to determine them independently. For ratio data, we have
\begin{equation}
\frac{F_2^d}{F_2^p}
= \left( \frac{F_2^d}{F_2^p} \right)^{LT} \left(1+\frac{H_2^{(d)}-H_2^{(p)}}{Q^2 + H_2^{(p)}}\right)\,.
\label{eq:pc_shift_rat}
\end{equation}
We have checked that the denominator in the HT correction can be replaced with
$Q^2$, consistent with our assumption that twist six terms (terms of
$O(\Lambda^4/Q^4)$) can be ignored.

Finally, DIS CC cross-sections are corrected similarly, by introducing a common
multiplicative shift for CHORUS, NuTeV, and HERA CC data regardless of the
nuclear target used in the experiment. Thus we will assume that CC $F_2$, $F_3$
and $F_L$ are all corrected by the same factor, and furthermore that nuclear
effects
at higher twist in CC data are similar to those included at leading twist
\cite{Ball:2018twp}. Hence, the
shift for any CC cross-section will be written as
\begin{equation}
  \sigma_{\rm CC}(x, Q^2)
  = \sigma^{\textrm{LT}}_{\rm CC}(x, Q^2)
  \left(1 + \frac{H_{\rm CC}(x)}{Q^2}  \right)\,.
  \label{eq:pc_shift_cc}
\end{equation}

The various functions $H(x)$ that model the shifts are parametrised as piecewise
linear functions, interpolating between a grid of points in $x$. For the present
analysis, points are placed at
$$x \in [0.0, 0.001, 0.01, 0.1, 0.3, 0.5, 0.7,0.9, 1]$$ for all structure
functions and targets. We further assume that $H(1)=0$, but make no assumption
about $H(0)$. We thus write
\begin{equation}
  H (x) = \sum_{\alpha=1}^{8} h_{\alpha} f_{\alpha}(x) \,,
  \label{eq:linear_shift}
\end{equation}
where the $f_{\alpha}(x)$ are piecewise linear interpolating functions. The
advantages of having a linear dependence on the parameters $h_{\alpha}$ is
twofold: it has a simple physical interpretation of the parameters used in the
interpolation, and it simplifies the error propagation since
Eq.~\eqref{eq:linearised_shift} becomes an exact result. The vectors
$T_\alpha^\prime$ are then simply proportional to the vectors $T$, that is
$T_\alpha^\prime = T f_\alpha(x)/Q^2$.

Finally, it is important to note that we consider the shifts to be fully
correlated amongst all the observables that share the same HT correction. So,
for example, a shift to $F_2^p$ will affect all the observables that depend on
$F_2^p$, whether they are pure structure functions, the NMC ratio, or the NC
reduced cross-sections.

\subsection{Modelling Linear Power Corrections for Jet and Dijet Cross-sections}

In NNPDF4.0 we include data on both single-inclusive jet and dijet
cross-sections, from ATLAS and CMS. These are compared to factorised
perturbative calculations of the form (for a single-inclusive jet cross-section)
\begin{equation}
  \left[\frac{d\sigma_{\T{jet}}}{dp_T dy}\right]_{\text{pert}} = \sum_{p,q} \int dx_1 dx_2
 f_p(x_1) f_q(x_2)
  \frac{d\hat{\sigma}_{pq \rightarrow \T{jet}}}{dp_T dy}.
  \label{eq:fctrs_jets_xsec}
\end{equation}
Here $f_p(x_1)$ and $f_q(x_2)$ are the PDFs of the incoming partons $p$ and $q$,
and $d\hat{\sigma}_{pq \rightarrow \T{jet}}/dp_T dy$ is the partonic
cross-section computed at NNLO in perturbative QCD. MHOUs estimated from scale
variations have been accounted for in a variant of NNPDF4.0
fits~\cite{NNPDF:2024dpb}. While it has been understood for some time that there
are linear power corrections to these processes, from a wide variety of sources,
in global PDF fits these are generally ignored.

There are many factors that influence the measurement of jet cross-sections,
mainly to do with the way the jets and their associated energy scales are
defined. For instance, the algorithm used to identify the jets introduces a
dependence on the radius of the jet $R$. Furthermore, assuming that jets have
been faithfully reconstructed, there is a variety of physical phenomena, not all
belonging to the domain of perturbative calculations, that can affect the
measured energy of the jet. In other words, the energy of the jet does not
necessarily match the energy of the hard parton that initiated the jet. For
example, the underlying event originates from uncontrolled interactions in
the initial proton-proton collision (multiple parton interactions and colour
interactions), as well as contributions from initial- and final-state radiation.
As a results, the measured energy of the jet is shifted due to the external
contribution from the underlying event that flows into the jet. Investigations to disentangle
the underlying event contribution from single-inclusive jets exploiting Monte Carlo event
generators have been performed by ATLAS~\cite{ATLAS:2014iez} and
CMS~\cite{CMS:2013ycn}. On the other hand, in the process of hadronization, some
of the jet energy leaks out from the jet cone, reducing the energy that is
eventually measured. This process is modelled through the use of final state
Monte Carlo generators. Experimentalists can correct for these effects in
their measurements, by computing a correction factor 
with associated systematic uncertainties.

However in addition to all this, the theoretical calculation of the perturbative
jet cross-section is also subject to non-perturbative uncertainties, resulting in
power corrections of $O(\Lambda/p_T)$ to the parton level transverse momentum
and thus to the perturbative cross-sections. These genuine non-perturbative
effects are by their nature very difficult to disentangle from the
quasi-partonic effects estimated by using final state Monte Carlos,
and thus, as with higher twist, an empirical approach is the most practical way
to assess them quantitatively.

Whatever the source of the effect, a change in energy results in a shift $\delta
p_T$ to the fixed parton level transverse momentum $p_T$ of the jet. The
correction $\delta p_T$ has been estimated in
Refs.~\cite{Dasgupta:2007wa,Dasgupta:2008di, Cacciari:2007aq}. Since jet
cross-sections fall very steeply with $p_T$ (typically as $\sim p_T^{-n}$, with
$n$ as large as six or seven), such a shift results in a multiplicative power
correction of the form $1+n\delta p_T/p_T$, which may be large simply because
$n$ is large~\cite{Olness:2009qd}.

We will thus model the shift in the single-inclusive jet cross-section induced
by all these various power corrections as
\begin{equation}
  \frac{d\sigma_{\T{jet}}}{dp_T dy} = \left[ \frac{d\sigma_{\T{jet}}}{dp_T dy} \right]_{\text{pert}}
  \biggl( 1 + \frac{H_j(y)}{p_T} \biggr) \,,
  \label{eq:shifted_jet_xsec}
\end{equation}
where the perturbative contribution is computed in perturbative QCD using the
NNLO factorization \eqref{eq:fctrs_jets_xsec} and the associated MHOU. We allow the
function $H(y)$ to depend on the jet rapidity $y$, but not on the hard scale
$p_T$, just as in the DIS case where the HT functions depend on $x$ but not
$Q^2$. Note that for jets a multiplicative parametrization is mandatory, since,
unlike for DIS, the perturbative cross-section is falling steeply
as the hard scale increases.

Similar arguments apply to dijet cross-sections, where the hard scale is now the
invariant mass $m_{jj}$ of the jets. The power corrections for the differential
cross-section for dijets can then be modelled as
\begin{equation}
  \frac{d\sigma_{\T{2jet}}}{d\eta dm_{jj}} = \left[ \frac{d\sigma_{\T{2jet}}}{d\eta dm_{jj}} \right]_{\text{pert}}
  \biggl( 1 + \frac{H_{jj}(\eta)}{m_{jj}} \biggr) \,,
  \label{eq:shifted_2jet_xsec}
\end{equation}
where $\eta$ is either the rapidity separation $y^*=|y_1-y_2|/2$ (in ATLAS) or
the maximum absolute rapidity $y_{\T{max}} = \rm{max}(|y_1|,|y_2|)$ in CMS.

The functions $H_j(y)$, $H_{jj}(\eta)$ that model the shifts are parametrised as
piecewise linear functions just as for DIS (see Eq.~\eqref{eq:linear_shift}), though
here we only use three points in $\eta$,
$$\eta \in [0.0,1.5,3.0]$$ since the rapidity dependence is expected to be
rather weak. We remark that while for single-inclusive jets we use the same
function $H_j(y)$ for both ATLAS and CMS, for dijets we must use different
functions, since the data are analysed using different rapidity variables.

Note that since the linear power corrections fall off much more slowly with the
hard scale than the HT corrections to DIS, they will be even more difficult to
disentangle from MHO perturbative corrections. This problem is compounded by
the fact that the N$^3$LO corrections to jet processes are unknown, to the
perturbative predictions are less precise than for DIS. On the other hand,
since as explained in~\cite{Olness:2009qd} the scale of the linear power
corrections may be relatively
large (several GeV rather than several hundred MeV), they may continue to be
significant up to relatively high scales. They can thus be significant, even
at the LHC.

%% file: sec-pctcm.tex
\section{Determination of Higher Twist and Jet Power Corrections}
\label{sec:pc-tcm}
In this section, we first discuss the determination of the HT corrections to
DIS, and then apply the same technique to the determination of PCs for single
inclusive jets and dijets. All these results will be obtained using the  theory covariance methodology described in the previous section.

The fits presented in this section and the next have been produced using the
open-source NNPDF framework~\cite{NNPDF:2021uiq} based on the same dataset and
theory parameters as used in NNPDF4.0~\cite{NNPDF:2021njg} with MHOUs as in
Ref.~\cite{NNPDF:2024dpb} and aN$^3$LO corrections as in
Ref.~\cite{NNPDF:2024nan}. Theory predictions have been generated with the
pipeline of Ref.~\cite{Barontini:2023vmr}, which supersedes the tools used for
NNPDF4.0. In particular, interpolation grids are generated using
\texttt{PineAPPL}~\cite{Carrazza:2020gss} with the DGLAP evolution computed
using EKO~\cite{Candido:2022tld} and DIS observables using the Yadism
package~\cite{Candido:2024rkr}. For a detailed discusion of the updated theory pipeline see App~A of Ref.~\cite{NNPDF:2024djq}.

\subsection{Determination of Higher Twist}
\label{sec:determination_ht}

The DIS measurements included in NNPDF4.0~\cite{NNPDF:2021njg} consist of
fixed-target NC structure function data from
SLAC~\cite{Whitlow:1991uw}, BCDMS~\cite{Benvenuti:1989rh}, and
NMC~\cite{Arneodo:1996kd,Arneodo:1996qe}, fixed-target inclusive and dimuon
CC cross-sections from CHORUS~\cite{Onengut:2005kv} and
NuTeV~\cite{Goncharov:2001qe,MasonPhD}, and collider NC and CC cross-sections
from HERA~\cite{Abramowicz:2015mha}.  In NNPDF4.0 (and indeed all earlier NNPDF
global fits) cuts are applied to all DIS data, $Q^2 > 3.49~{\rm GeV}^2$, $W^2 >
12.5~{\rm GeV}^2$, in order to avoid the kinematic regions where the HT is
expected to be significant.

All calculations of DIS processes in NNPDF4.0 include target mass corrections~\cite{Schienbein:2007gr},
following the implementation of Refs.~\cite{Georgi:1976ve,Ball:2008by}, heavy quark mass corrections
implemented using FONLL~\cite{Forte:2010ta,Ball:2011mu,Barontini:2024xgu}, fitted charm
\cite{Ball:2015tna,Ball:2015dpa,Ball:2016neh}, and nuclear corrections and
uncertainties, both for heavy nuclei~\cite{Ball:2018twp} and deuteron targets
\cite{Ball:2020xqw}.

\begin{figure}[t!]
  \centering
  \includegraphics[width=0.6\textwidth]{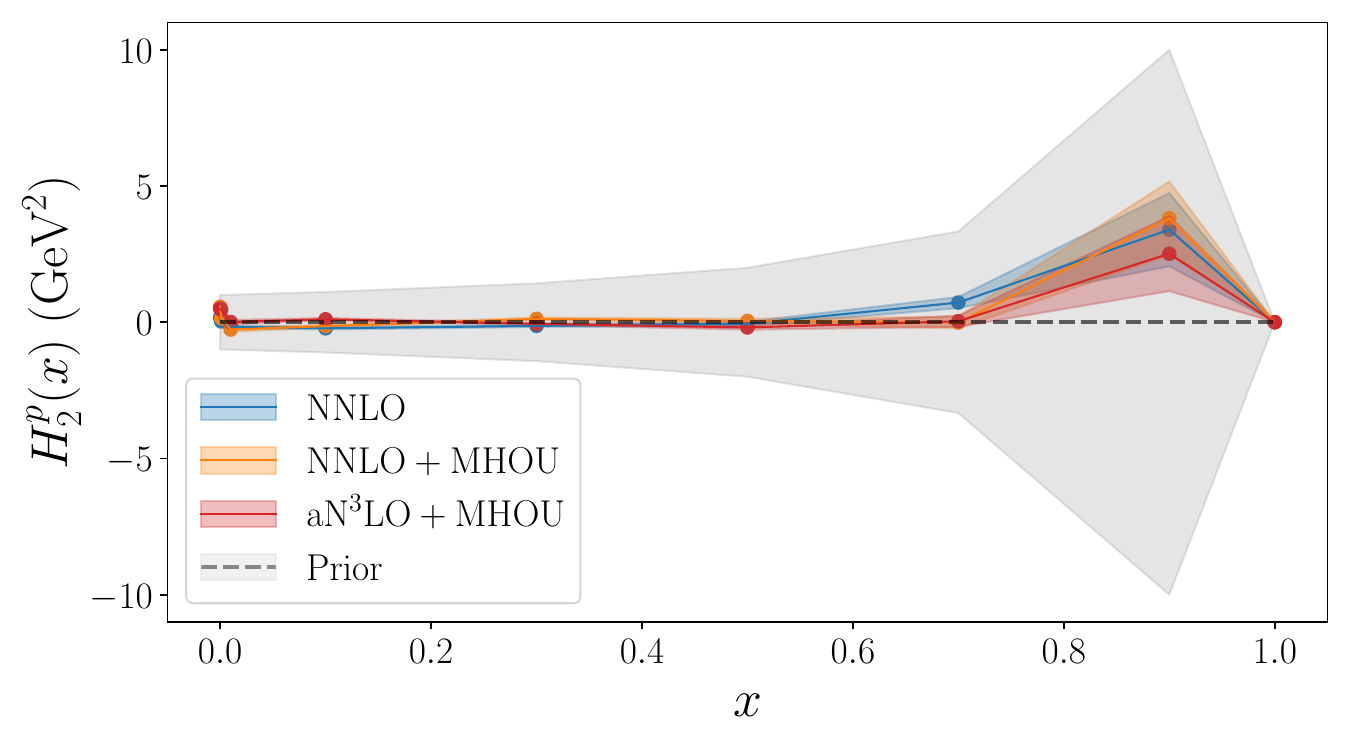}\\
  \includegraphics[width=0.6\textwidth]{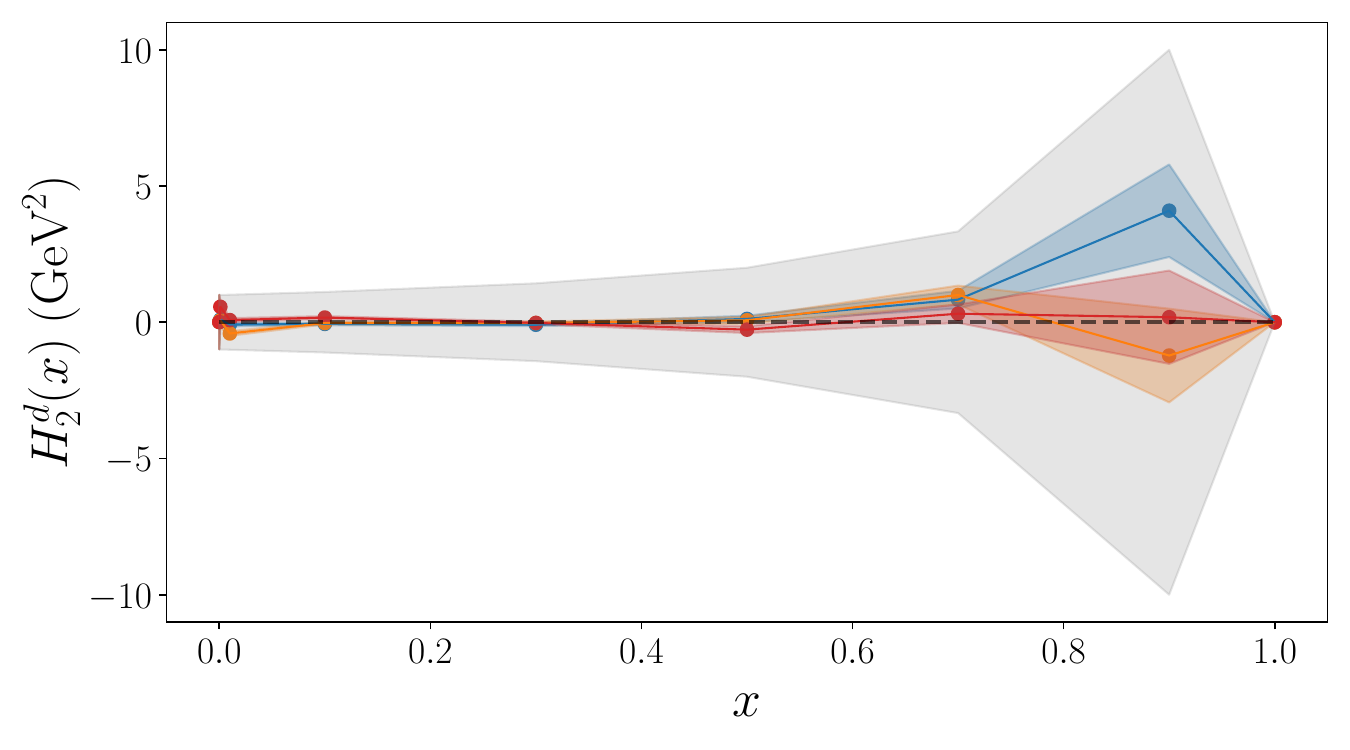}\\
  \includegraphics[width=0.6\textwidth]{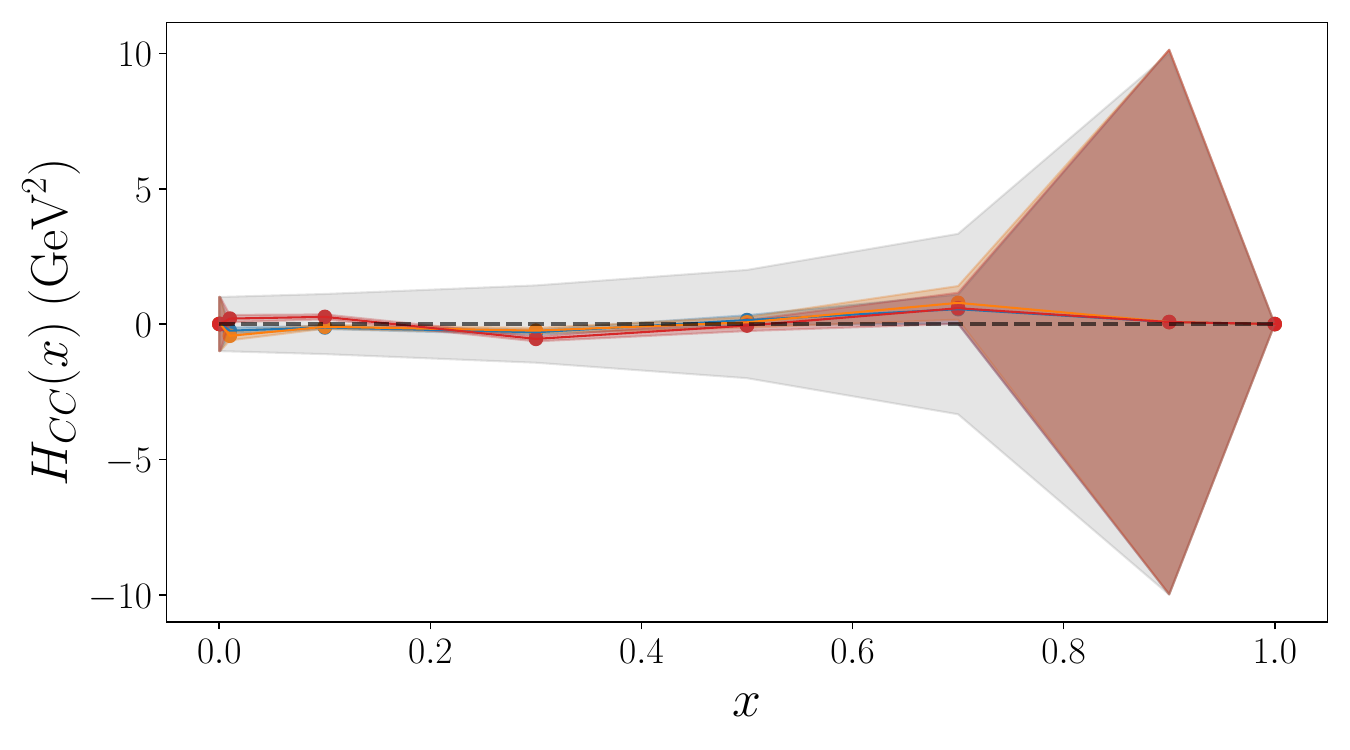}\\
  \caption{The three HT functions $H_2^p(x)$ (top), $H_2^d(x)$ (middle) and
    $H_{CC}(x)$ (lower). The prior uncertainty is shown in gray. The posterior
    distribution is then shown for HT determinations where the leading twist is
    computed at NNLO (blue), NNLO+MHOU (green) and aN$^3$LO+MHOU (red).}
  \label{fig:posterior_dis}
\end{figure}

We estimate the higher twist using the theory covariance methodology, as
described in \cref{sec:methodology}, constructing a prior theory covariance
matrix by computing the shifts in DIS predictions for a prior variation of the
HT parameters $h_\alpha$. Since our prior knowledge of the HT is limited, we
assume that the prior is centred on zero and allow the higher twist nodes
$h_\alpha$ to vary independently in the range $\pm 1~\rm{GeV}^2/(1-x)$. This
choice is motivated by the expectation that HT corrections are enhanced at
large-$x$ due to the fact that the hadronic mass of the final system approaches
threshold, which amplifies non-perturbative effects in this region. The prior
distributions of the three HT functions $H_2^p(x)$, $H_2^d(x)$ and $H_{CC}(x)$
are shown in Fig.~\ref{fig:posterior_dis}.

\begin{figure}[b!]
  \centering
  \includegraphics[width=1.0\textwidth]{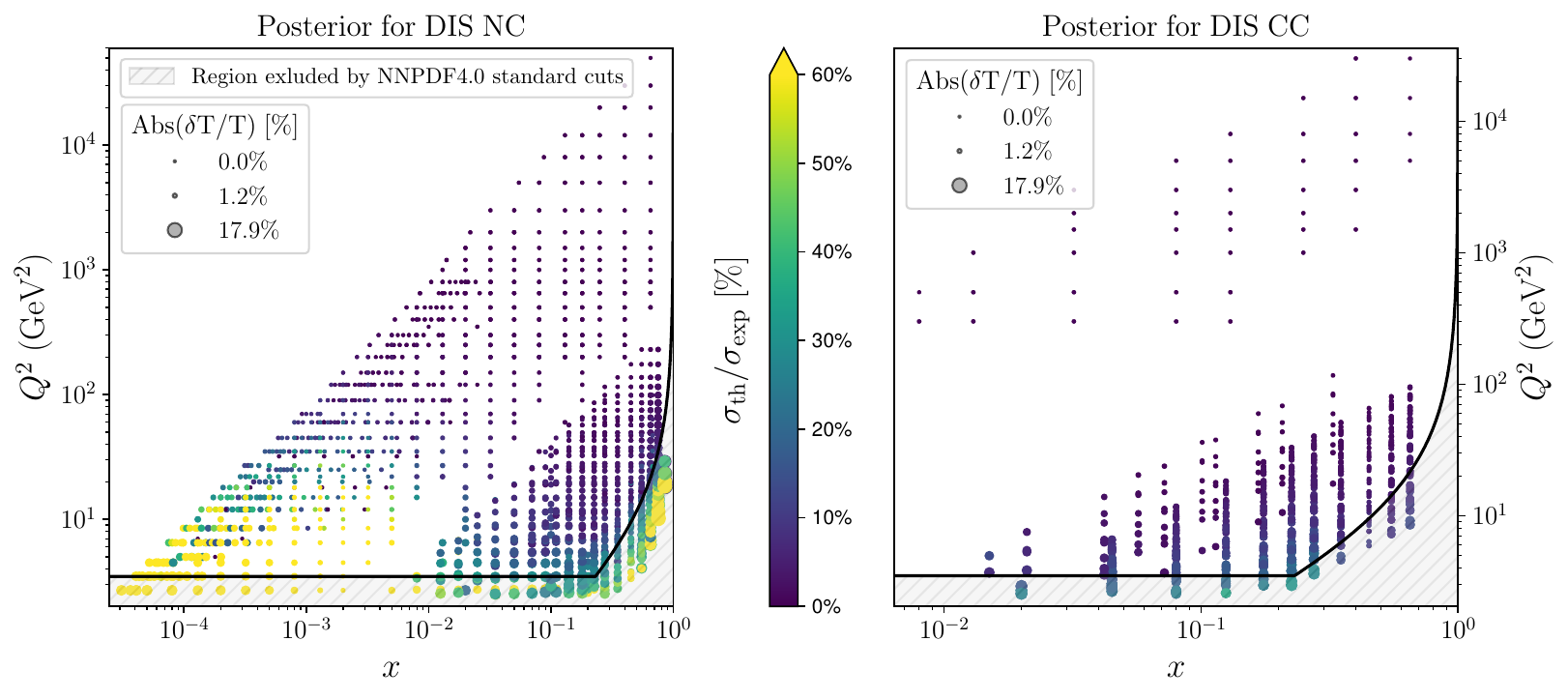}
  \caption{The $x,Q^2$ plane of the DIS data points used in the fits: the NC
    data (left) and CC data (right). The solid line indicates the position of the
    standard cut used in NNPDF4.0. The additional data points below this line
    are the data added when the cuts are lowered to determine the HT. The size
    of the points indicates the size of the the absolute value of the relative
    shift due to HT, determined in the NNLO fit. The colour code represents the
    ratio between the posterior uncertainty due to HT and the experimental
    uncertainty.}
  \label{fig:kin_dis_nnlo}
\end{figure}

We then add this prior theory covariance matrix to the covariance matrix
$C_{ij}$ used in NNPDF4.0, perform a new fit, and use the results of this fit to
extract posterior HT parameters. We then check that when the prior is varied (by
a factor of two in either direction), the posterior HT remains unchanged. To
maximise the sensitivity to HT corrections, we perform the fit in an extended
kinematic region that includes DIS data that we expect to be sensitive to the
non-perturbative effects that we are trying to capture. For that reason, we
lower the standard cuts to $Q^{2}> 2.5$~GeV$^{2}$ and $W^2 > 3.24$~GeV$^2$. The
impact of this lowered cut on the DIS dataset is shown in
Fig.~\ref{fig:kin_dis_nnlo}.

The HT determined in this way is shown in Fig.~\ref{fig:posterior_dis} for three
different sets of PDFs: NNLO, NNLO with MHOUs, and \annnlo with MHOUs. At NNLO
we find that the posterior distribution is significantly more constrained than
the prior, justifying the choice of prior. For the proton structure functions,
the HT correction $H_2^p$ is constrained to be very small at low and
intermediate $x$, consistent with zero within uncertainties, but increases
towards larger values of $x$, reaching a peak of about $4$~GeV$^2$. The deuteron
HT correction $H_2^d$ shows a similar pattern, although with slightly larger 
uncertainties. Note that, in both cases, the low-$x$ region is well
constrained by the data. The HT for CC observables, $H_{CC}$, shows a similar
pattern, but is less well constrained, particularly at very small $x$ and very
large $x$, where there is no CC data. In particular, at $x=0.9$ there is no
constraint at all, with the posterior equal to the prior. This said, we see
no evidence here for the expectation~\cite{Dasgupta:1996hh}, based on
renormalons, that HT effects for CC processes are larger than
for NC processes, or that they have a different shape (although it should be remembered that in our analysis sum rule constraints have been imposed only at leading twist).

\begin{figure}[t!]
  \centering
  \includegraphics[width=0.49\textwidth]{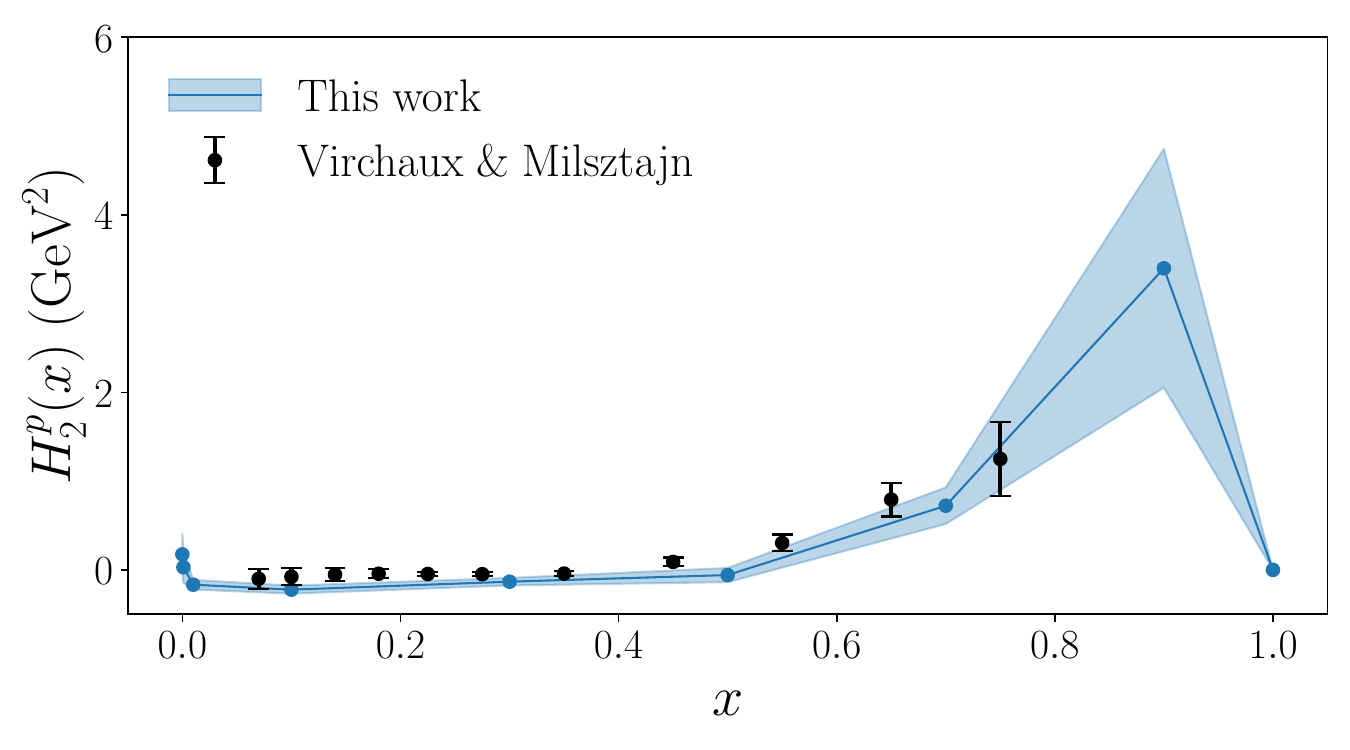}
  \includegraphics[width=0.49\textwidth]{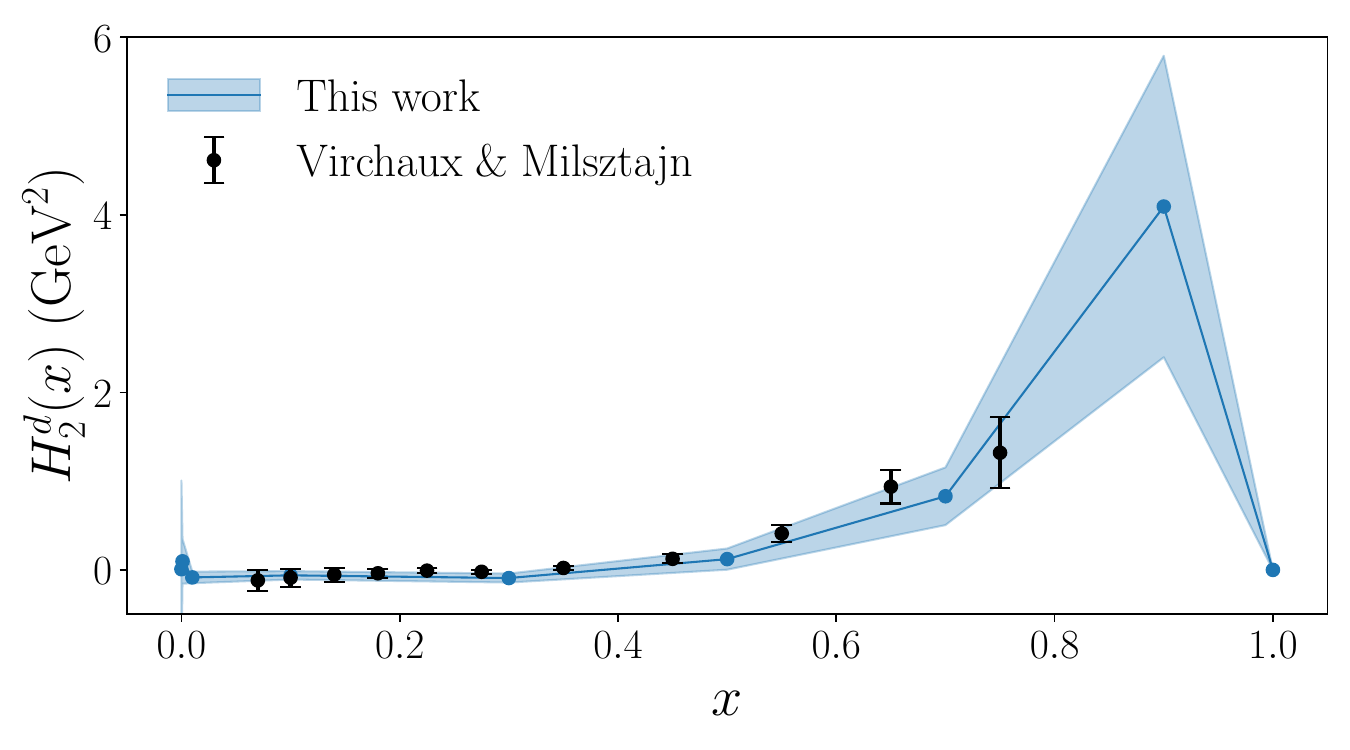}
  \caption{Comparison of our NNLO $F_2$ higher twist to that determined
    in protons and
    deuteron by Virchaux and Milsztajn in a NLO PDF fit to SLAC and BCDMS
    data~\cite{Virchaux:1991xp}. They include TMCs, and use the same
    multiplicative
    parametrization of the HT, fitting proton and deuteron data separately, so
    the results are directly comparable to our own.}
    \label{fig:abmp_bcdms_compare}
\end{figure}

In Fig.~\ref{fig:abmp_bcdms_compare} we show a comparison of the HT for the
$F_2$ structure function in proton and deuteron determined in our global NNLO
fit, with results obtained in an early PDF fit to just SLAC and BCDMS data by
Virchaux and Milsztajn~\cite{Virchaux:1991xp}. While this fit was only at NLO,
it used a similar multiplicative parametrization to our own, and target mass
corrections were included, so the comparison is meaningful. Considering the
fact that this fit was
performed before the advent of NMC and HERA data, and all the
technical improvements made over the intermediate thirty years, the
agreement is rather impressive,
though it should  not come as too much of a surprise since it is the same data
(SLAC and BCDMS) that dominates the large-$x$, low-$Q^2$ region today. Direct
comparisons with more recent determinations which use an additive HT
parametrization
\cite{Alekhin:1998df,Alekhin:1999af,Blumlein:2008kz,Alekhin:2012ig,Alekhin:2017kpj,Alekhin:2022tip}
are more difficult, since the results necessarily depend to some extent on the
choice of scale. However for these too, the results for HT correction to
$F_2^p$ are qualitatively consistent with our own, within uncertainties.

When we add MHOUs to our NNLO fit, the HT correction is generally reduced in size. The same
happens again when we go to \annnlo, where the MHOUs are significantly reduced. This
shows that some of the supposed HT being seen at NNLO is in fact due to MHOUs,
principally the N$^3$LO perturbative correction to the leading twist contribution.
This shows the difficulty in distinguishing phenomenologically between $Q^2$
dependence which is power like, and $Q^2$ dependence which is merely a power of
a logarithm. This effect was also seen when going from NLO to NNLO in earlier
attempts to determine HT~\cite{Martin:2003sk}. One is forced to conclude that
even the HT determined in the \annnlo fit may still be contaminated by a
perturbative contribution. Similarly, $H_2^d$ may be tangled with the deuteron
nuclear correction~\cite{Ball:2020xqw}, and $H_{CC}$ with the nuclear
corrections applied to much of the CC data~\cite{Ball:2018twp}.

The size and significance of the HT we determine is also displayed in the
kinematic plots in Fig.~\ref{fig:kin_dis_nnlo}, where the per-datapoint impact
of the posterior distribution for NC and CC datasets is shown (indicated by the
size of each data point), and its uncertainty compared to the overall
experimental uncertainty (indicated by the colour of each data point). For both
NC and CC corrections, we see that the relative shift of the HT exhibits the
expected behaviour, being largest at low $Q^2$ values where the non-perturbative
effects dominate, and also increasing in size at large $x$. Most importantly,
the size of the relative shifts increases significantly for both NC and CC DIS
datasets, particularly in the cut region, up to about ~10\%. This justifies the
cuts chosen in NNPDF4.0, designed precisely to eliminate data
significantly contaminated with HT contributions. For the
large-$x$ data that remain above the cut, the effect of the HT is small, and in
particular the uncertainty due to the HT is rather smaller than the (quite
large) experimental uncertainties. However, for the NC data from HERA at small
$x$, although the HT is small, the uncertainty due to the HT is a significant
fraction of the experimental uncertainty, even for some data above the cut. This
is because the HERA experimental uncertainties are themselves small. This effect
is reduced when MHOUs are added at NNLO, and further reduced at \annnlo. This
last observation should be borne in mind when considering the effect of HT on
the PDFs in the next section.

\subsection{Determination of Linear Power Corrections to Jets and Dijets}
\label{sec:determination_pc}

We now turn to the determination of power corrections for jet observables. The
jet cross-sections included in the NNPDF4.0 fit are double differential
single-inclusive jet data at $\sqrt{s} = 8$ TeV and dijet cross-sections at
$\sqrt{s} = 7$ TeV from ATLAS~\cite{ATLAS:2017kux,ATLAS:2013jmu} and
CMS~\cite{CMS:2016lna,CMS:2012ftr}, following a detailed study of the reliability and impact of the datasets~\cite{AbdulKhalek:2020jut}. The jets are
constructed in both cases using the anti-$k_T$ algorithm, and the data used
in the fit were those
with $R=0.6$ for ATLAS, $R=0.7$ for CMS.

The published datasets are at hadron
level. Hadronization, underlying event and multi-parton corrections were
estimated by the
experimentalists from the spread in several runs of final state Monte Carlos
with different
tunes and input PDFs, was included in the covariance matrix. However the
spread in the ATLAS estimates is very wide (consistent with zero), while
that from CMS is much narrower, and significantly nonzero, particularly at
low scale. Given this fact, it was decided in NNPDF4.0 not to correct the
central values of the data, but to implement only an overall hadronization
uncertainty as a contribution to the experimental covariance matrix.  
An internal cut on the jet $p_T$ was applied to the single-jet
measurements of CMS in order to avoid points for which $p_T \lesssim 70$~GeV,
where fixed-order calculations are not reliable. This cut is not necessary for
ATLAS single-jet data since the lowest $p_T$ bin starts at $70$~GeV. No
additional cuts are applied to dijet data, as the lowest value of the hard
scale is one
order of magnitude larger than that for the single-inclusive jet data.
All these considerations will become relevant in the discussion below.

As discussed in Sec.~\ref{sec:methodology}, we use the same methodology as for
the higher twist determination, but with a different prior. For jet observables,
the prior is still centred on zero, but given the expectation that $H(\eta)$
might be relatively large, we allow the shift to vary in the range $[-20.0,
20.0]$~GeV. As for the HT, we check that the results are independent of the
prior by varying it by a factor of two in either direction.

\begin{figure}[t!]
  \centering
  \includegraphics[width=0.6\textwidth]{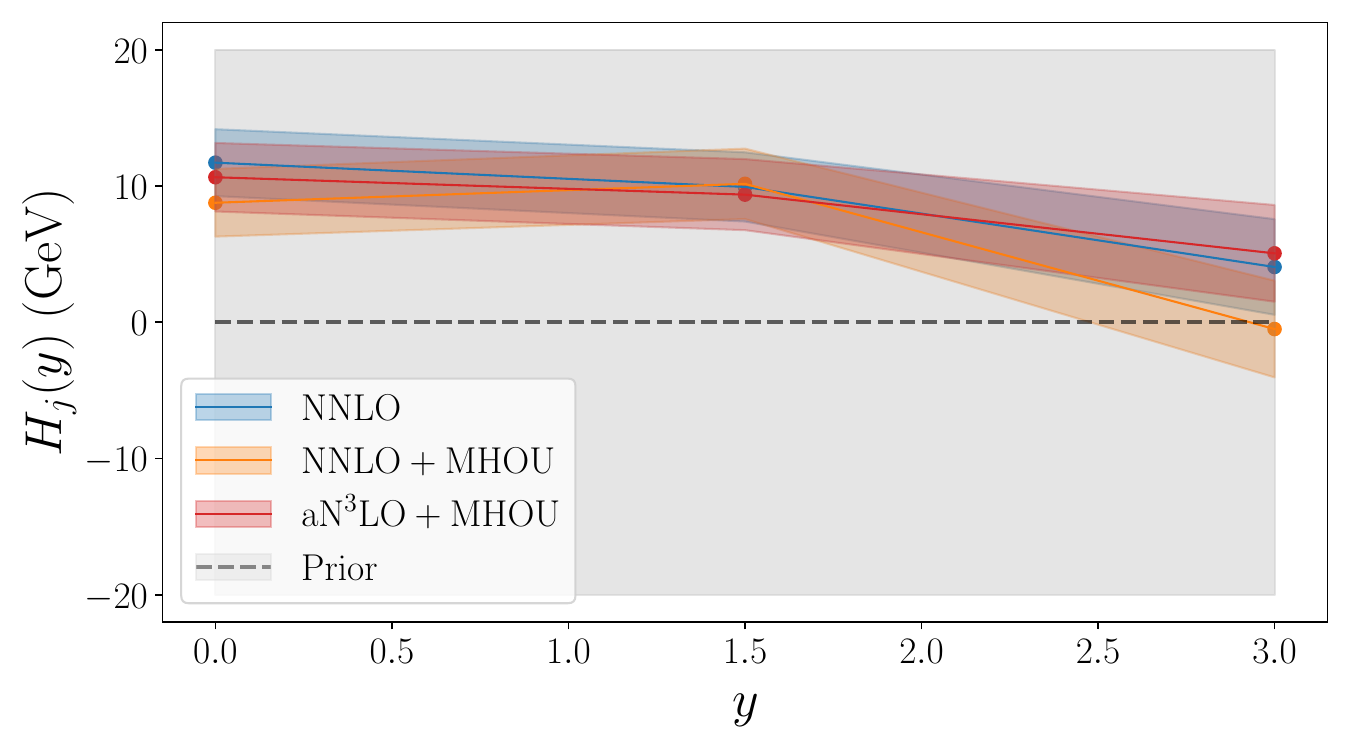}\\
  \includegraphics[width=0.45\textwidth]{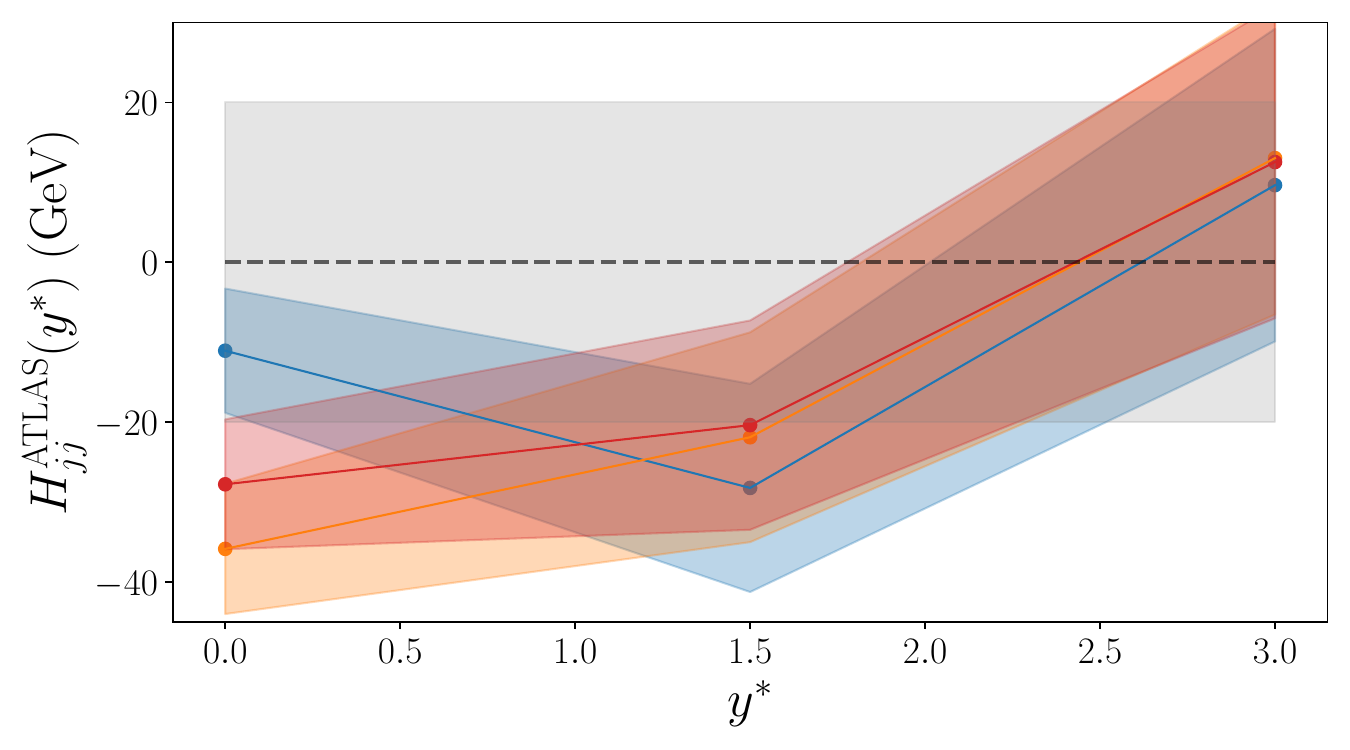}
  \includegraphics[width=0.45\textwidth]{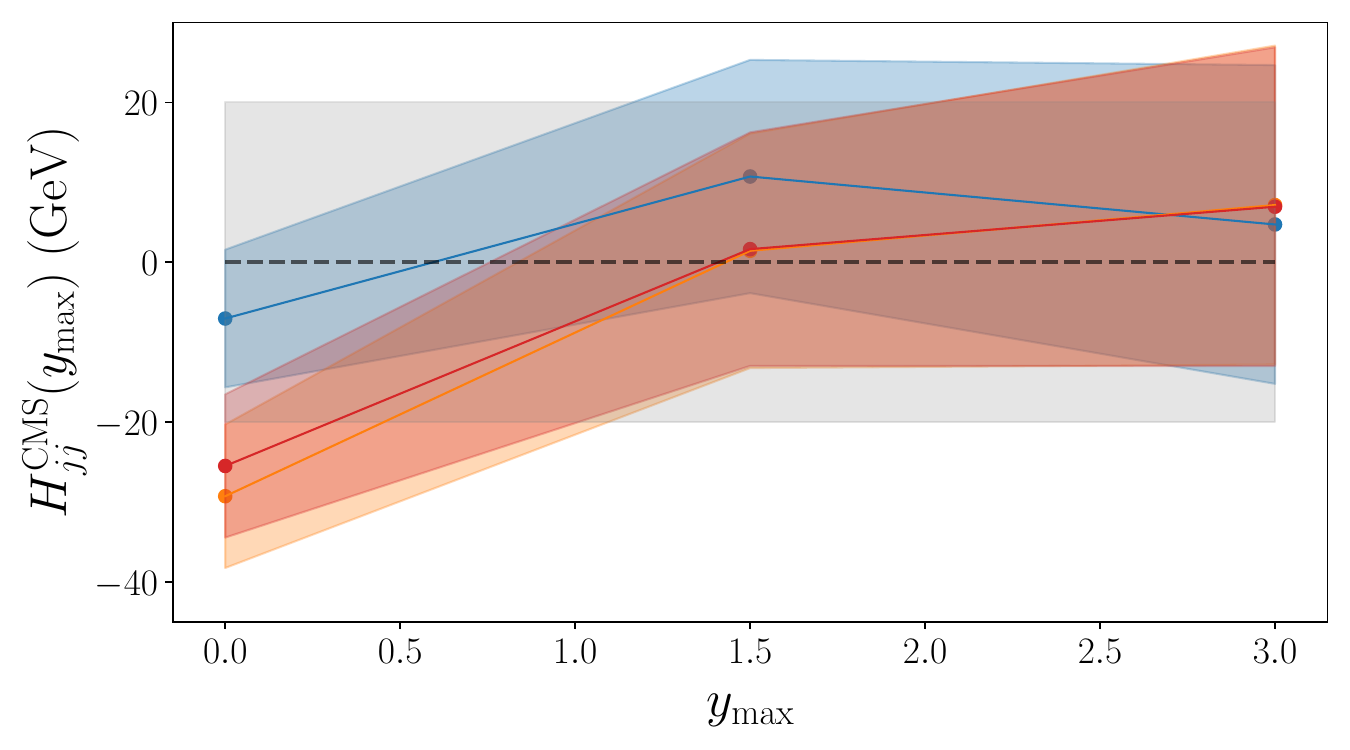}\\
  \caption{Same as Fig.~\ref{fig:posterior_dis}, but for single-inclusive jets
    (top), ATLAS dijet (lower left) and CMS dijet (lower right) observables.}
  \label{fig:posterior_jet}
\end{figure}

\begin{figure}[h!]
  \centering
  \includegraphics[width=1.0\textwidth]{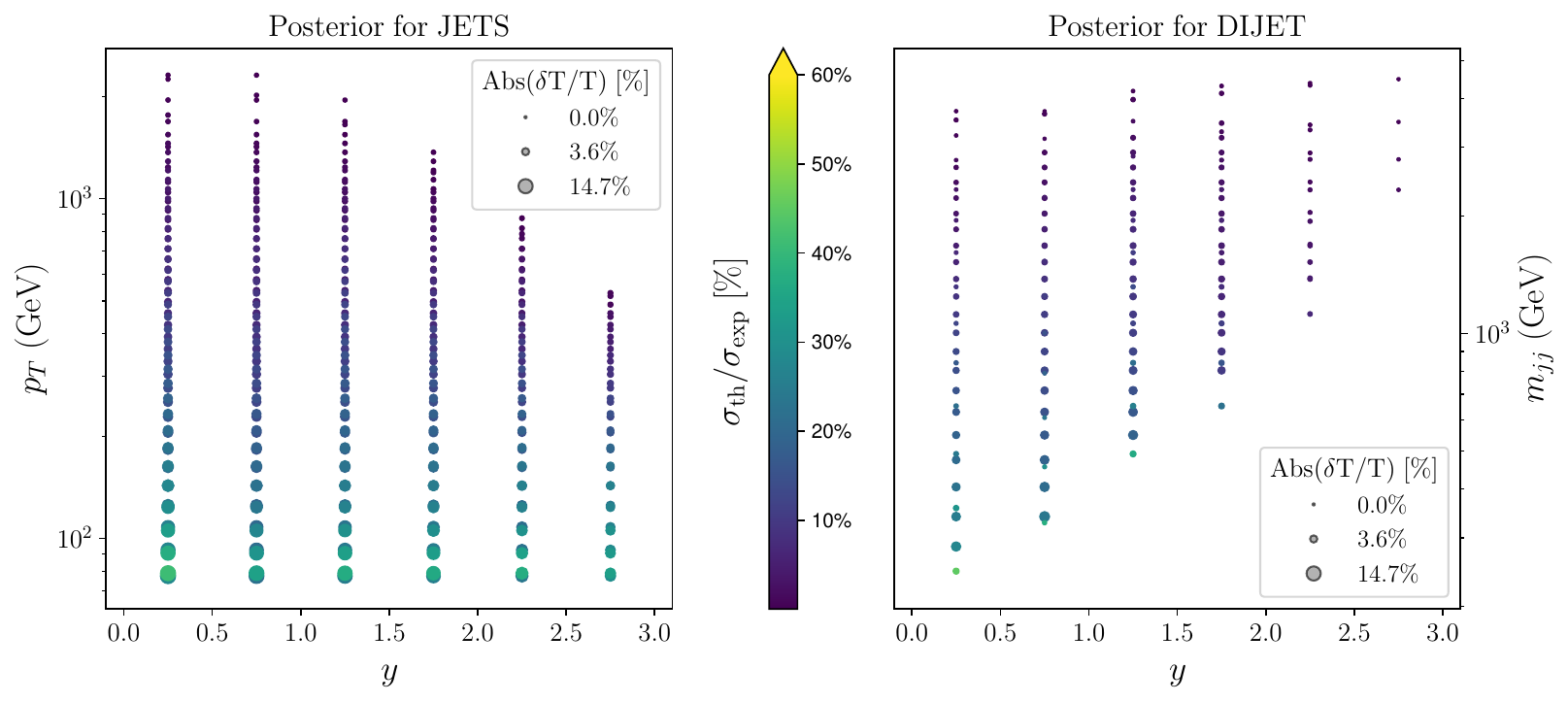}
  \caption{The $y,p_T$ plane of the single-inclusive jet data points (left) and
  the $y, m_{jj}$ plane of the dijet data points (right). For the dijet data, the ATLAS and CMS definitions of rapidity are both used on the same plot. As in
  Fig.~\ref{fig:kin_jet_nnlo}, the size of the points indicates the size of the
  the absolute value of the relative shift due to the PCs, determined in the NNLO
  fit. The colour code represents the ratio between the posterior uncertainty
  due to PCs and the experimental uncertainty.}
  \label{fig:kin_jet_nnlo}
\end{figure}

In Fig.~\ref{fig:posterior_jet} we present the posterior distribution for the
power corrections to single inclusive jets and dijets, and in
Fig.~\ref{fig:kin_jet_nnlo} the related kinematic distribution of the
data points. We
observe two distinct patterns for single-inclusive jets and dijets. Starting
with the single-inclusive jets, we find that not only is the prior significantly
constrained, but also that the posterior distribution is peaked around $10$~GeV.
Further, the posterior shows a weak dependence on the rapidity. It should be
noted that the previous study in Ref.~\cite{Olness:2009qd} estimated a
positive power
correction of about $7$~GeV for single-inclusive jet cross-sections, independent
of rapidity. Our result is consistent with this estimate.

For the dijets, we
find rather weaker evidence for a linear PC at central rapidity in the NNLO fit,
with uncertainties
remaining large. Note
that for the dijet
kinematics the bins with lowest invariant mass (and thus most likely to
be affected by PCs) are at central rapidity, and even here the lowest
invariant mass bin is still at a rather
higher scale than the lowest $p_T$ scale of the single inclusive jet data.

\begin{figure}[h!]
  \centering
  \includegraphics[width=0.45\textwidth]{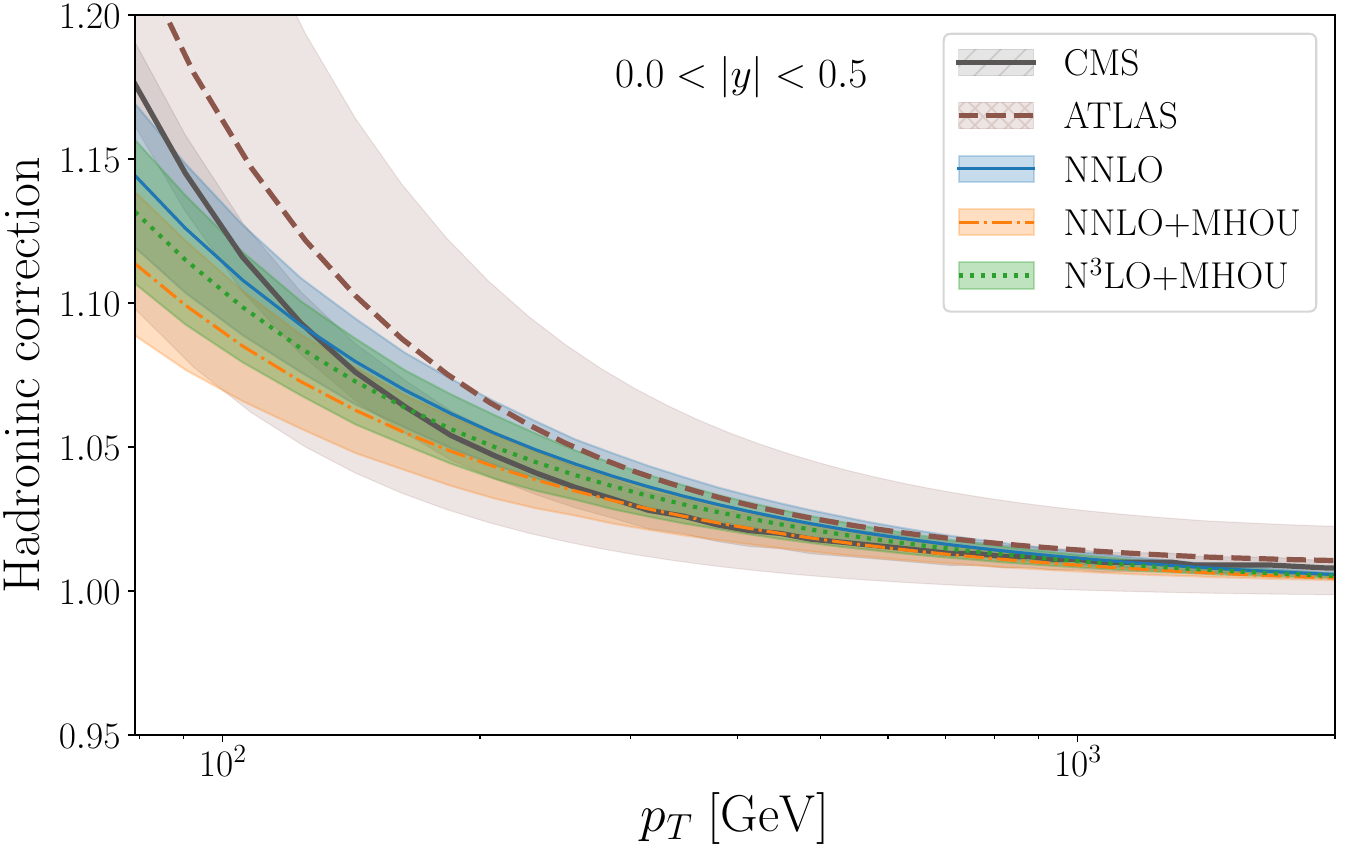}
  \includegraphics[width=0.45\textwidth]{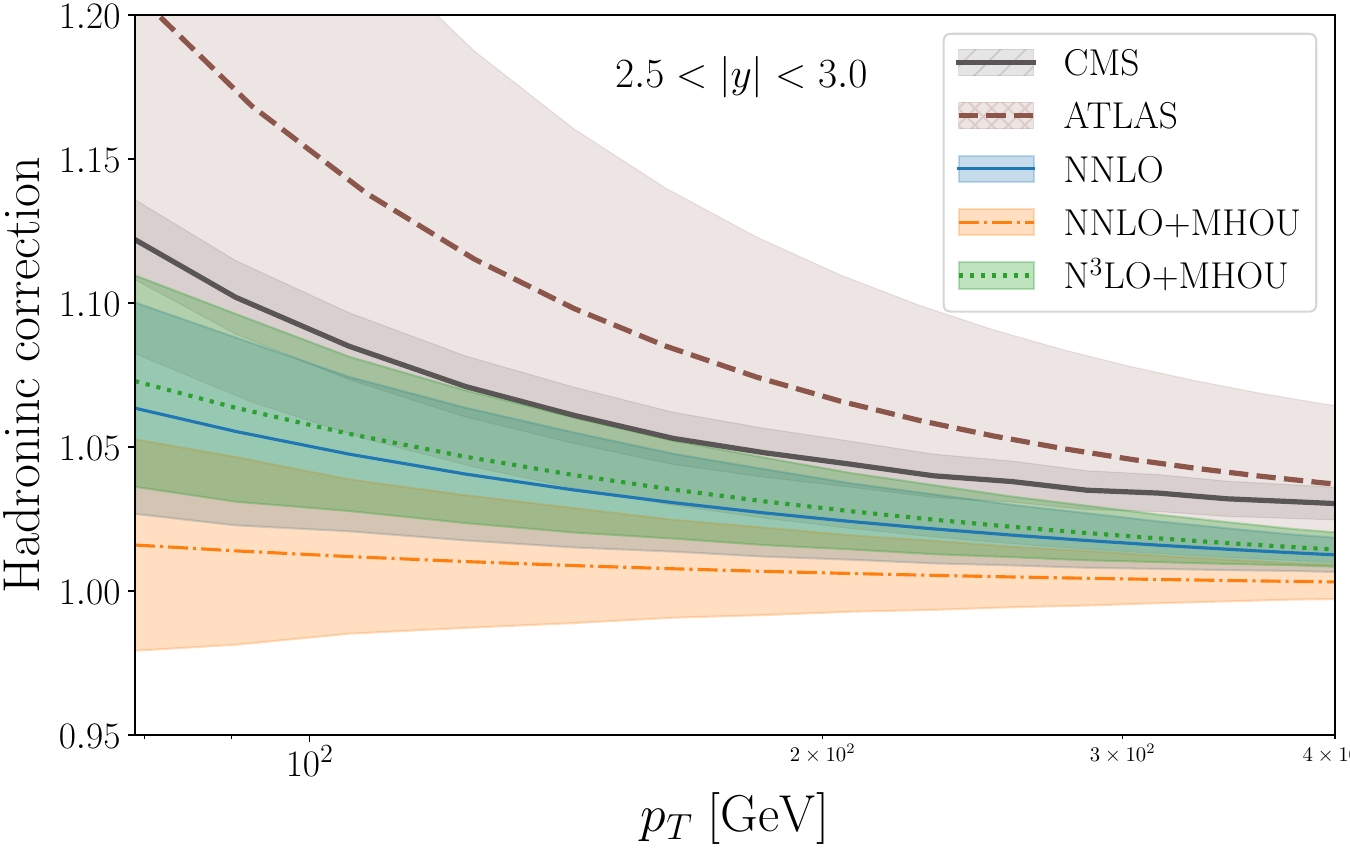}
  \caption{The hadronization correction as a function of $p_T$ for the $8$ TeV single inclusive jet data, for $0<|y|<0.5$ (left) and  $2.5<|y|<3.0$ (right). Our empirical estimates for the linear power correction at NNLO, NNLO+MHOU and N3LO+MHOU are compared to the ATLAS and CMS estimates obtained using final state Monte Carlo generators~\cite{ATLAS:2017kux,CMS:2016lna}.}
  \label{fig:NP_corrn}
\end{figure}

There is perhaps a small reduction in the size of the PC to single-inclusive
jets when MHOUs are added at NNLO, particularly at large rapidities, but in the
\annnlo fit the result is entirely consistent with NNLO. This is not too
surprising, since the \annnlo fit does not include an N$^3$LO correction to the
perturbative cross-section, since such corrections are at present unknown.
However for the dijets, adding MHOUs or going to an \annnlo fit seems to give a
more negative power correction at central rapidities (for both ATLAS and CMS),
even though the uncertainties remain large. It is possible that this large
negative PC at central rapidity will be cancelled, at least in part, by the (as
yet unknown) N$^3$LO correction to the perturbative cross-section.

The size of the shifts due to the linear PC are shown in
Fig.~\ref{fig:kin_jet_nnlo} by the size of the data points. Clearly the power
corrections to jets fall off slowly with $p_T$, as expected from the $1/p_T$
scaling behaviour, but can be as large as about $14\%$ for the lowest $p_T$
values (around $75$ GeV) of the single-inclusive jet data, across all
rapidities. The uncertainty due to the linear PC is quite a modest fraction of
the experimental uncertainty, even at these low scales, since the experimental
uncertainty is already quite large. The dijet data start at invariant dijet
masses of around $240$ GeV, and this only for the bins at central rapidity: for
the large rapidity bins the invariant mass is always above a TeV. This explains
why at the LHC dijet data are less vulnerable to power corrections than
single-inclusive jet data. The underlying reason for this is kinematic: it is
easier to identify unambiguously the hard scale for dijets than for
single-inclusive jets.

Finally, it is interesting to compare these empirical estimates of linear power
corrections Eq.~\ref{eq:shifted_jet_xsec} with the estimates of hadronization,
underlying event and multi-parton corrections made by ATLAS and CMS using final
state Monte Carlos, which make no assumption regarding the $p_T$ dependence of
the correction. The comparison for the CMS $8$ TeV single inclusive jet data
\cite{ATLAS:2017kux,CMS:2016lna}, for both central rapidity and high rapidity
bins, is shown in Fig.~\ref{fig:NP_corrn}. At central rapidity the agreement is
very good, though at low $p_T$ the Monte Carlo estimates rise more steeply than
our own, while at high $p_T$ they fall less quickly. At high rapidity the
agreement is less good, the Monte Carlo estimate being rather higher, though
still consistent within uncertainties. The consistency between these two very
different approaches is quite encouraging, suggesting that if the Monte Carlo
estimates of the hadronization (and multi-particle interactions) had been used
to correct the data, converting it from hadron level to parton level, the linear
power corrections found in our analysis would have been rather smaller, though
more dependent on rapidity. Similar remarks apply to the dijet datasets.

%% file: sec-results.tex
\section{The NNPDF4.0HT Parton Sets}
\label{sec:pheno}
We evaluate the phenomenological consequences of including the HT corrections
and PCs discussed in the previous sections at the level of a new global PDF fit,
and discuss the impact on relevant LHC observables and on the determination
of $\alpha_s$.

\subsection{Global PDF fits with Power Corrections}
\label{sec:pdf_construction}

The HT and power corrections discussed above were determined by performing
global PDF fits to the NNPDF4.0 dataset, but lowering the kinematic cuts to
increase sensitivity to the HT. However, we do not suggest performing a fit with
these lowered cuts to determine the PDFs, since such a fit would include data
which is significantly affected by HT. Instead, we use the HT and the PCs
determined above to give corrections to the theory predictions in a fit with the
conventional cuts used in NNPDF4.0~\cite{NNPDF:2021njg}, i.e. $Q^{2} >
3.49$~GeV$^{2}$ and $W^2 > 12.5$~GeV$^2$.

The multiplicative factors are constructed following the definition of the
shifts in \cref{eq:pc_shift_sf,eq:pc_shift_nc,eq:pc_shift_rat,eq:pc_shift_cc}
for DIS and in \cref{eq:shifted_jet_xsec,eq:shifted_2jet_xsec} for jet
observables, using the central values of the posterior parameters, $h^{\rm
post}_{\alpha}$, \cref{eq:pc_ht_post}. Thus the shift in theoretical predictions
is given by
\begin{equation}
  \delta T = \sum_\alpha \bar{\lambda}_\alpha\beta_\alpha.
  \label{eq:shift_pstr}
\end{equation}
Likewise the uncertainties and correlations of these parameters are encoded in
the posterior covariance matrix $H^{\rm post}_{\alpha\beta}$,
\cref{eq:pc_ht_post}, which propagates through \cref{eq:linearised_shift} into a
covariance matrix in the space of data which gives the uncertainties and
correlations due to the HT and PCs:
\begin{equation}
  C^{\rm post} = 2 \sum_{\alpha, \beta} \beta_{\alpha}\beta_{\beta}^T \bar{Z}_{\alpha \beta}.
  \label{eq:theory_cov_pstr}
\end{equation}
Here we have introduced a factor of two to provide a conservative estimate of
the posterior uncertainty: we do not want to be too aggressive with the
uncertainty on the HT and PCs, in part because some data have been used twice
(first to determine the PCs, then again to determine the PDFs), but also because
of the inherent difficulties discussed previously in finding clear evidence for
HT and PCs. The total covariance matrix $C$ to be used in the PDF fit is then
obtained by adding this contribution to the existing covariance matrix
\cref{eq:ctot}, which already includes the experimental uncertainties, and other
sources of theoretical uncertainty (nuclear and MHO). We use this setup to
perform a global PDF fit to the NNPDF4.0 dataset for each perturbative theory
setting considered in this work. For each setting, the multiplicative shifts and
the posterior uncertainty determined in \cref{sec:pc-tcm} are chosen
consistently with the theory setting employed in the PDF fit. We discuss the
results in the following sections.

\subsection{Fit Quality}
\label{sec:fit_quality}

\begin{table}[t!]
  \scriptsize
  \centering
  \renewcommand{\arraystretch}{1.4}
  \input{tables/tab_total_chi2.tex}
  \caption{The $\chi^2$ per data point for fits to the global NNPDF4.0 dataset
    with conventional cuts, with and without HT and PCs, and with the addition
    of the HT and PC uncertainty. The total is broken down into the
    contributions from the various processes included in the fit.}
  \label{tab:chi2_TOTAL}
\end{table}
We start by discussing the fit quality. In Tab.~\ref{tab:chi2_TOTAL} we report
the values of the experimental \chisq~per data point for the global fit with and
without HT and PCs, and for fits with perturbative calculations at NNLO,
NNLO+MHOU, and aN$^3$LO+MHOU. Together with the global \chisq, values are also
displayed for each separate process type. In order to disentangle the effect of
the shifts in the theory predictions from the effect of the additional
uncertainty due to the theory covariance matrix in \cref{eq:theory_cov_pstr}, we
performed a series of fits where only the shifts are included in the theory
predictions. The corresponding \chisq~values are also reported in
Tab.~\ref{tab:chi2_TOTAL}. All sources of correlations are fully accounted for,
and the $t_0$ method~\cite{Ball:2009qv} is used for multiplicative
uncertainties. We refer to Ref.~\cite{NNPDF:2024nan} for further details on the
inclusion of the experimental uncertainties in the \chisq.

Inspecting Tab.~\ref{tab:chi2_TOTAL}, we observe a uniform improvement of the
global \chisq~upon including the HT corrections across all theory settings.
Furthermore, the improvement is retained when the theory covariance matrix as in
\cref{eq:theory_cov_pstr} is removed, showing that the posteriors determined in
\cref{sec:pc-tcm} are indeed moving the theory closer to the experimental data.
Of course, the lowest global \chisq~is obtained when both the shifts and the
theory covariance matrix are included, as expected. A closer inspection of
Tab.~\ref{tab:chi2_TOTAL} reveals that at \annnlo, when the shift is applied to
the inclusive jet and dijet data, the description of the top pair data also
improves, indicating that the power corrections resolve some of the tension
between the data for these processes. A similar effect can be seen in the NC DY
data, the description of which is improved following the shift in NC DIS from
the HT.

\subsection{PDFs and Parton Luminosities}
\label{sec:impact_pdfs}

\begin{figure}[t!]
  \centering
  \includegraphics[width=0.32\textwidth]{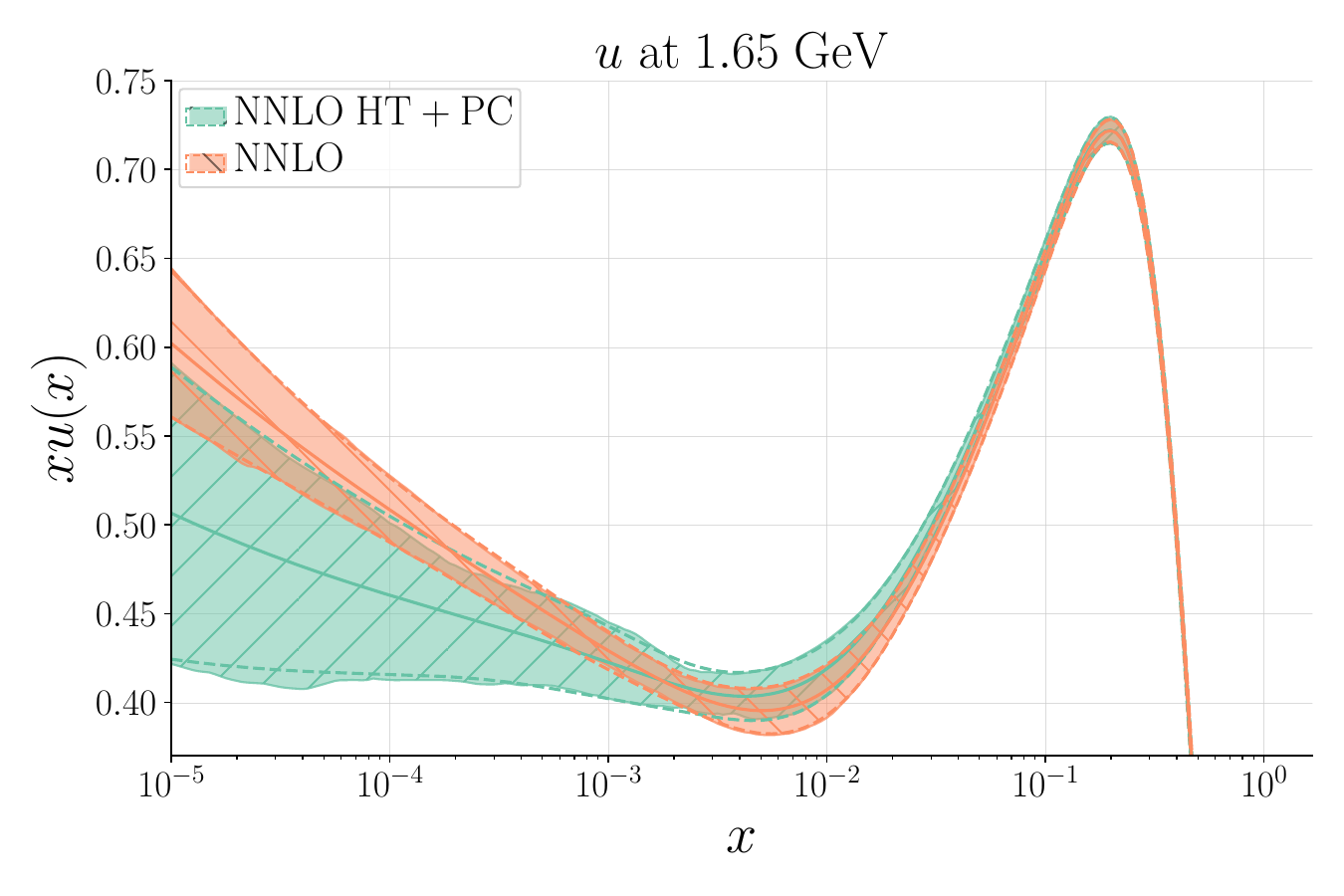}
  \includegraphics[width=0.32\textwidth]{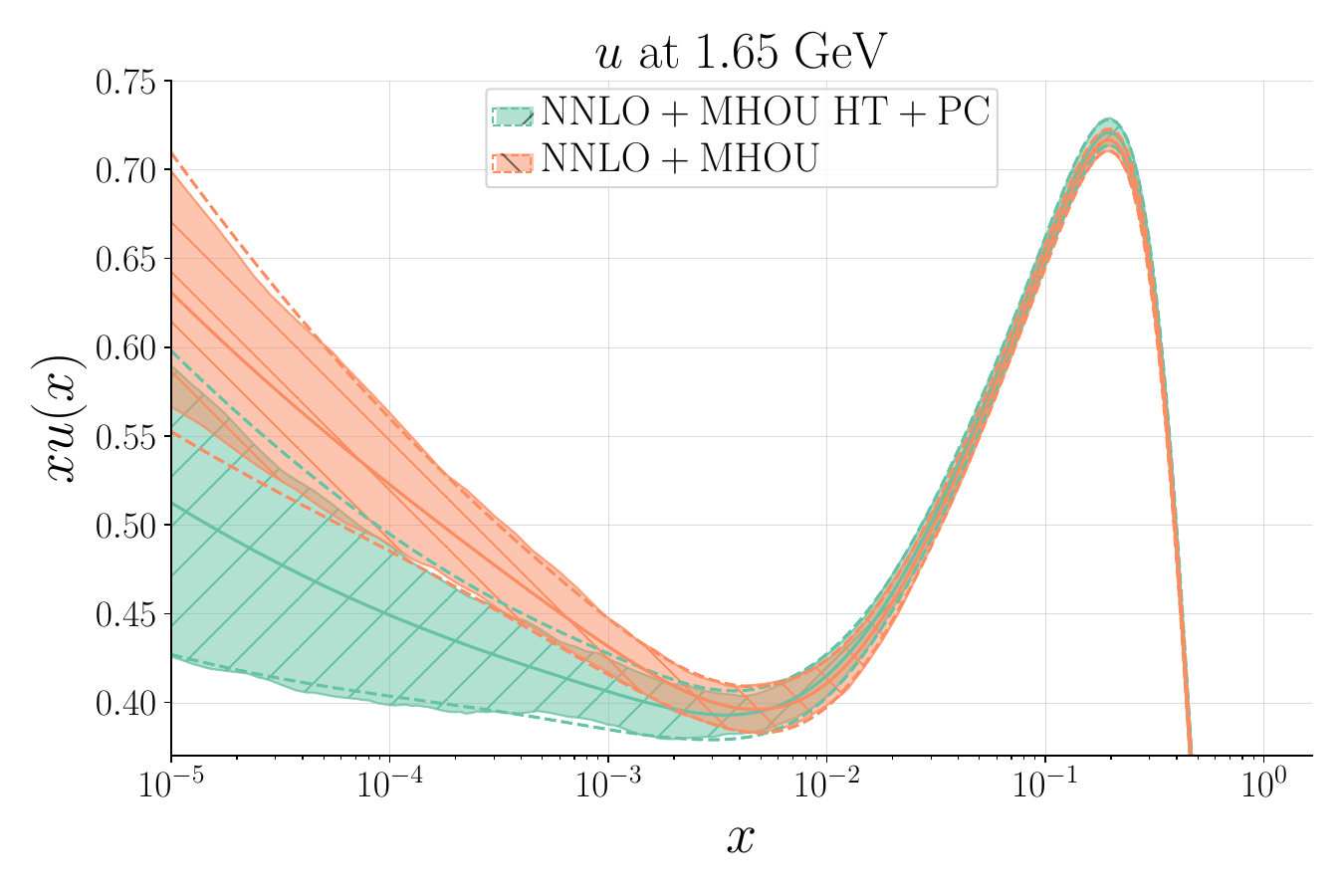}
  \includegraphics[width=0.32\textwidth]{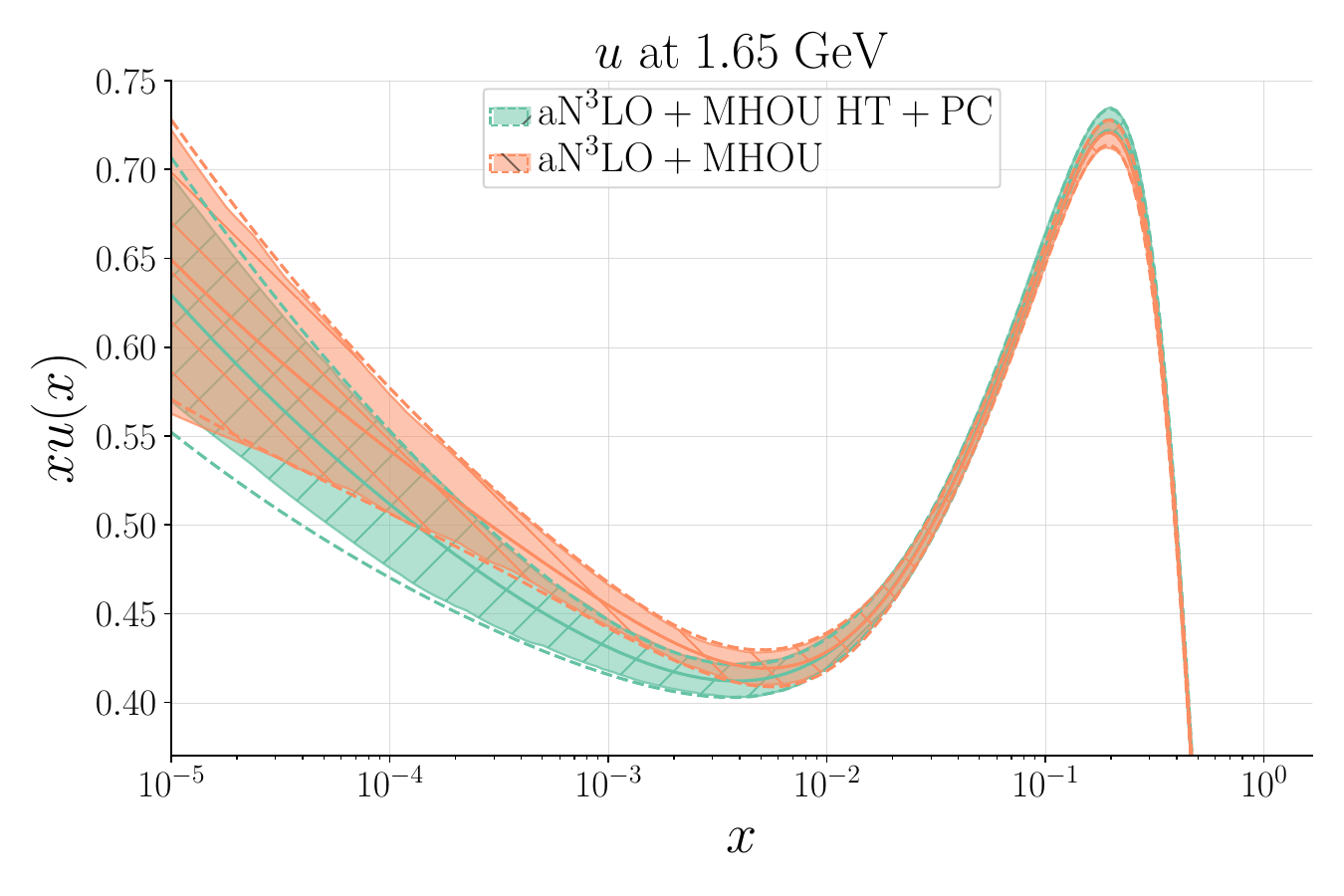}
  \includegraphics[width=0.32\textwidth]{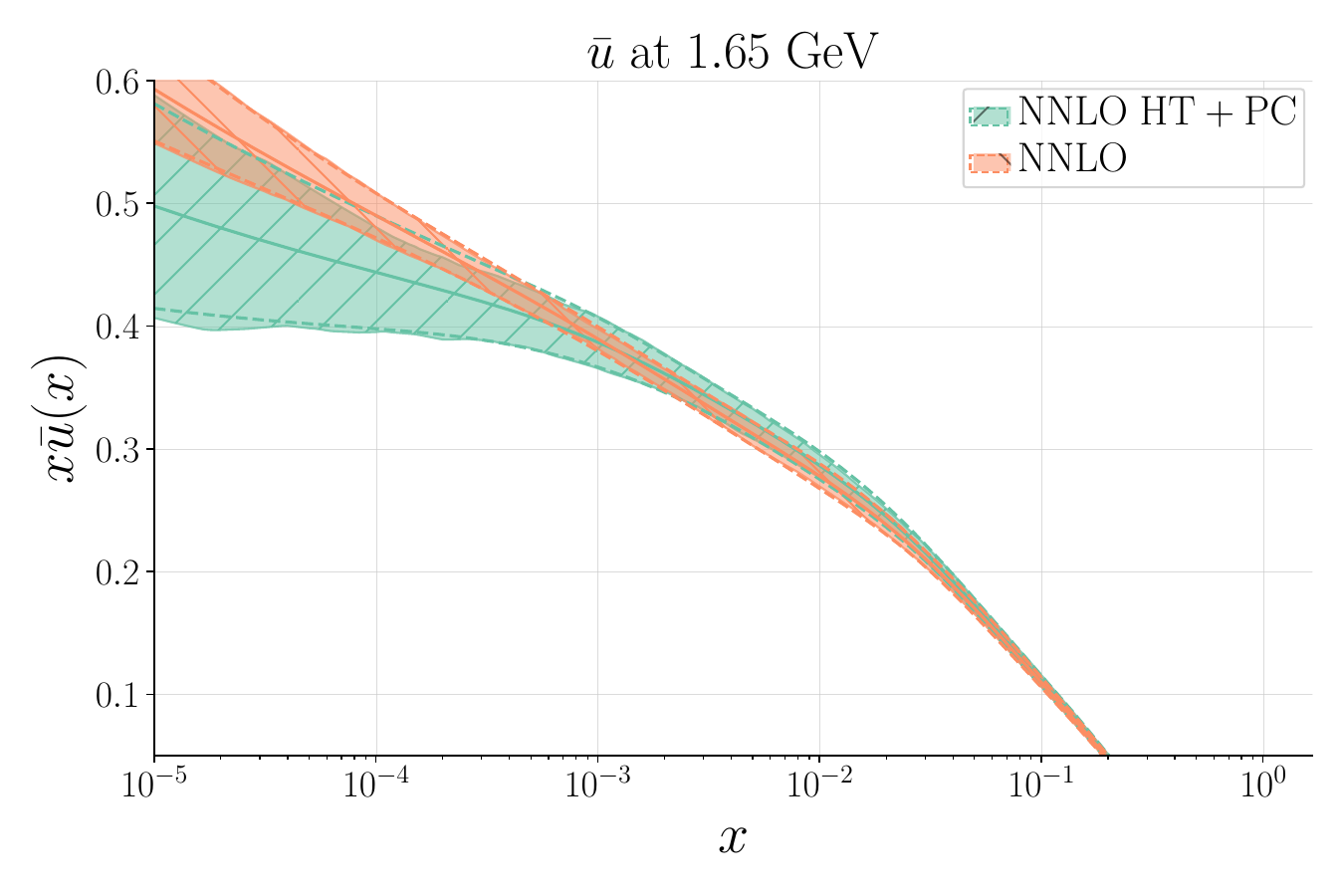}
  \includegraphics[width=0.32\textwidth]{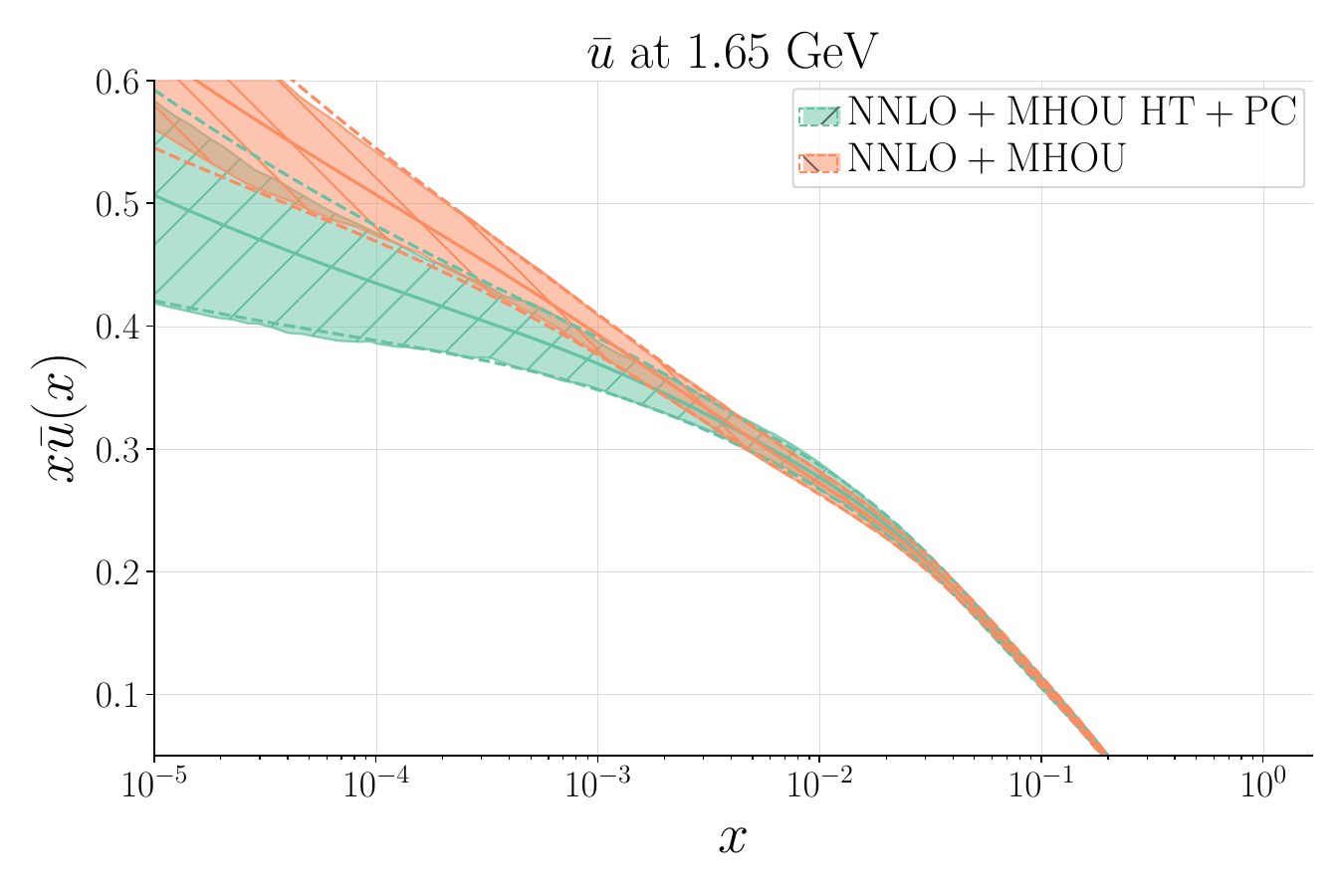}
  \includegraphics[width=0.32\textwidth]{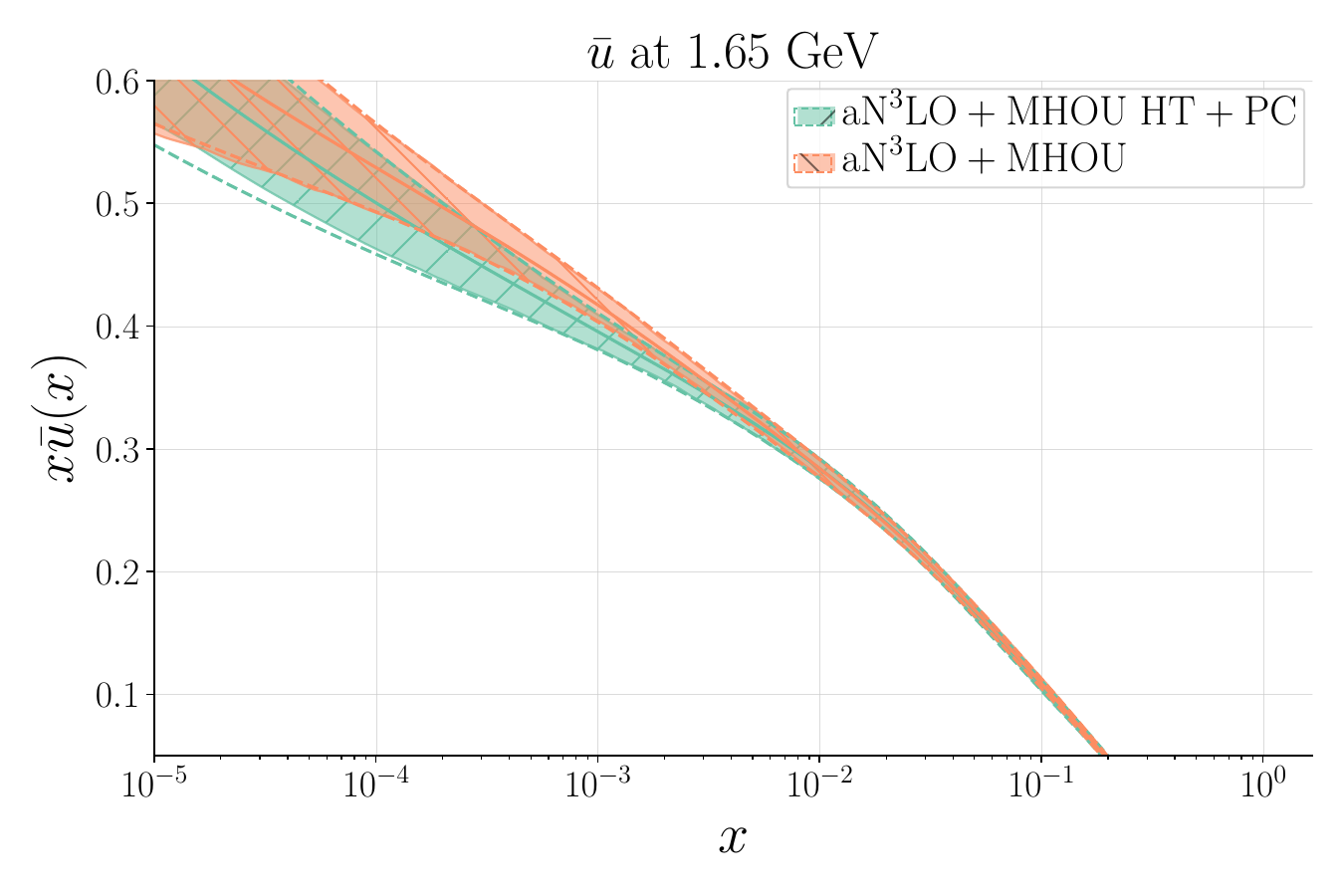}
  \includegraphics[width=0.32\textwidth]{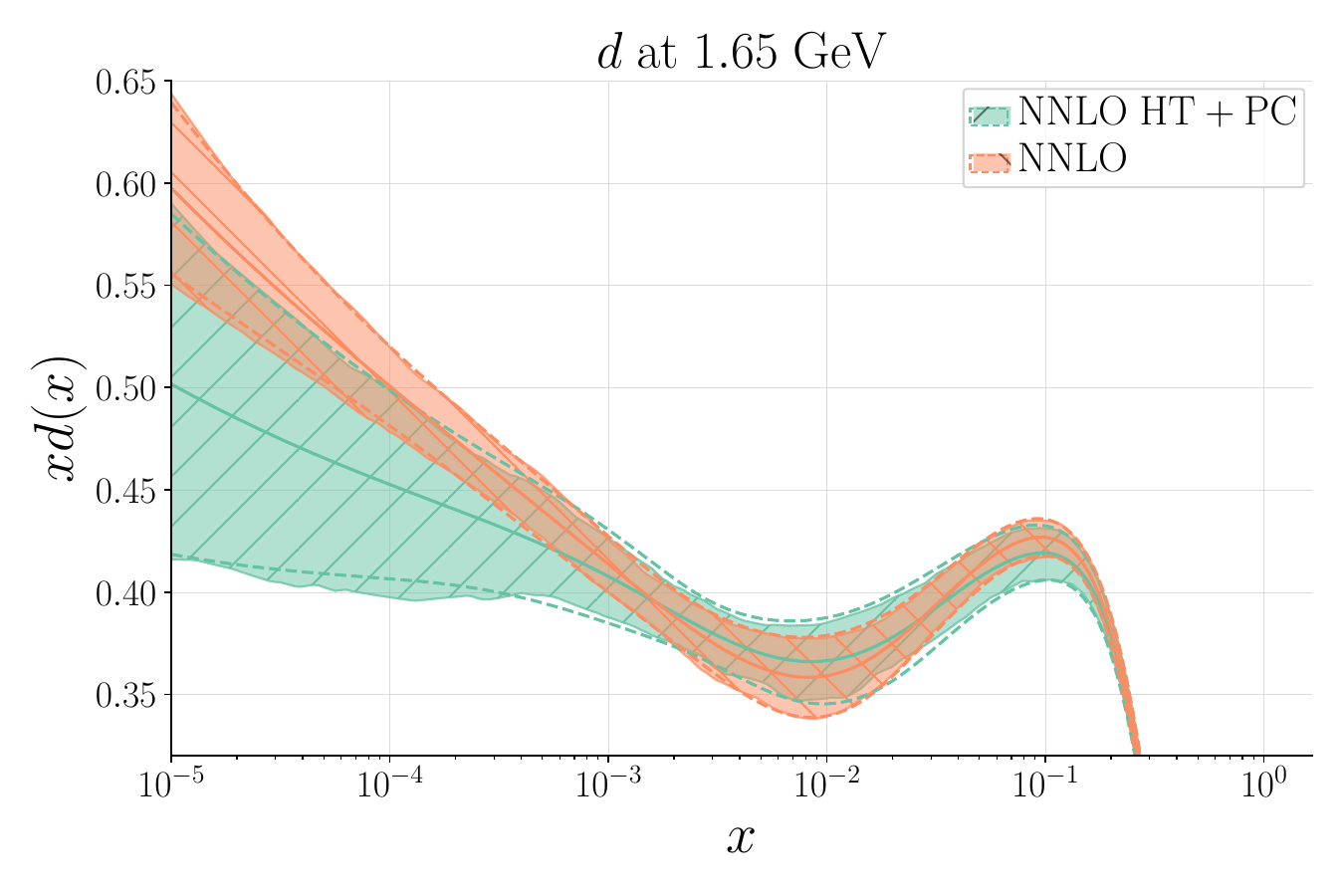}
  \includegraphics[width=0.32\textwidth]{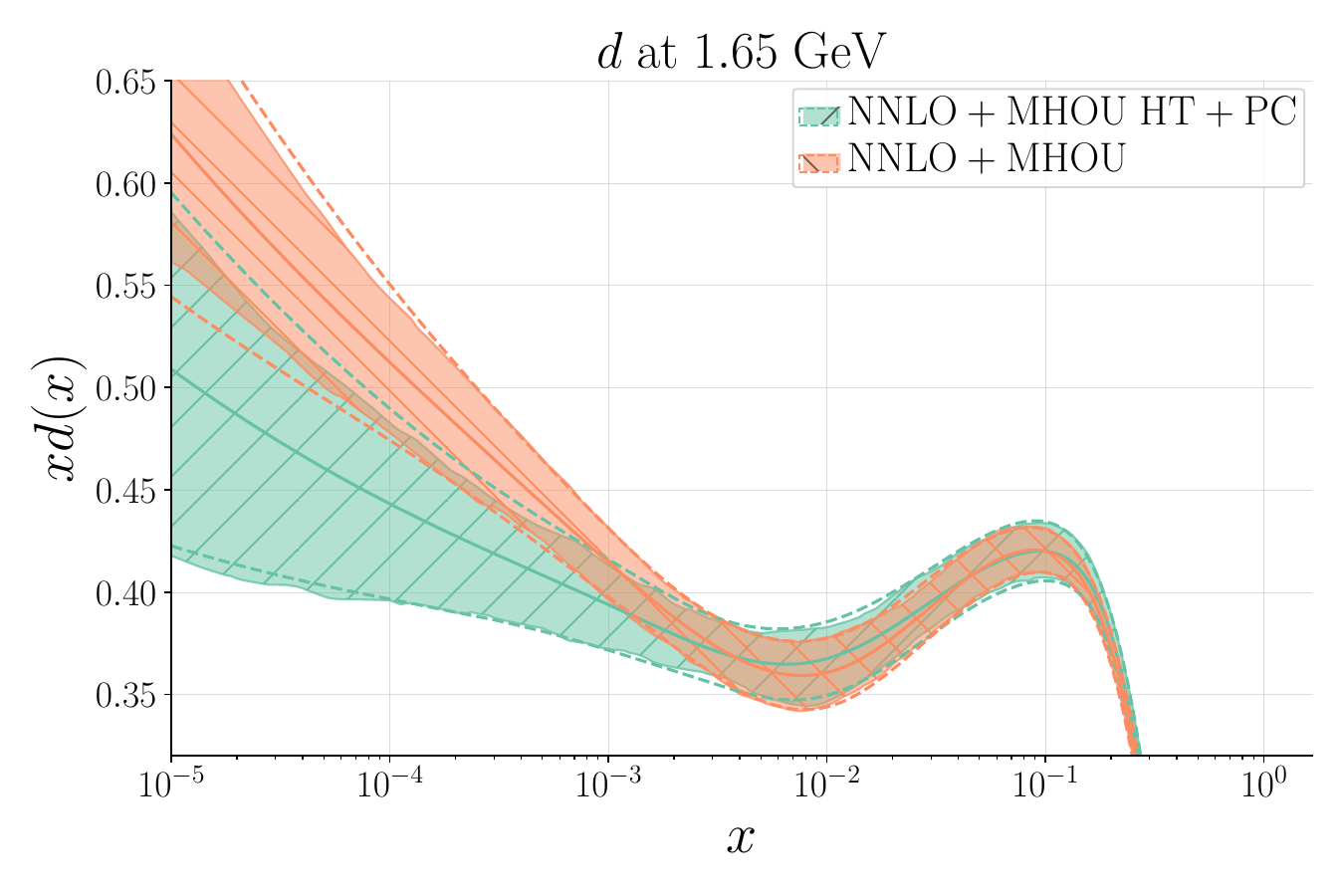}
  \includegraphics[width=0.32\textwidth]{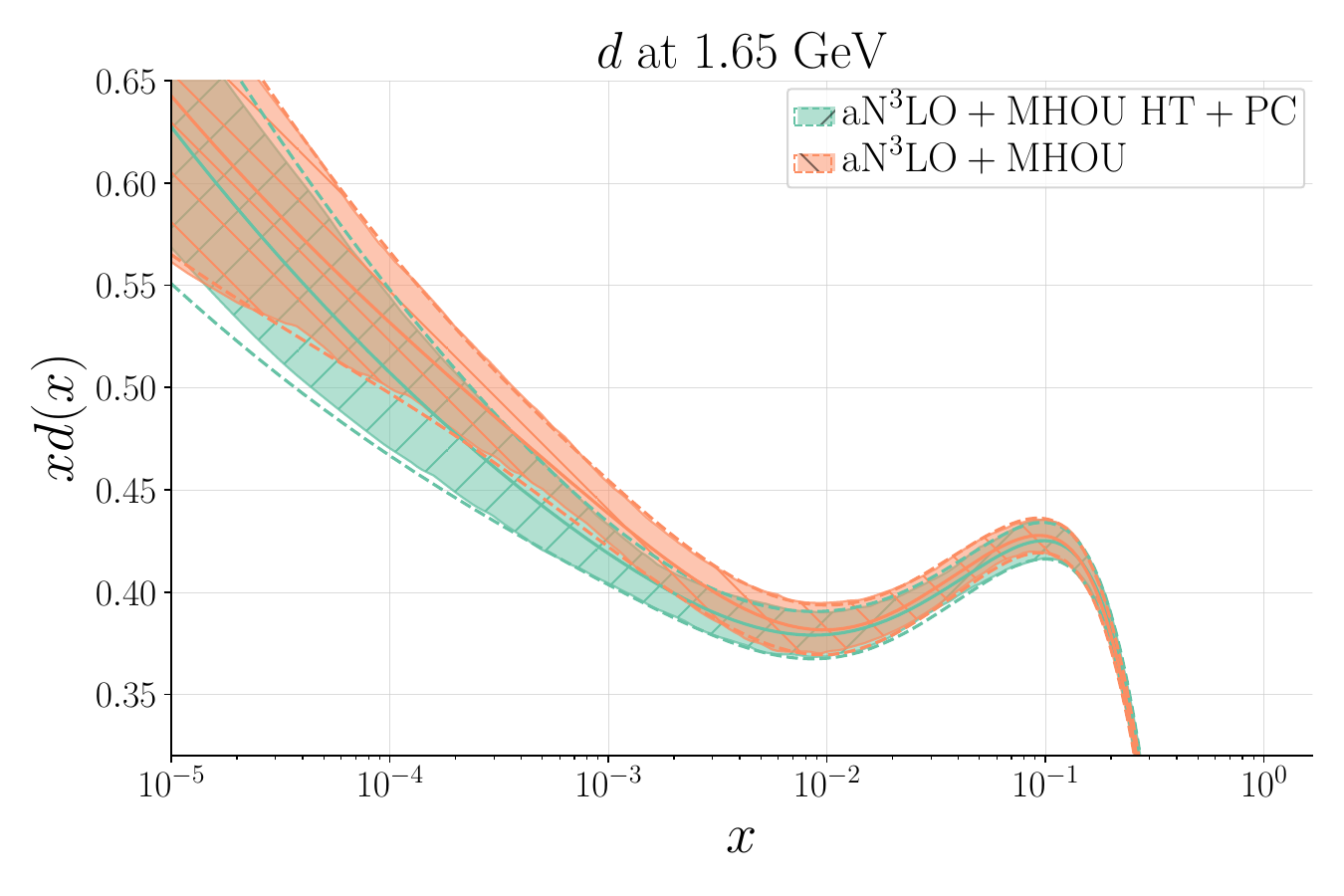}
  \includegraphics[width=0.32\textwidth]{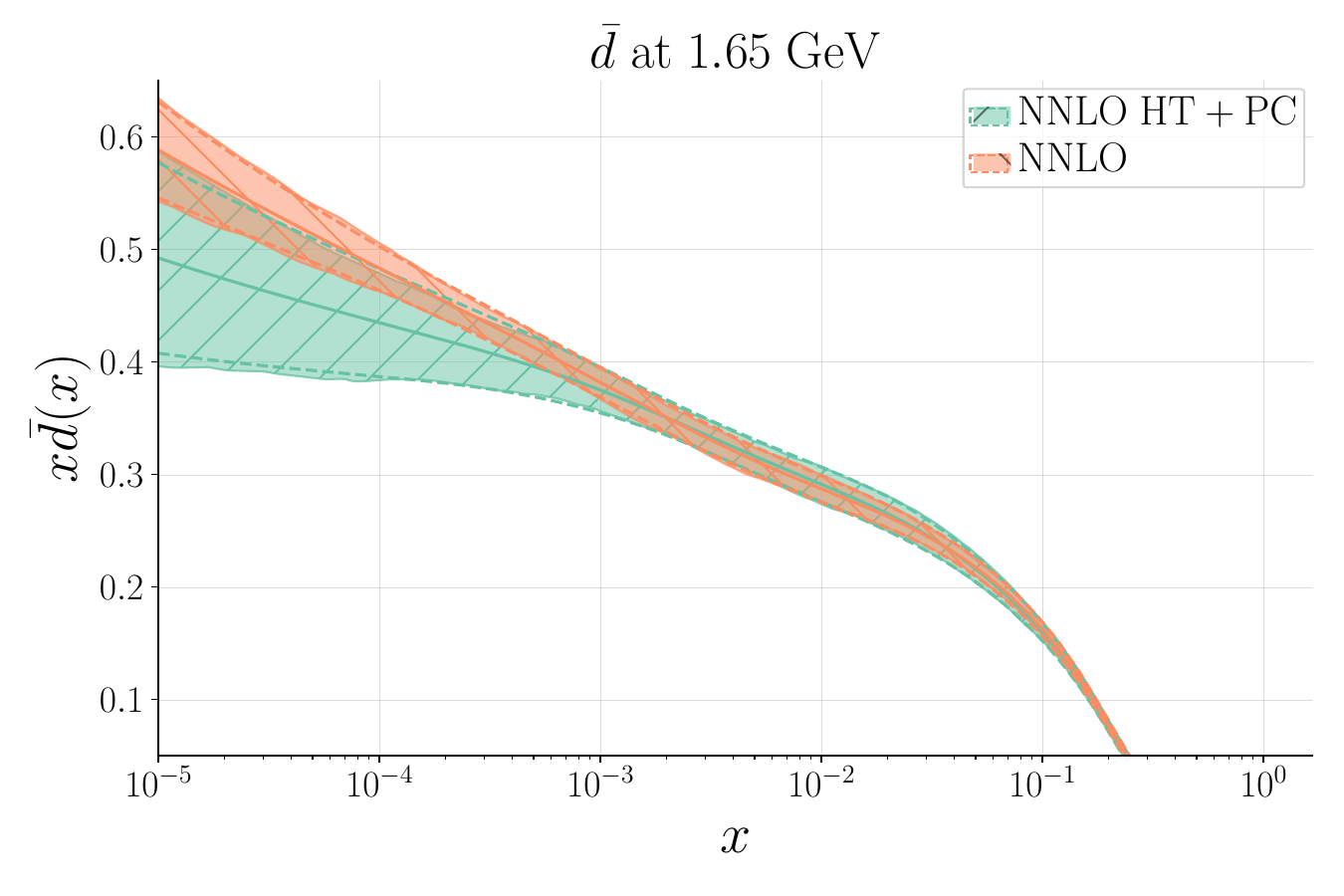}
  \includegraphics[width=0.32\textwidth]{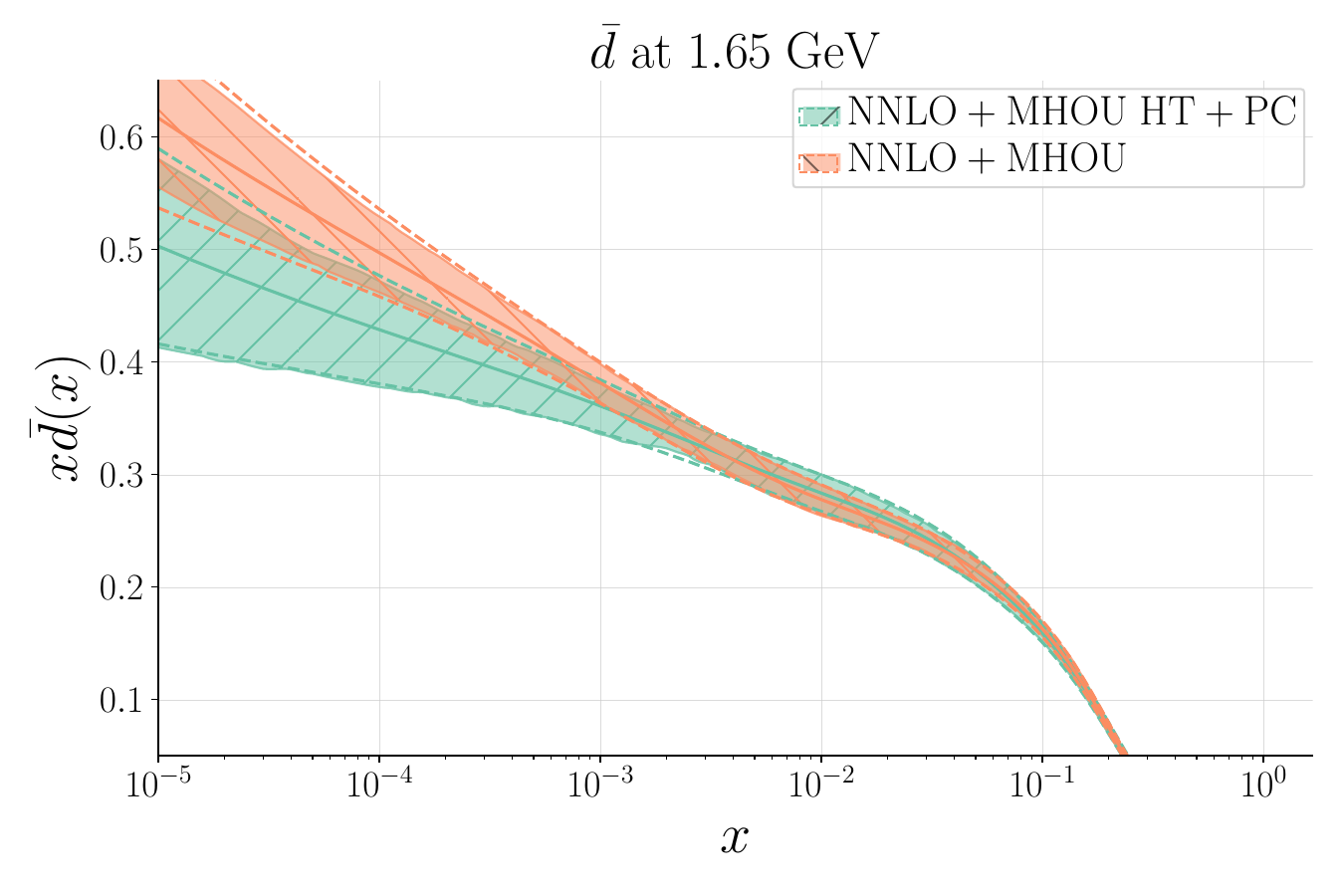}
  \includegraphics[width=0.32\textwidth]{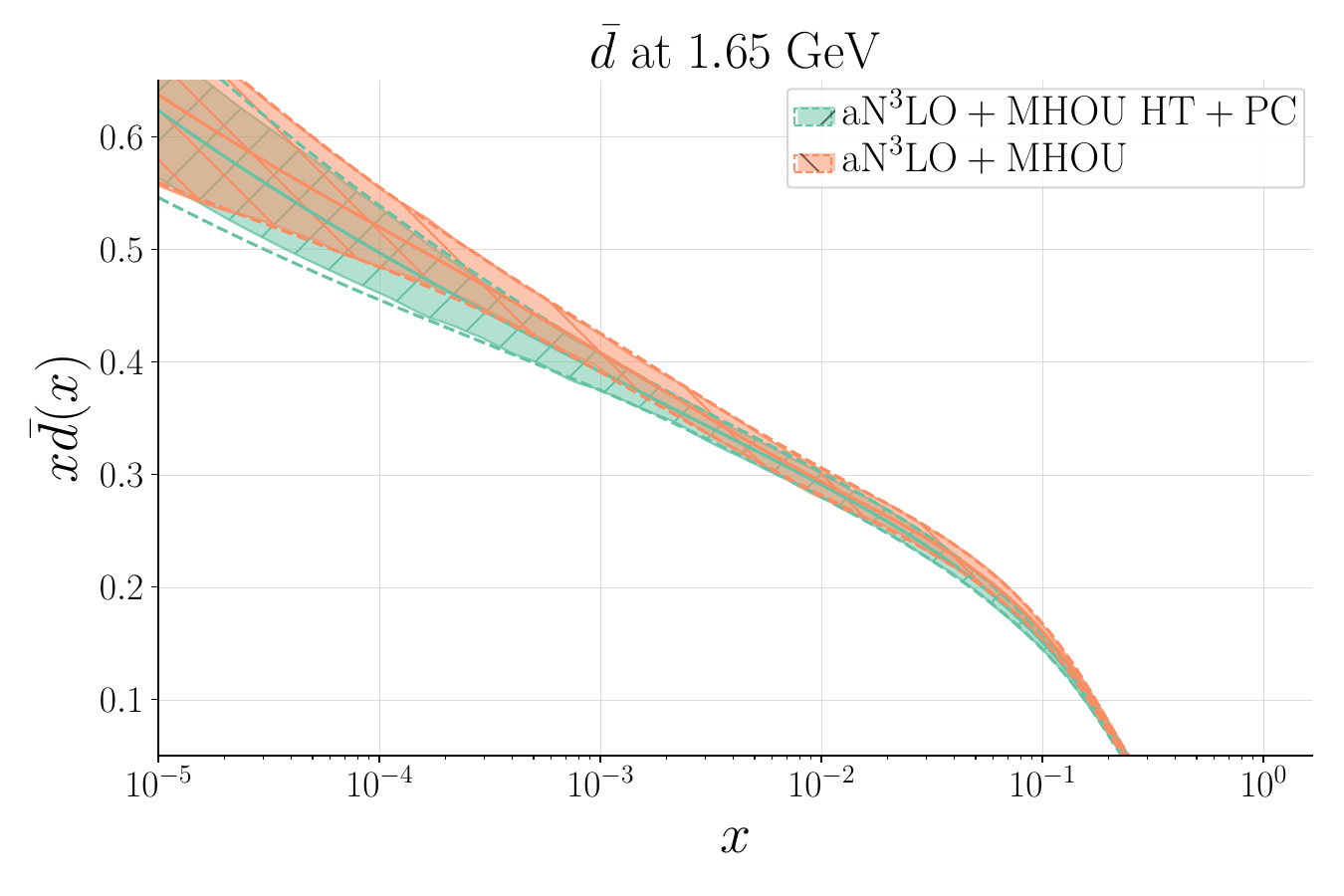}
  \caption{The PDFs with HT and PCs (in blue) for perturbative calculations at NNLO (left
    column), NNLO+MHOU (central column) and aN$^3$LO+MHOU (right column), for
    up, anti-up, down, and anti-down. These are compared to the corresponding PDFs
    without HT and PCs (in red). Bands with dashed contours correspond to one-sigma
    uncertainties, whereas bands with continuous contours correspond to 68\%
    confidence-level uncertainties.}
  \label{fig:pdfs}
\end{figure}
\begin{figure}[t!]
  \centering
  \includegraphics[width=0.32\textwidth]{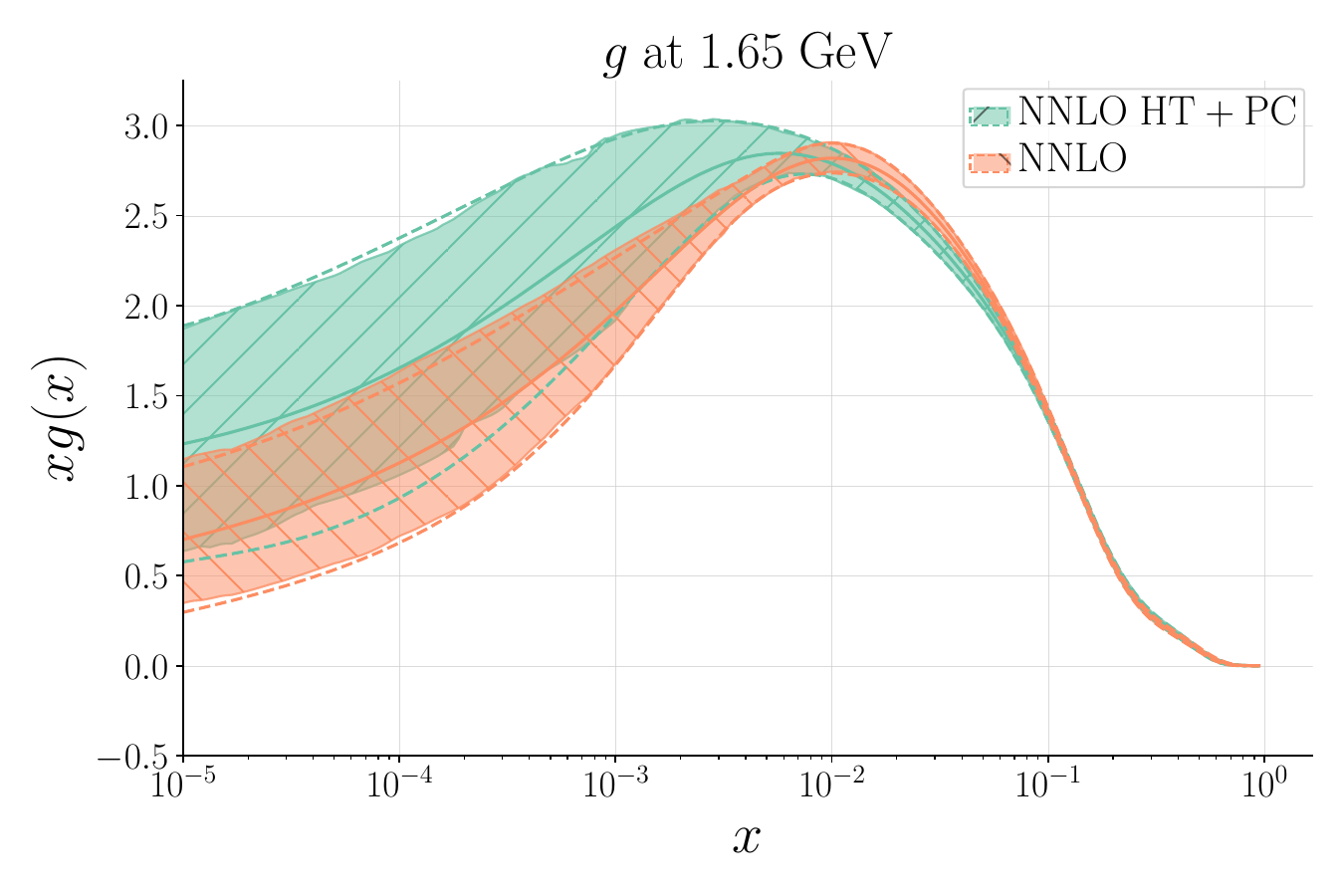}
  \includegraphics[width=0.32\textwidth]{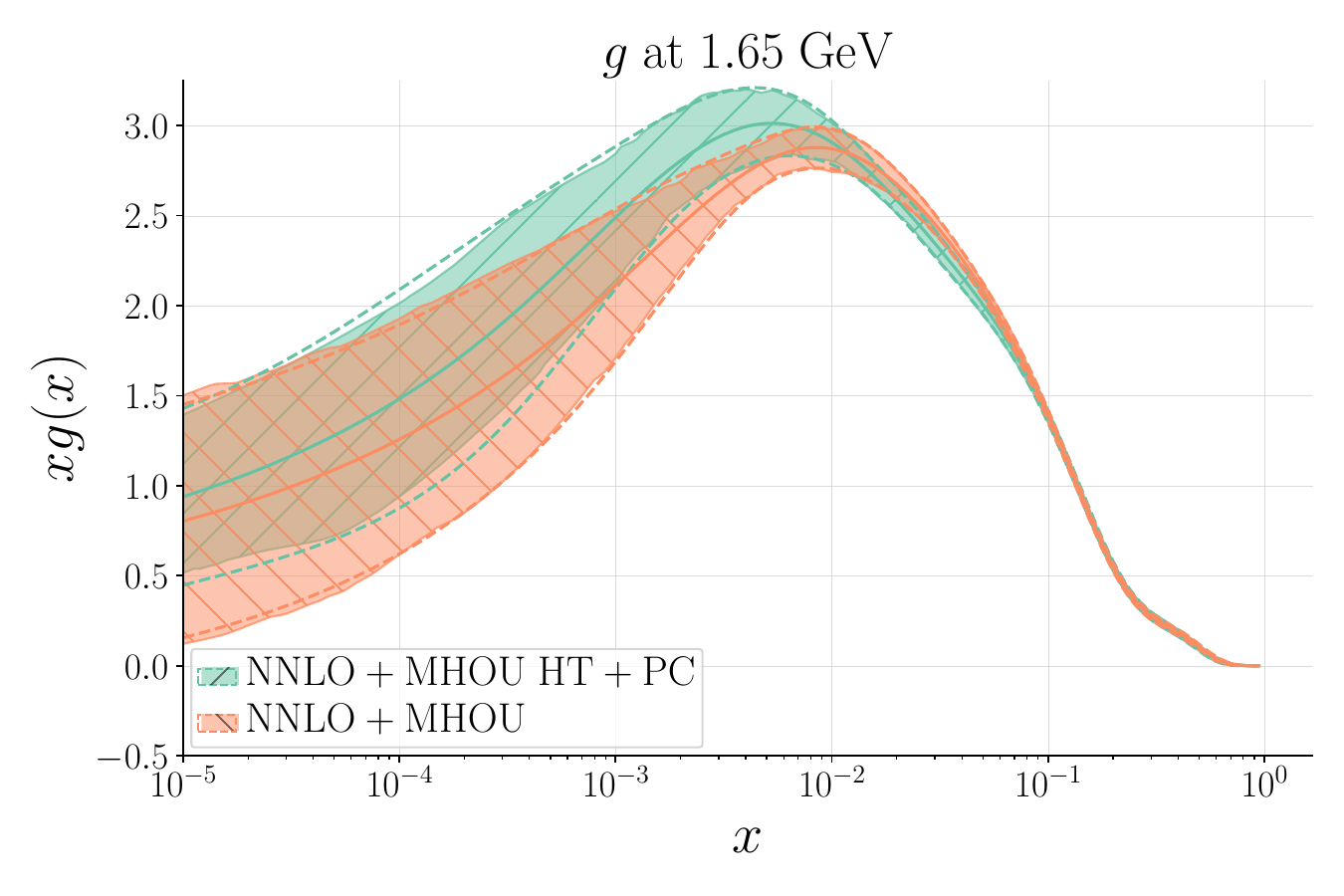}
  \includegraphics[width=0.32\textwidth]{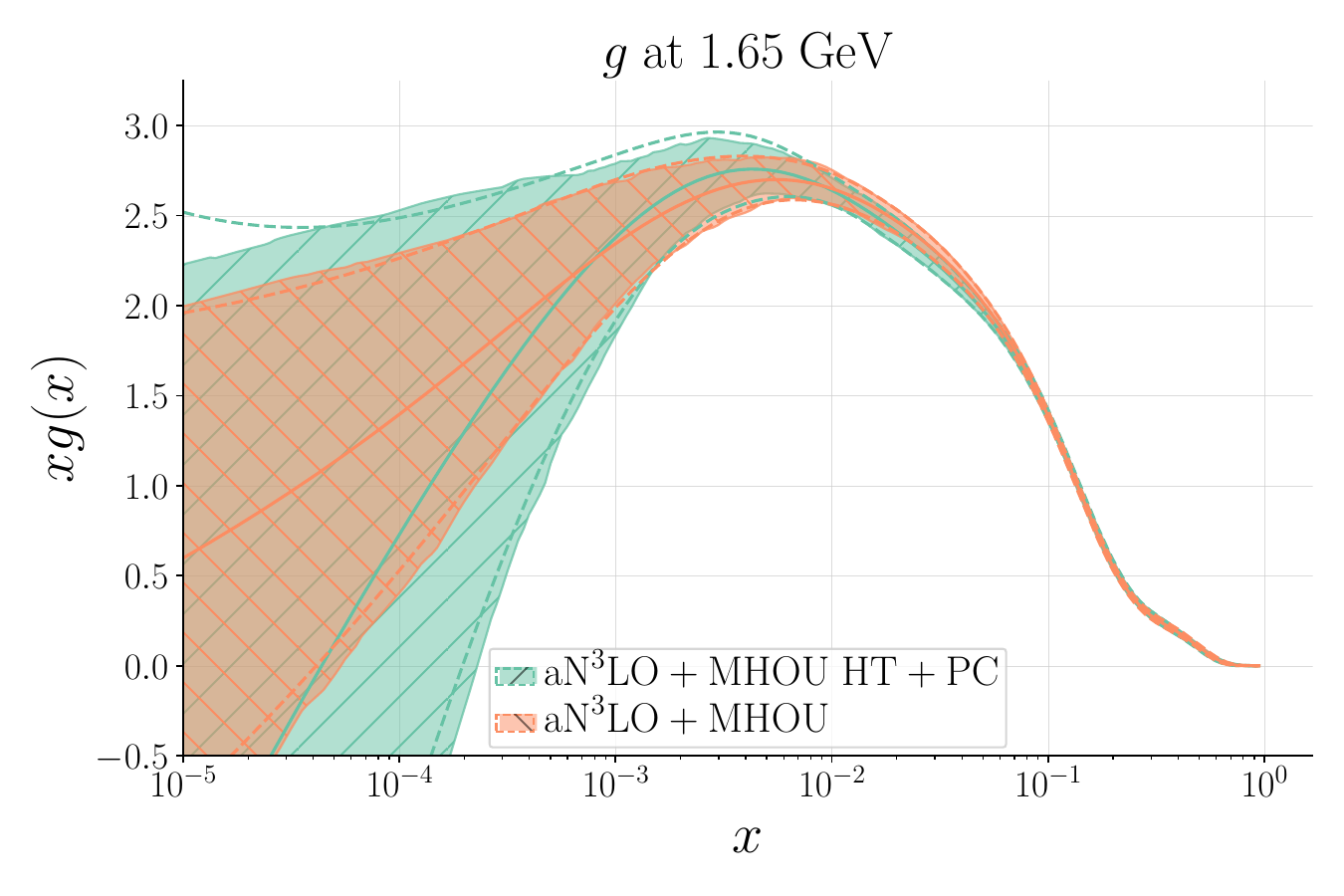}
  \includegraphics[width=0.32\textwidth]{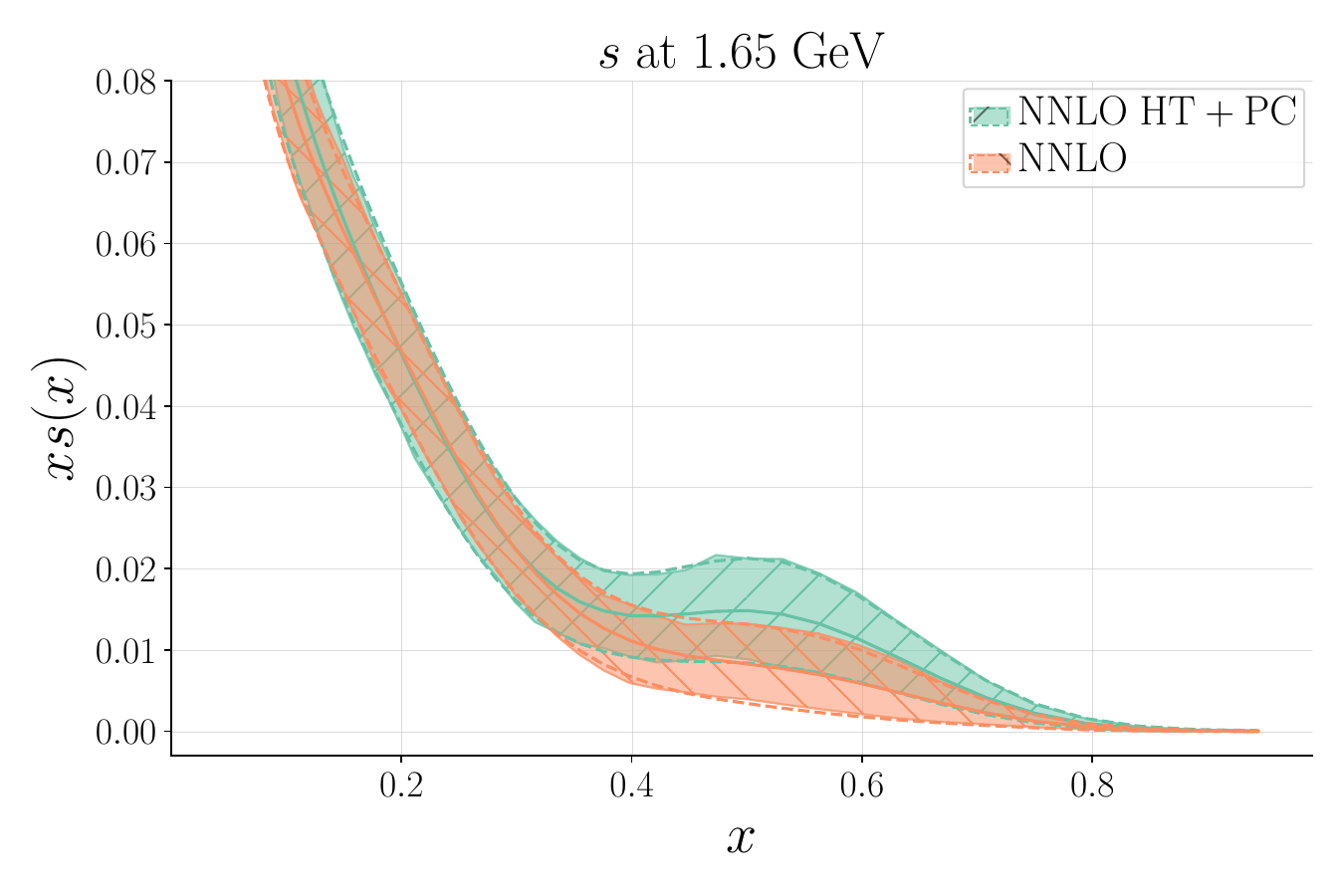}
  \includegraphics[width=0.32\textwidth]{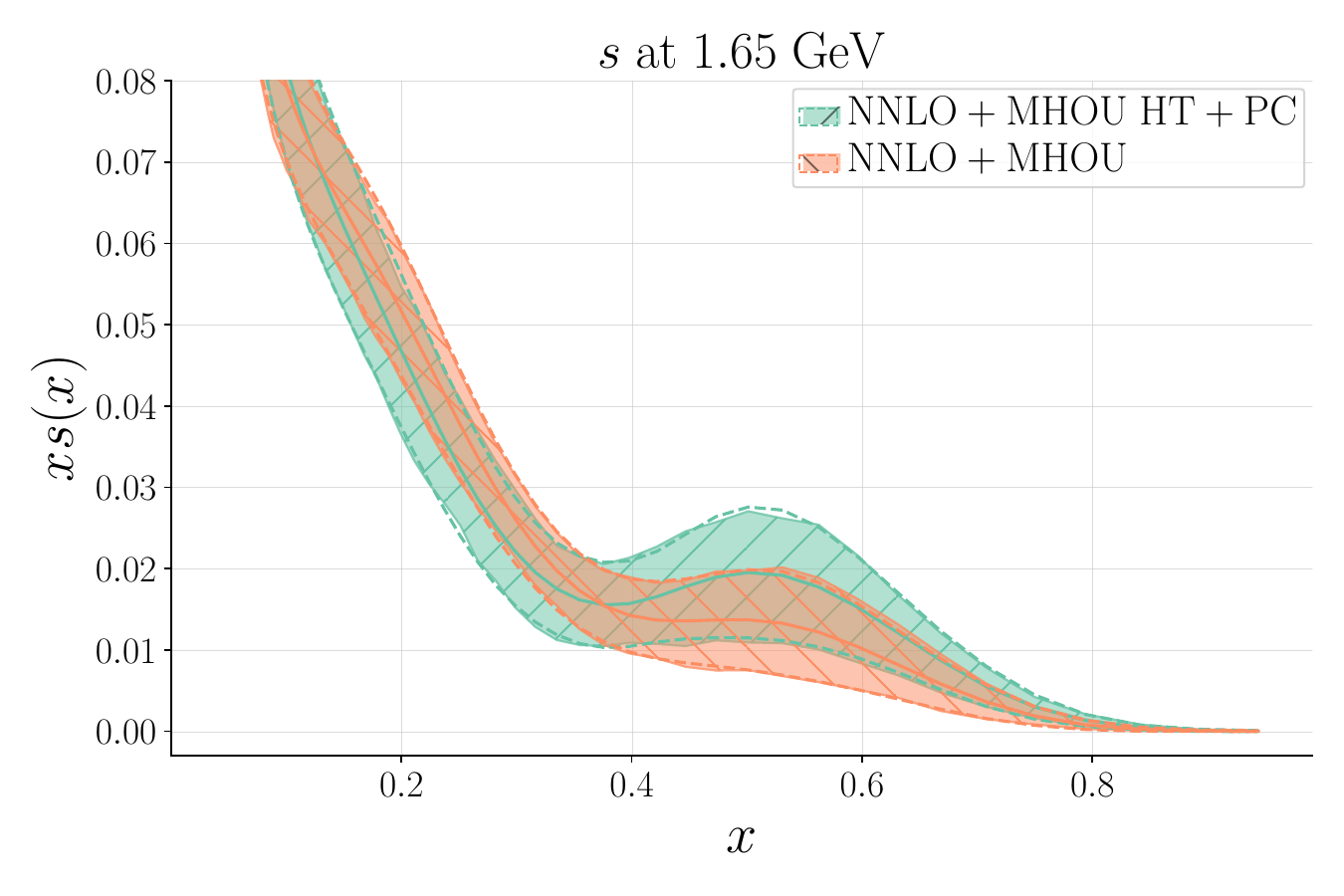}
  \includegraphics[width=0.32\textwidth]{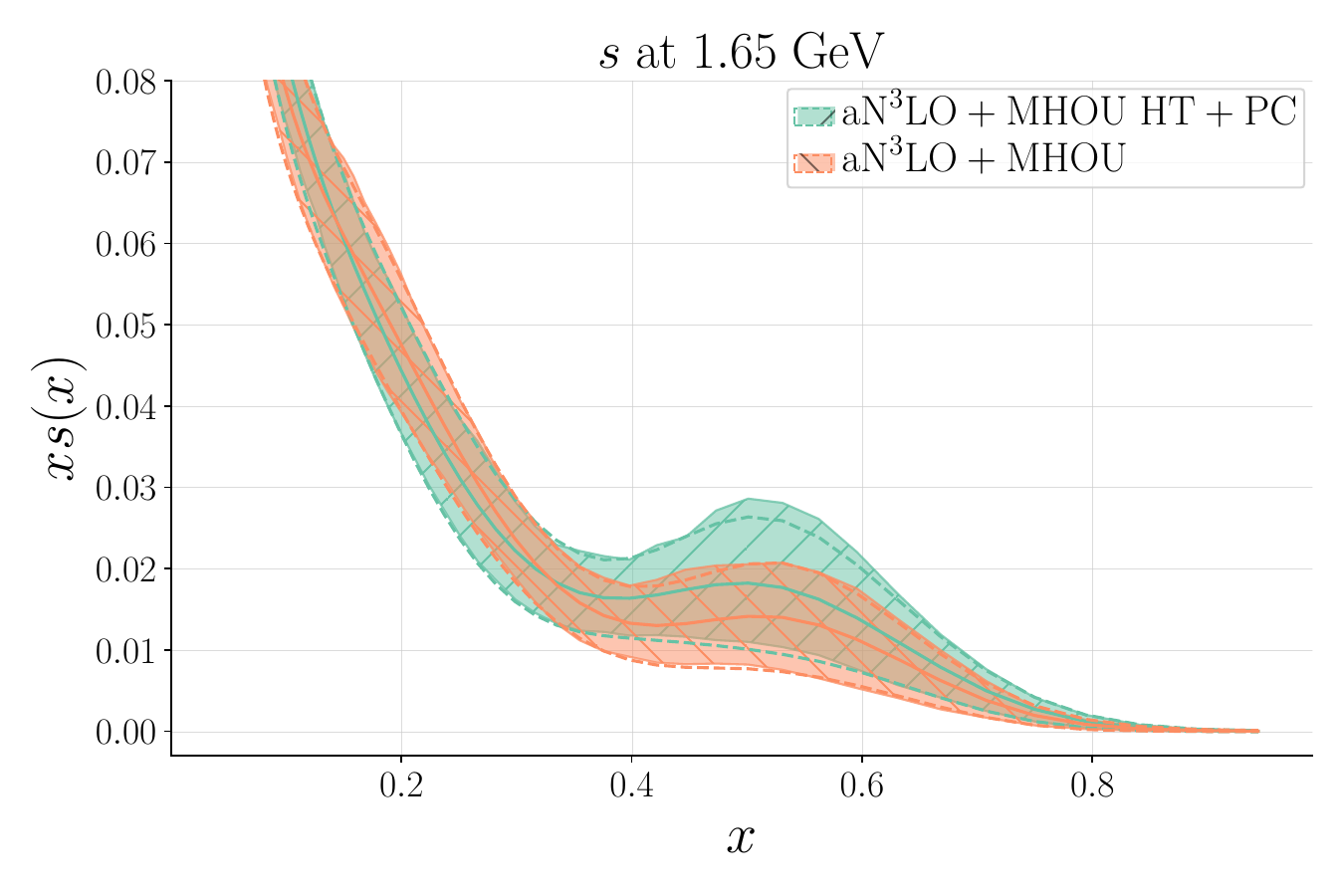}
  \includegraphics[width=0.32\textwidth]{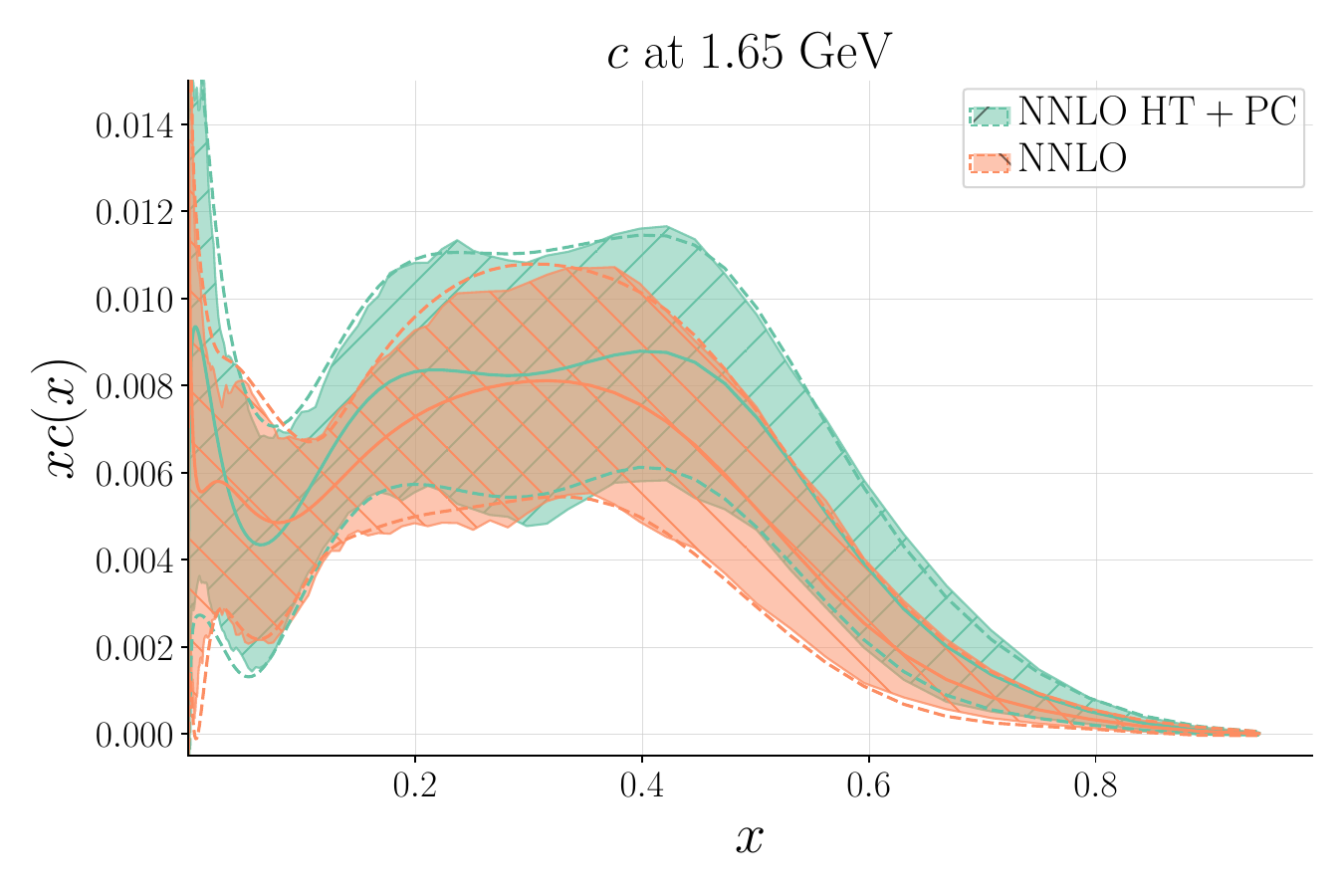}
  \includegraphics[width=0.32\textwidth]{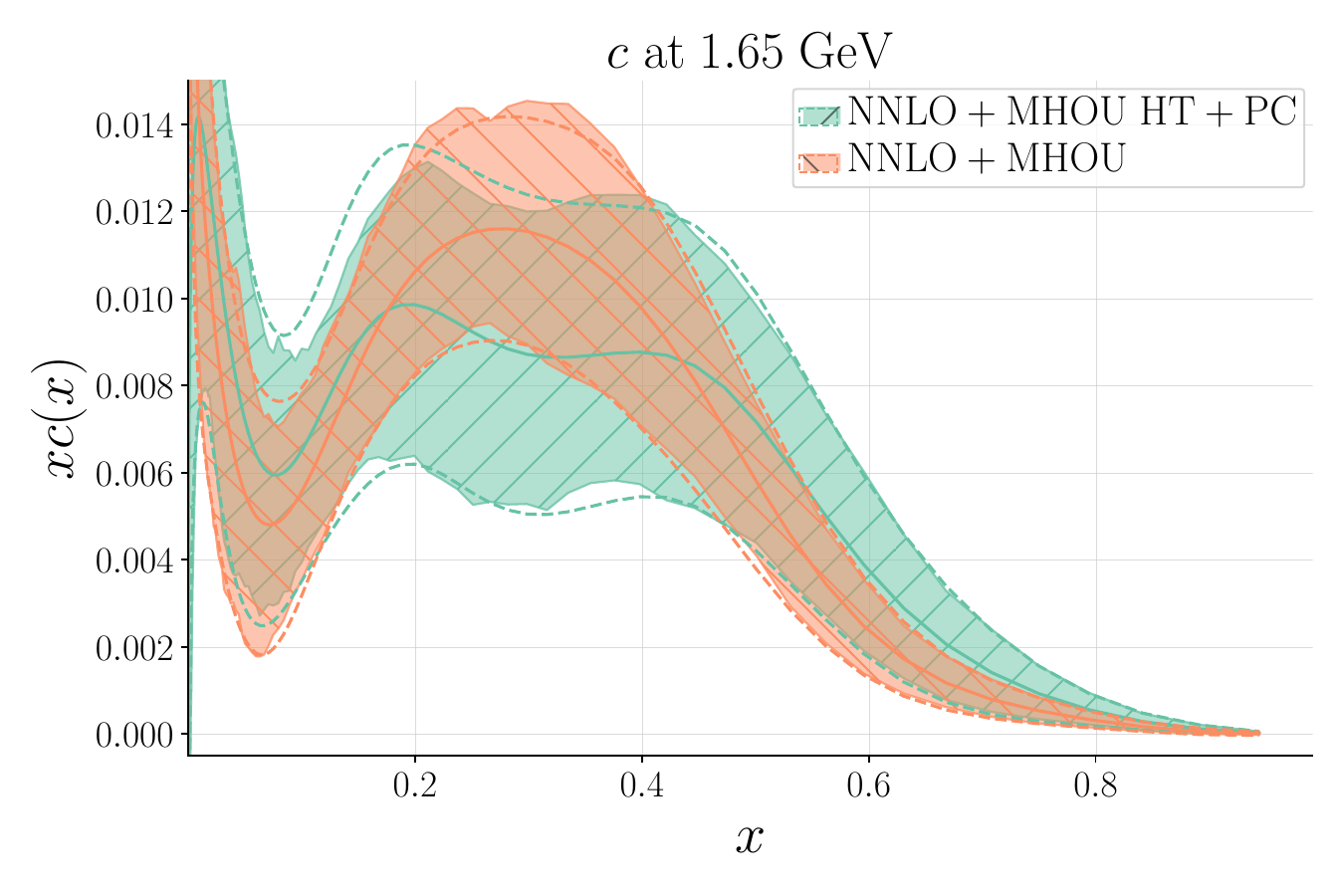}
  \includegraphics[width=0.32\textwidth]{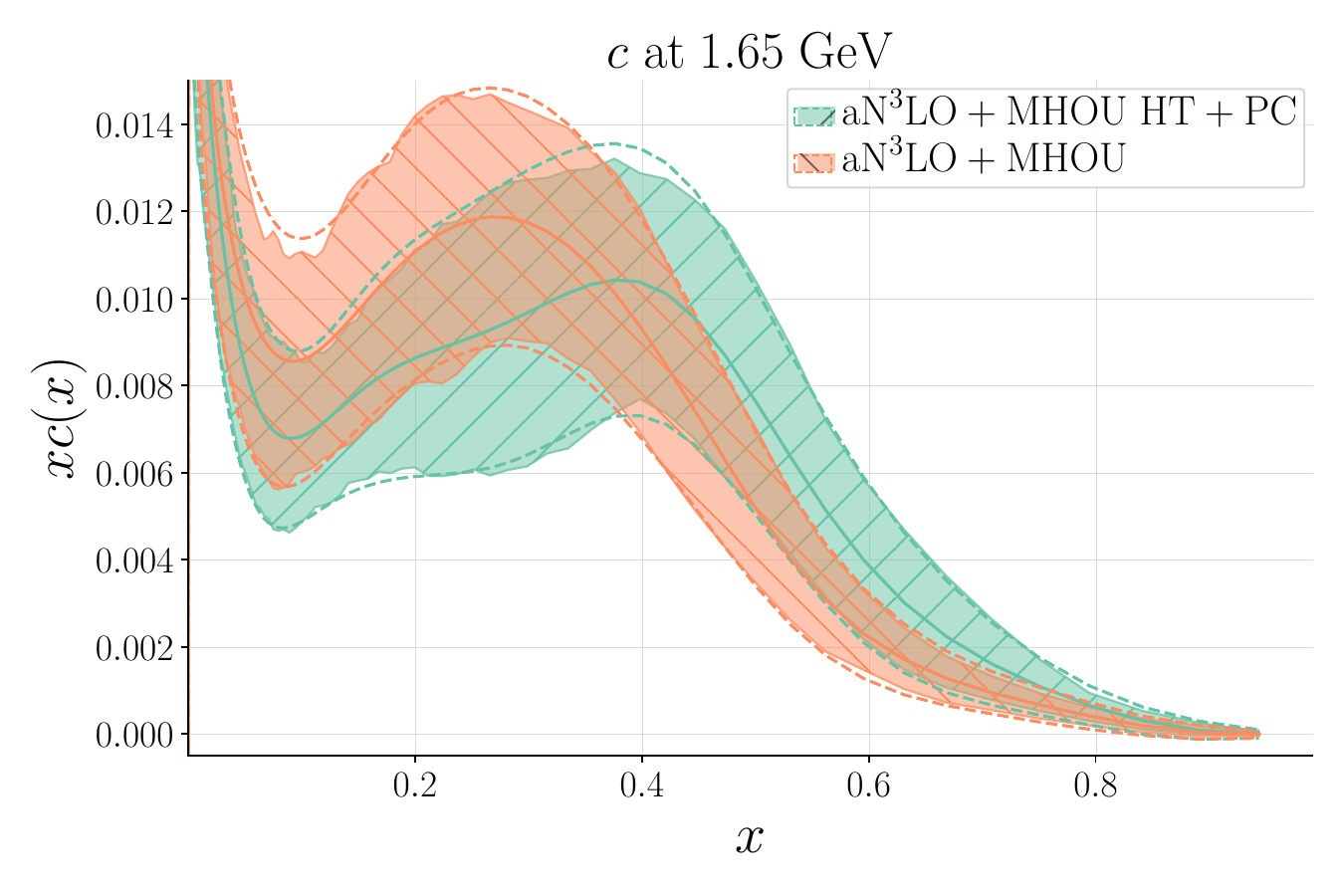}
  \caption{Similar to \cref{fig:pdfs}, but for the gluon (upper row), strange
  (middle row) and charm (lower row). Note the linear $x$-scale used for the
  strange and charm PDFs. Bands with dashed contours correspond to one-sigma
  uncertainties, whereas bands with continuous contours correspond to 68\%
  confidence-level uncertainties.}
\label{fig:pdfs2}
\end{figure}

We now turn to the impact of the HT and PCs on the PDFs, which are shown in
\cref{fig:pdfs} for the light quarks, and in \cref{fig:pdfs2} for
the gluon, strange and charm. Results are displayed as functions of $x$ at
$Q=1.65$~GeV, incorporating both shifts for the HT and PCs, and their
(correlated) uncertainties, for the three perturbative theory settings. Overall,
the impact of the HT and PCs on the PDFs is very modest, almost always within
one sigma. In particular, the HT has very little impact on the large-$x$ PDFs,
other than a slight increase in strangeness, and a shift in the valence charm
peak to higher values of $x$. This is reassuring: the cuts to DIS data employed
in NNPDF4.0 are very effective at reducing the sensitivity of the PDFs to HT.
The main effect of the HT is at small $x$, where it increases PDF uncertainties
due to the deweighting of the low-$x$ HERA data. This is due to the increased
theoretical uncertainty from HT in the predictions for these very precise data,
as was noted in Fig.~\ref{fig:kin_dis_nnlo}. However, the increased uncertainty
seen at NNLO is already reduced when MHOUs are added, and is smallest in
the \annnlo fit: most of the effect seen at NNLO is not due to genuine HT, but
rather to the N$^3$LO corrections, which are substantial at small $x$.

The most interesting changes are in the gluon distribution. Again there is very
little impact of the HT at large $x$, and at NNLO an increase in the uncertainty
at small $x$ which is reduced once MHOUs are included, but increases again
at \annnlo. More interesting is the change in the gluon at intermediate
values of $x$, from roughly $0.05$ down to $0.005$, due mainly to the
PCs in the single
inclusive jet data (as we saw previously, the effect of PCs in the dijet data is
relatively small). As remarked in the \chisq~discussion, the PCs reduce tensions
between these data and the top data. At NNLO this leads to a reduction by almost
two sigma in the gluon at $x\sim 0.02$, resulting in a smoother profile, less
sharply peaked. Once MHOUs are included this effect is reduced, to around one
sigma, since the MHO already allows the smoother profile even before the PCs are
added. At N$^3$LO the effect of the PC is even less: the flattening of the gluon
is achieved without the PC. This second observation shows that much of the
tension between single inclusive jets and top at NNLO is resolved by using
N$^3$LO evolution, since both the NNLO+MHOU fit and the aN$^3$LO+MHOU fit use
the same NNLO result for the single inclusive jet partonic cross-section, with
the same estimate for its uncertainty, and a similar linear PC Fig.~\ref{fig:posterior_jet}.

\begin{figure}[t!]
  \centering
  \includegraphics[width=0.32\textwidth]{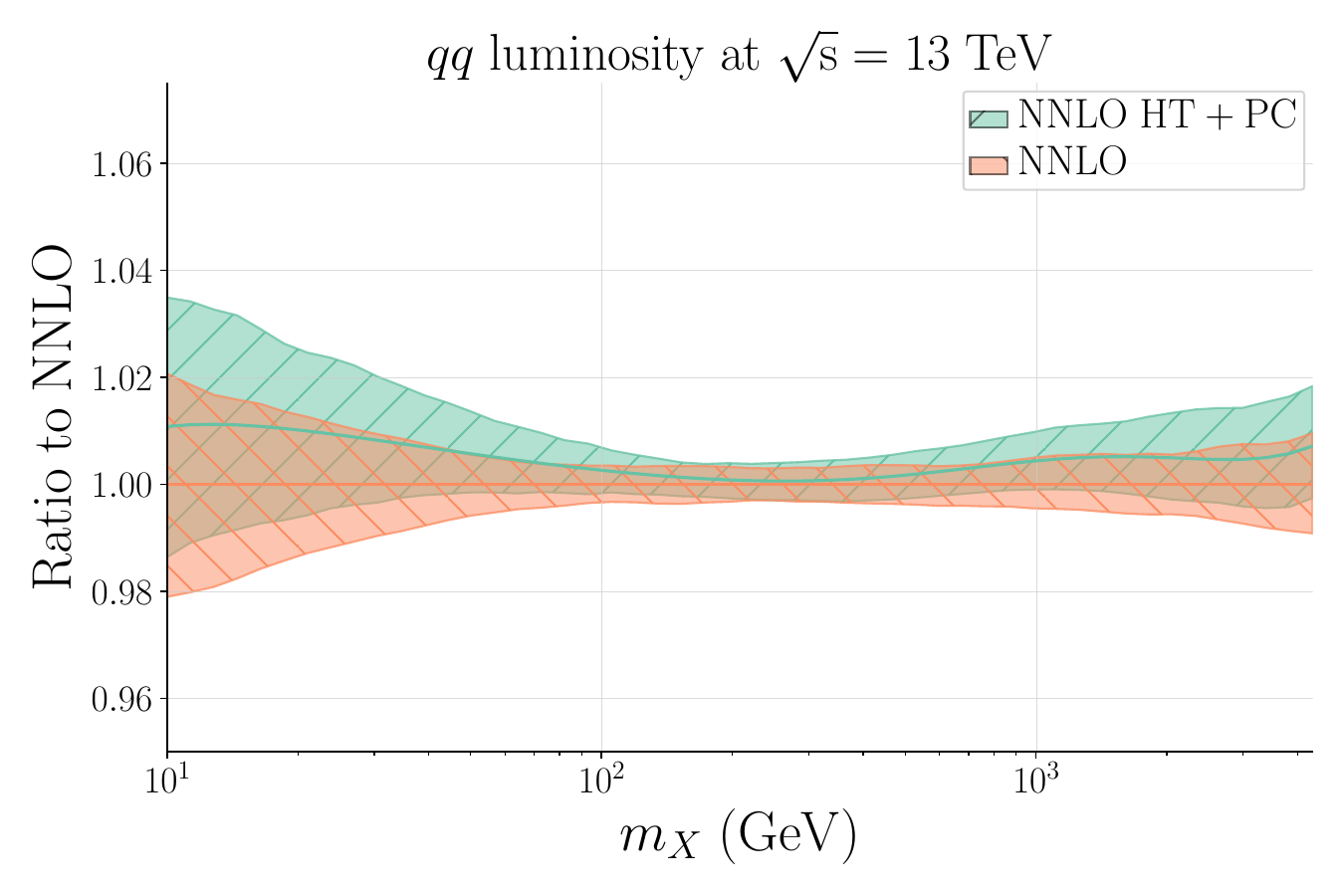}
  \includegraphics[width=0.32\textwidth]{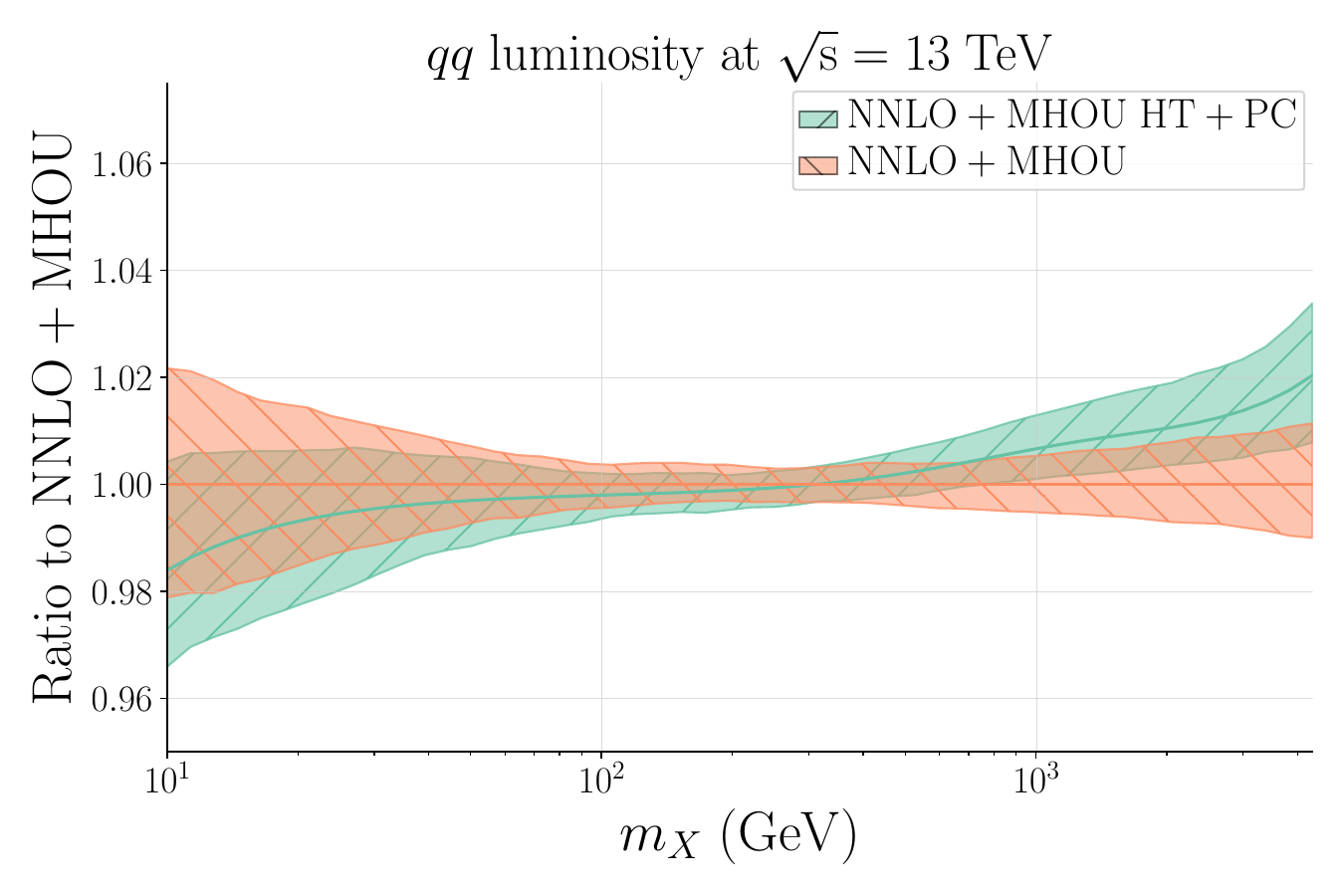}
  \includegraphics[width=0.32\textwidth]{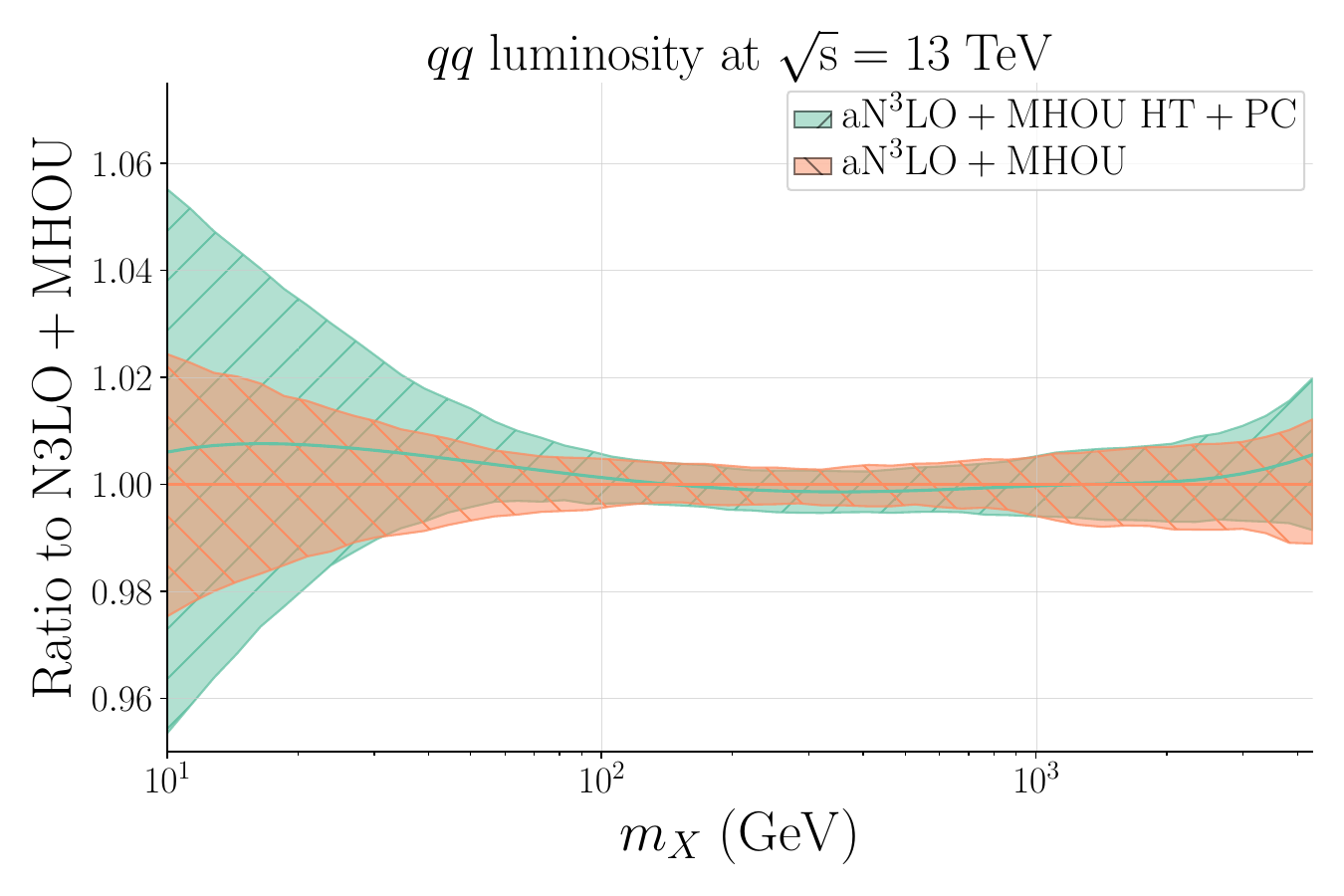}
  \includegraphics[width=0.32\textwidth]{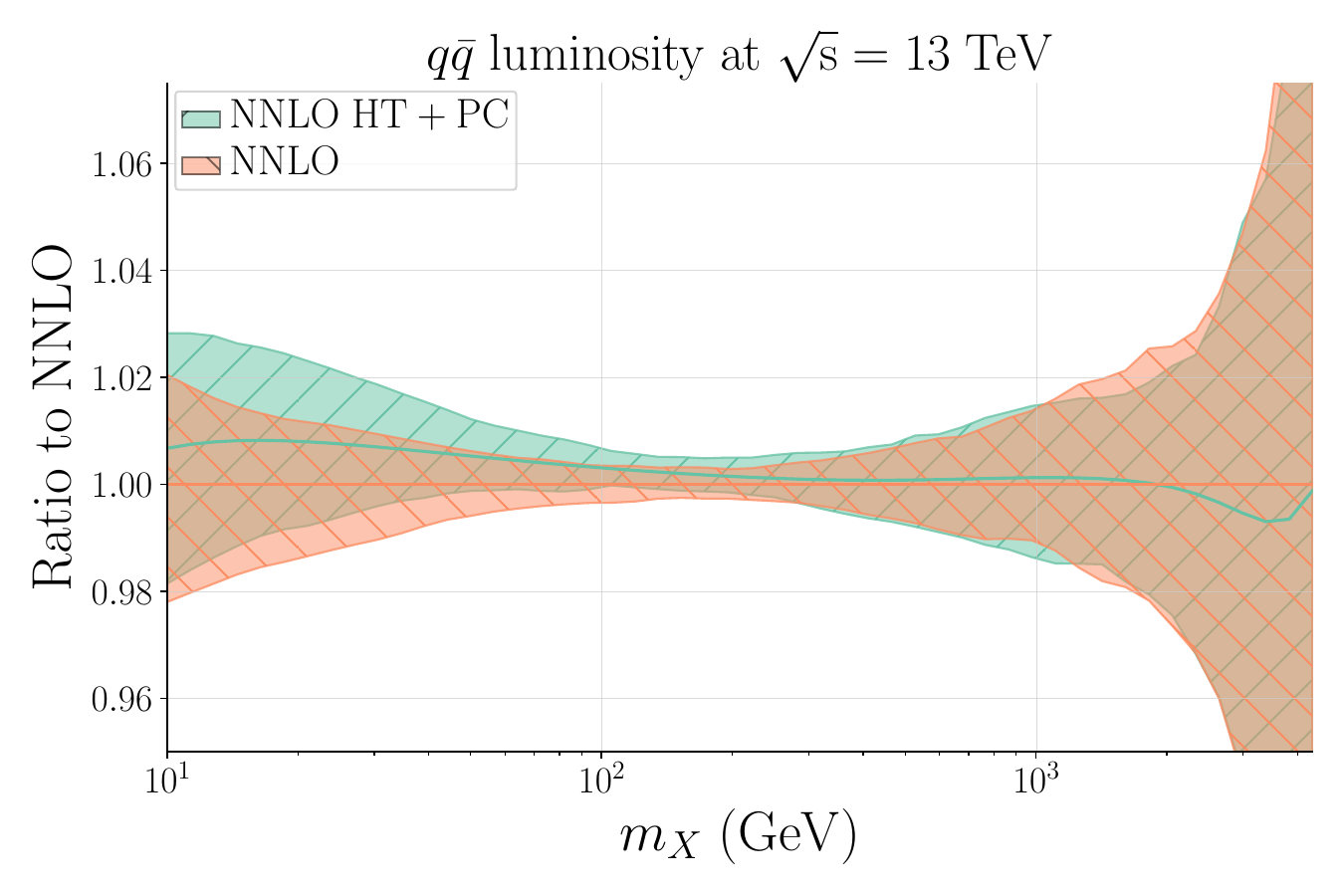}
  \includegraphics[width=0.32\textwidth]{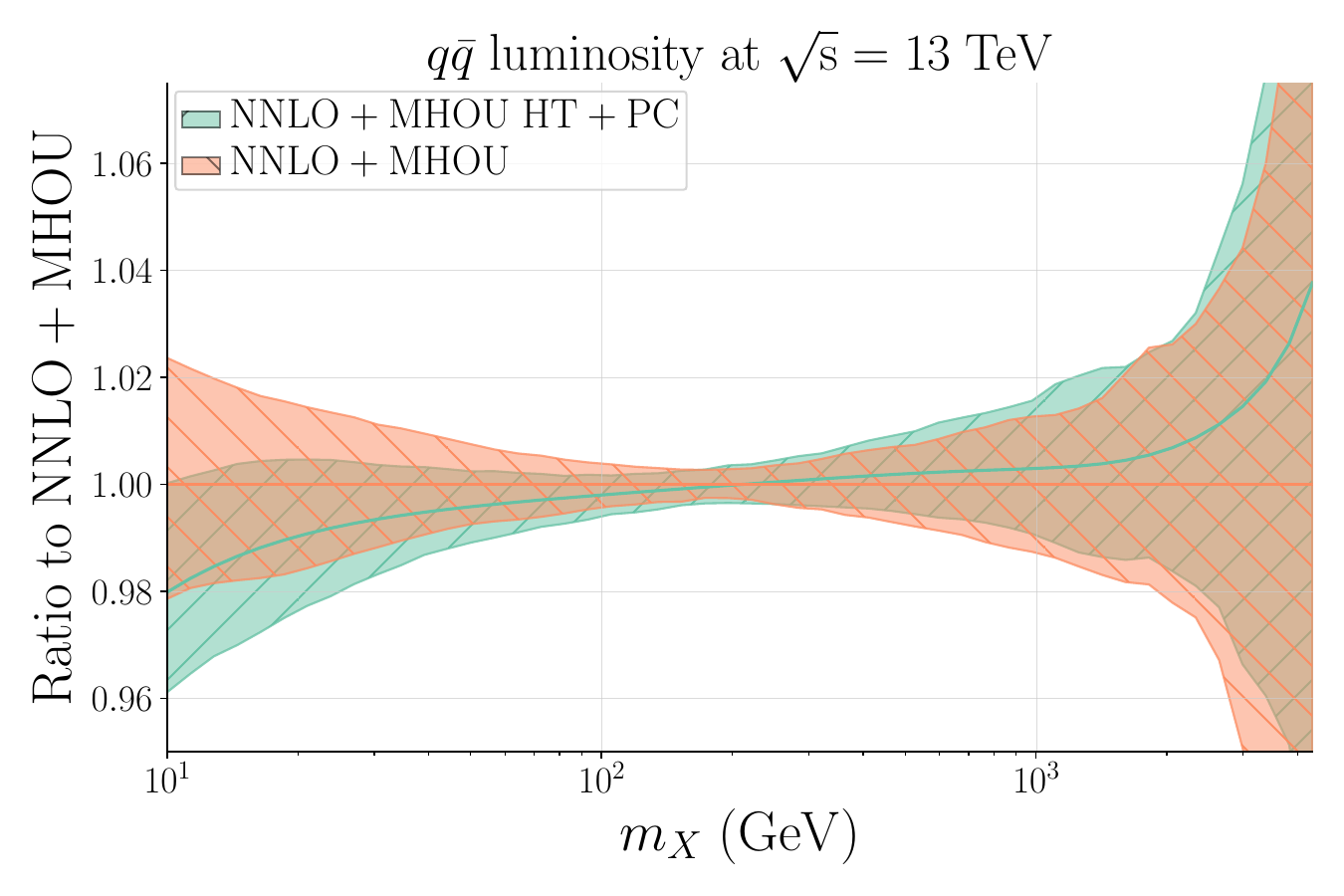}
  \includegraphics[width=0.32\textwidth]{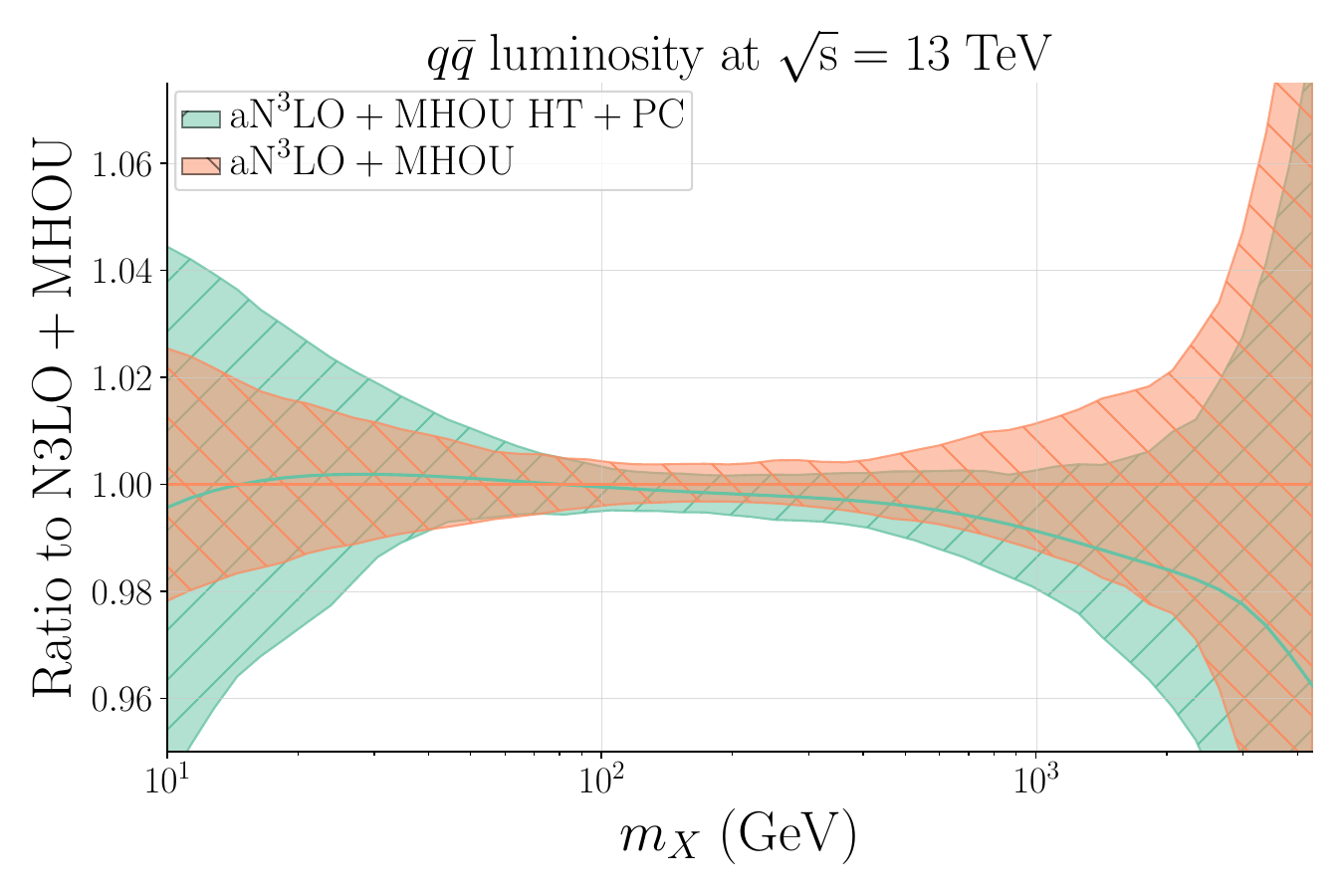}
  \includegraphics[width=0.32\textwidth]{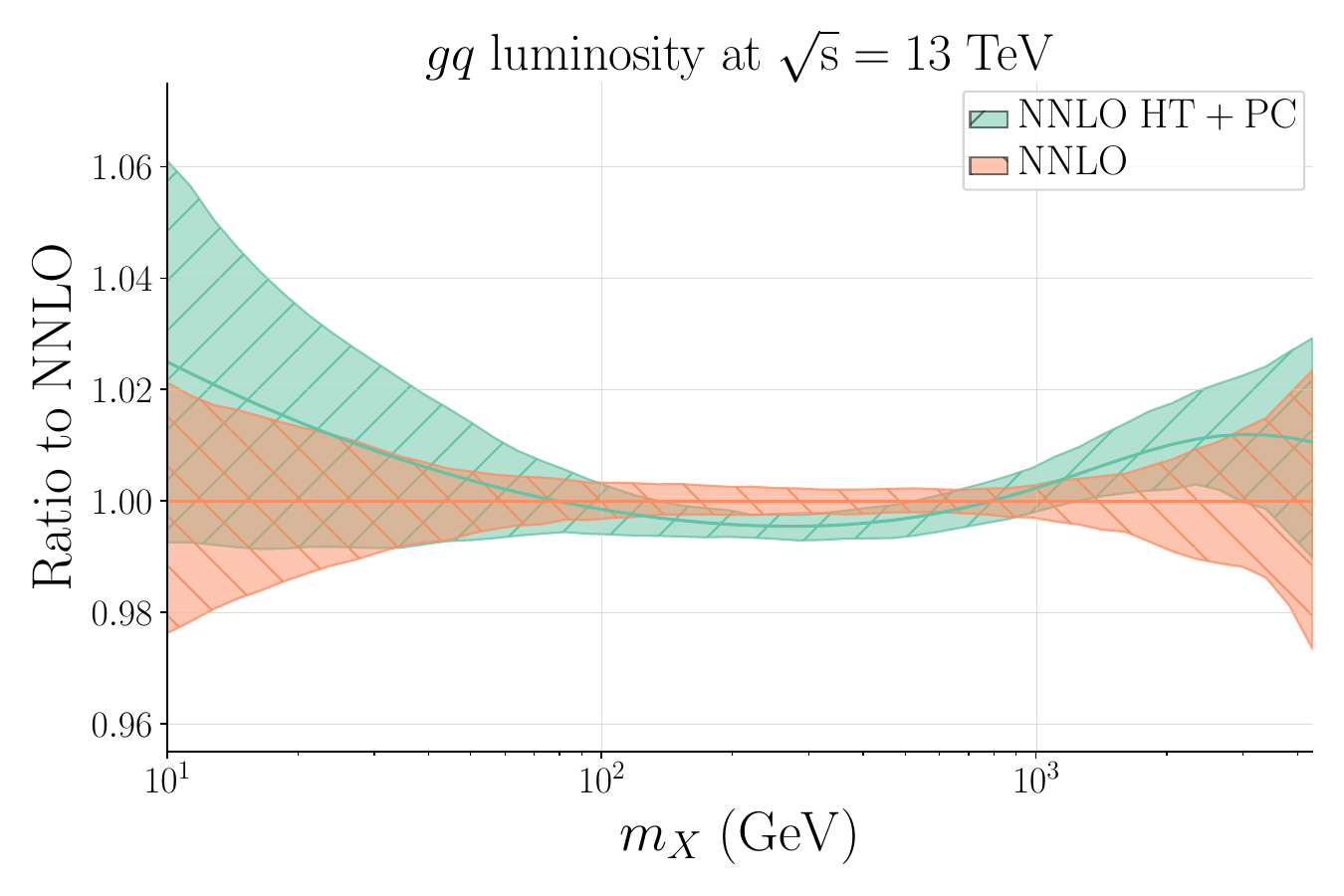}
  \includegraphics[width=0.32\textwidth]{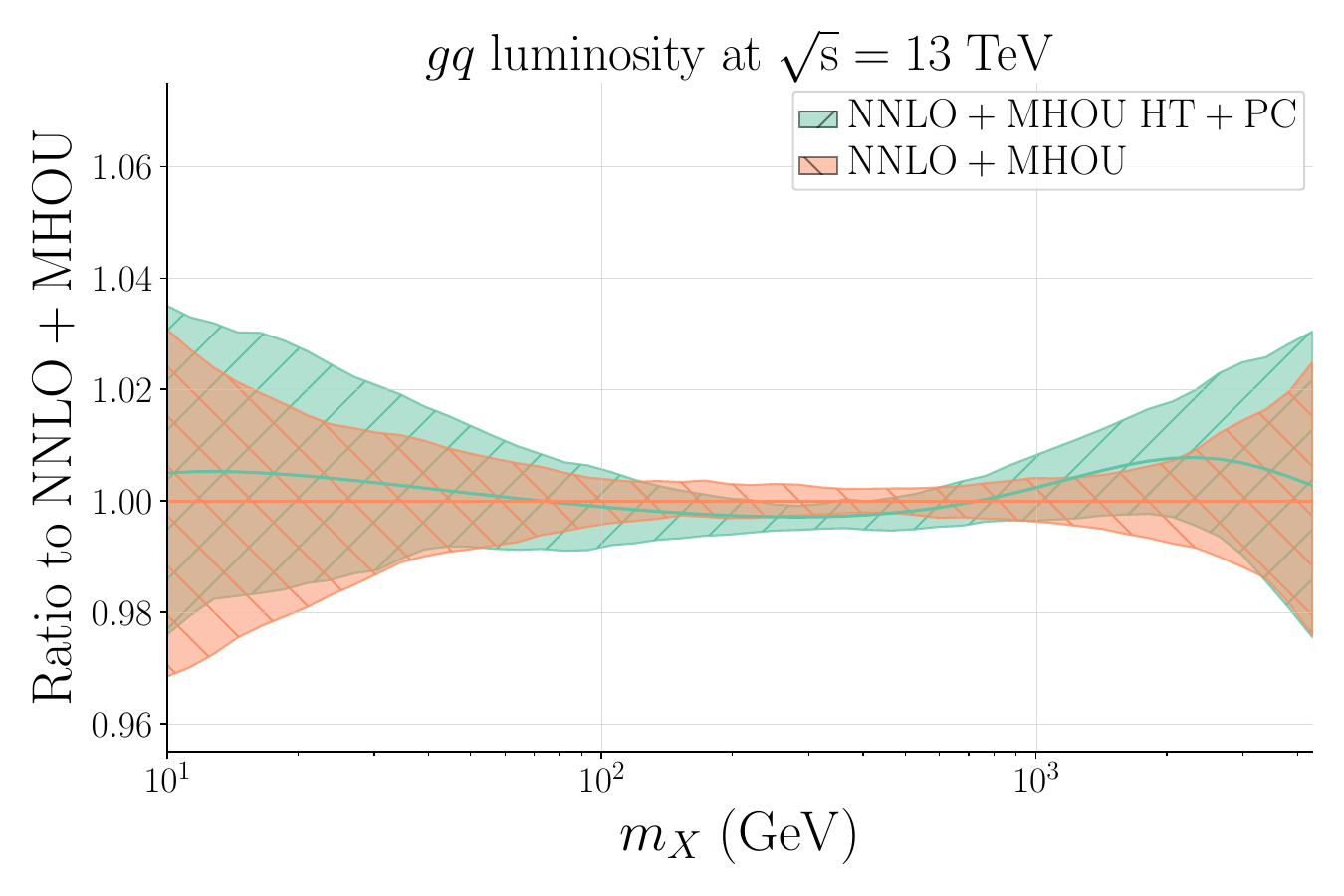}
  \includegraphics[width=0.32\textwidth]{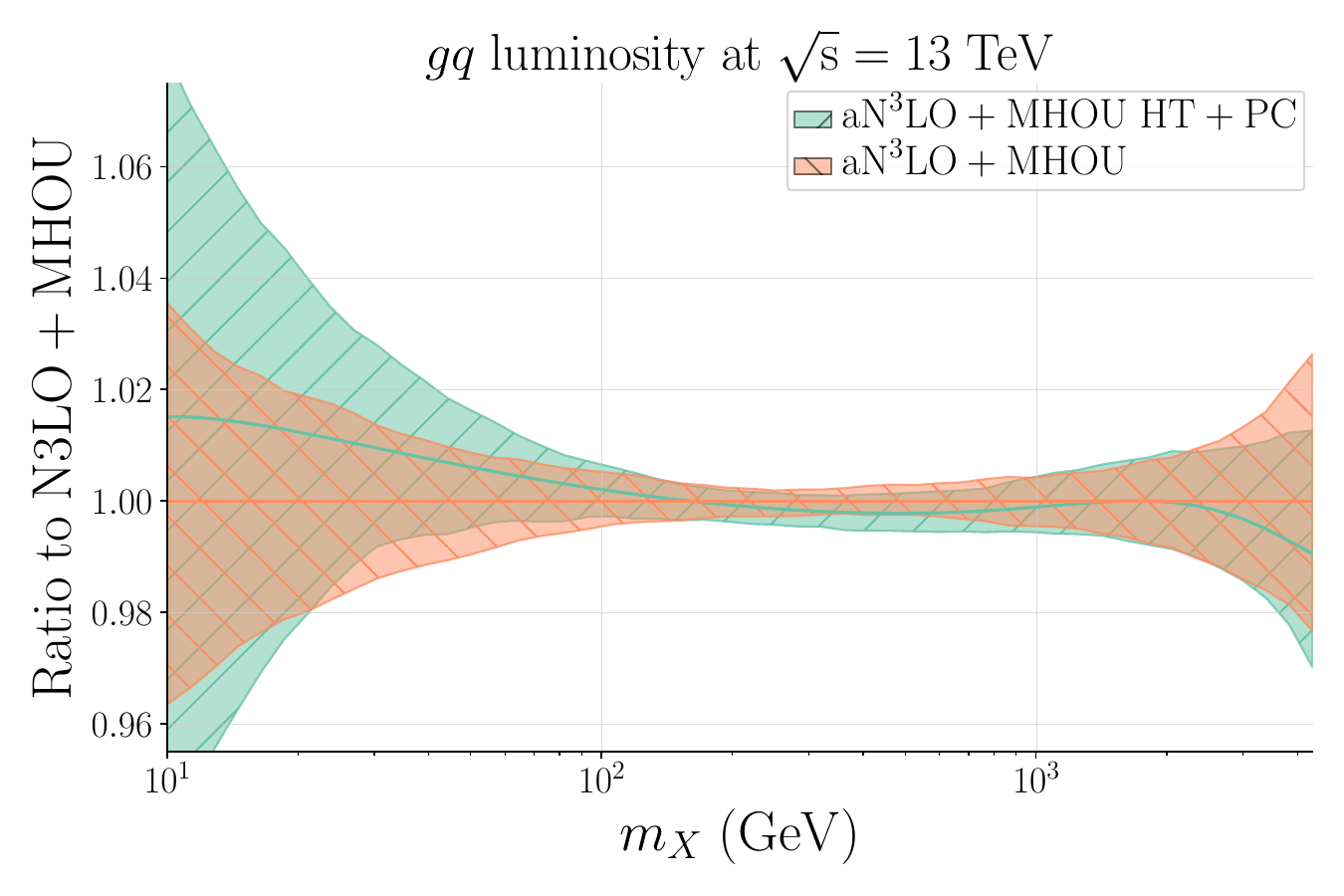}
  \includegraphics[width=0.32\textwidth]{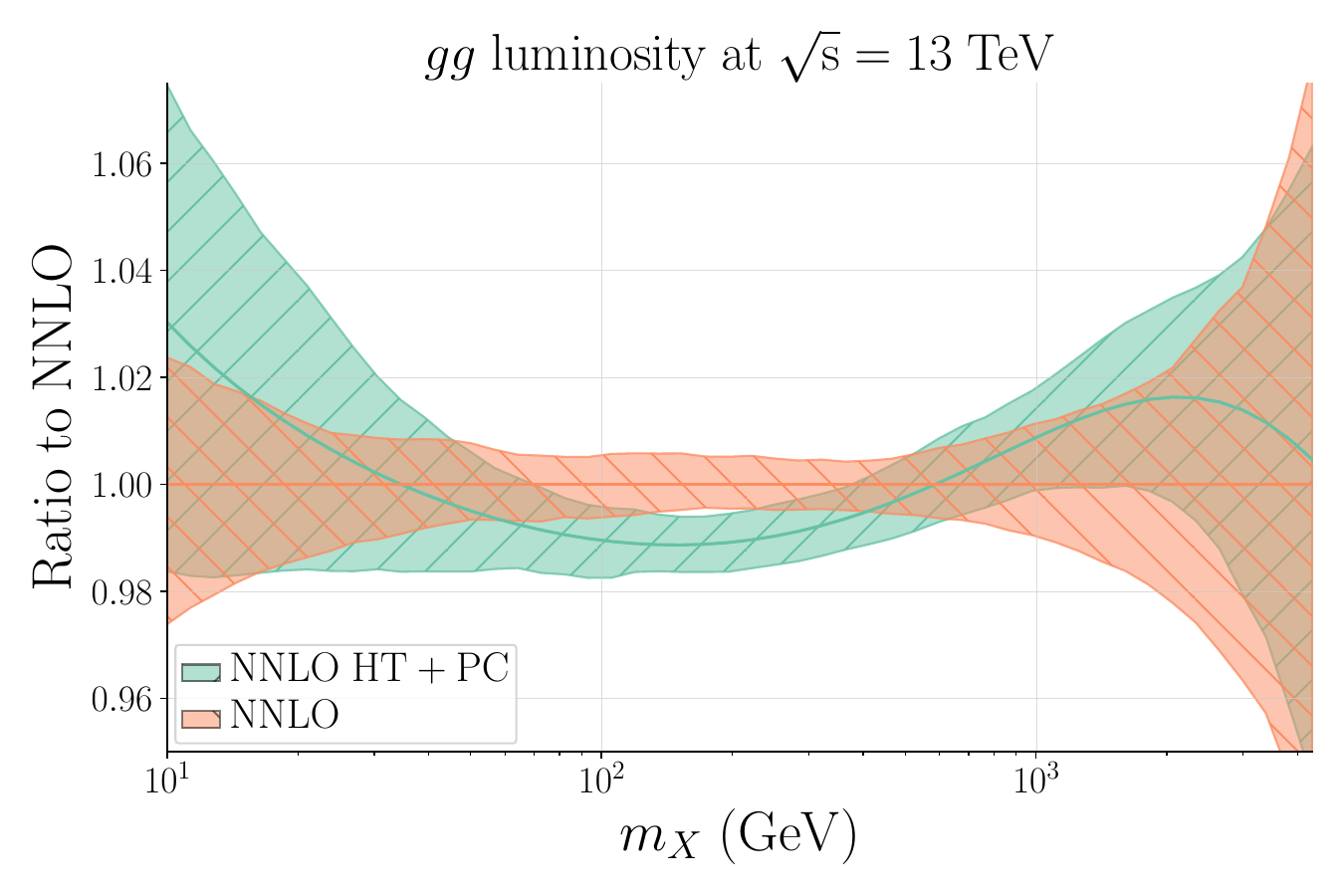}
  \includegraphics[width=0.32\textwidth]{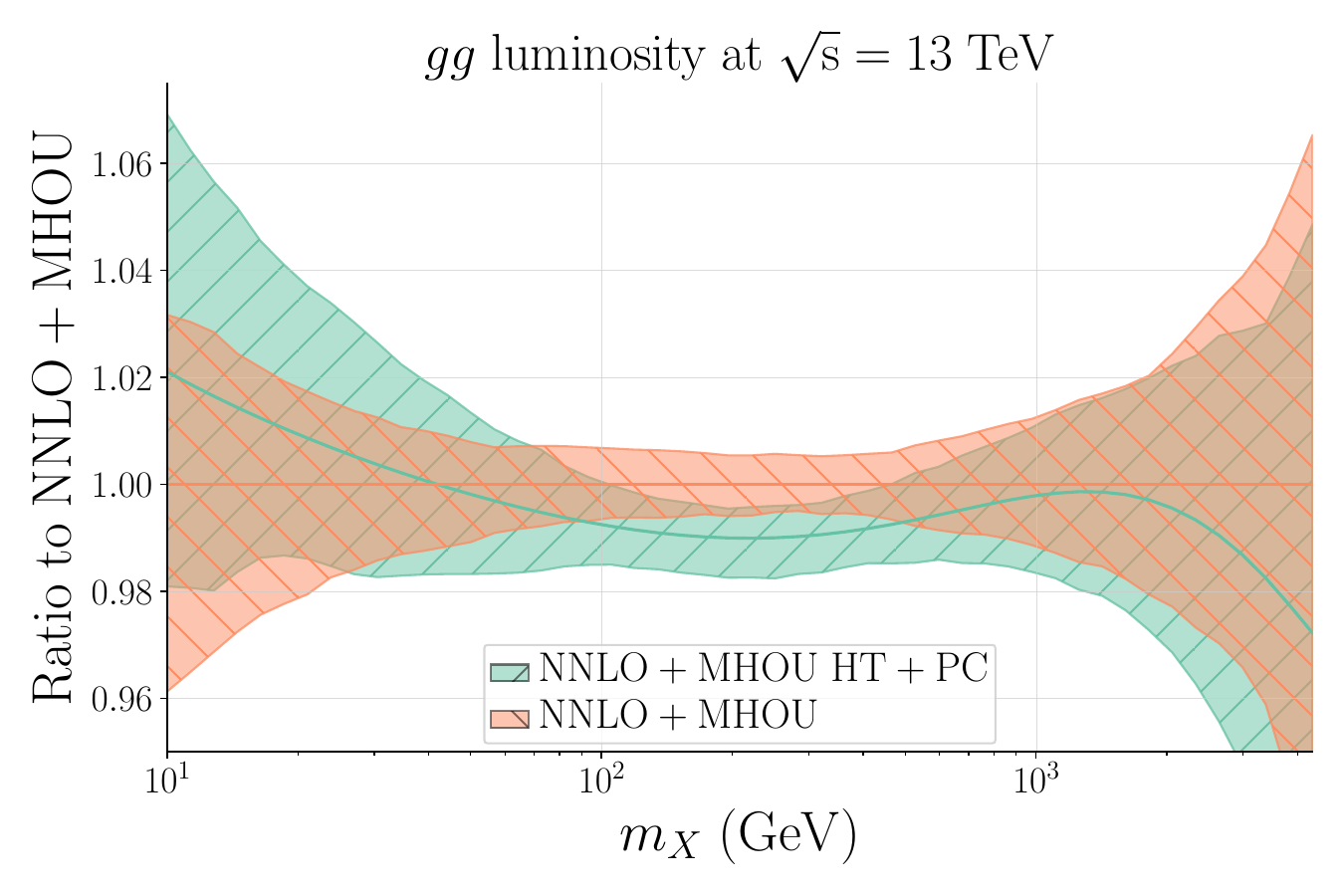}
  \includegraphics[width=0.32\textwidth]{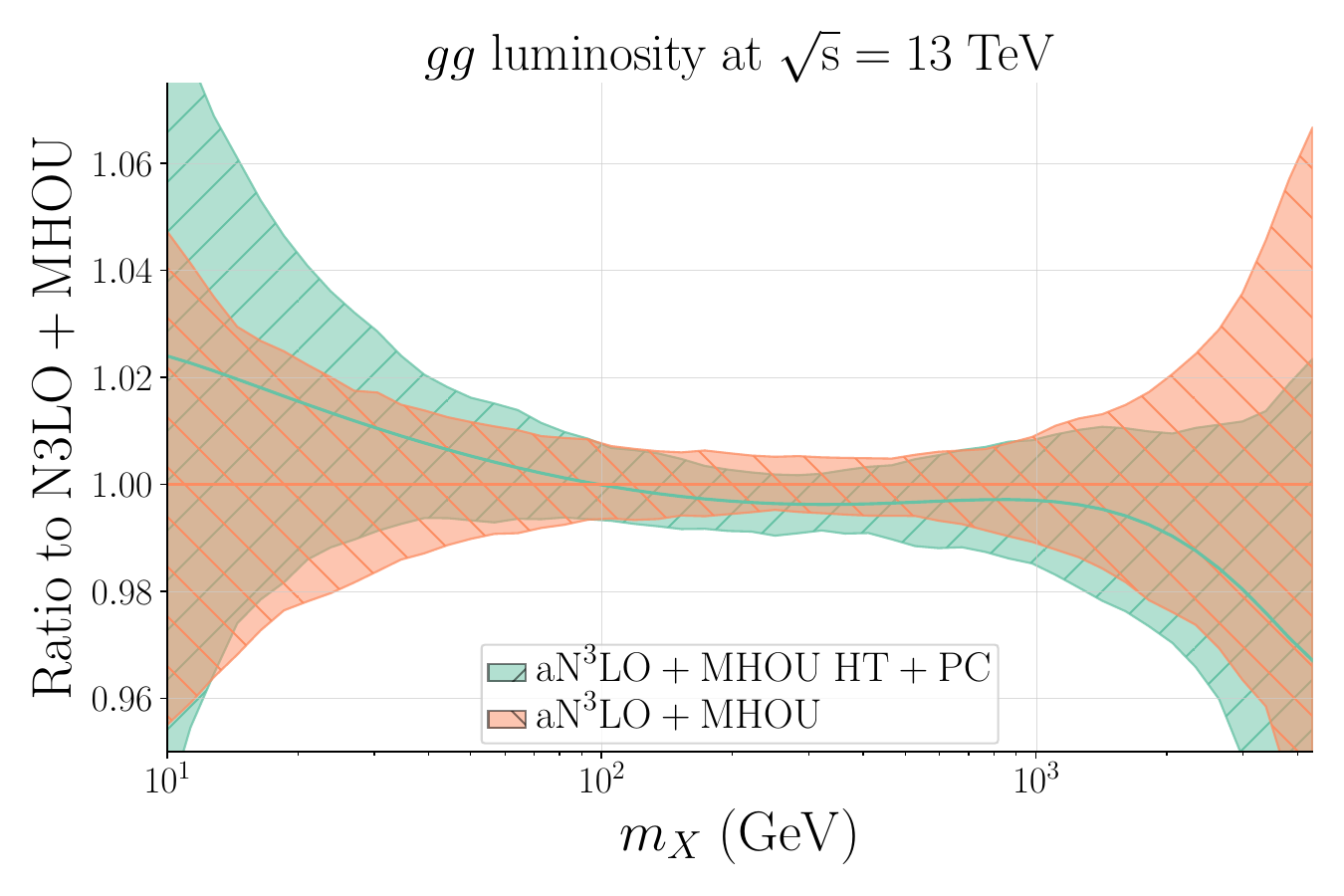}
  \caption{Comparison of luminosities at $\sqrt{s}=13\tev$ for quark-quark,
  quark-antiquark, gluon-quark, and gluon-gluon channels (top to bottom) for
  NNLO, NNLO+MHOU, and aN$^3$LO+MHOU (left to right), showing the ratio of
  the results from fits with shifts
  in the theory predictions due to HT and PCs (blue) to those without (red). Bands with
  correspond to 68\% confidence-level uncertainties.}
  \label{fig:lumi1d}
\end{figure}

In Fig.~\ref{fig:lumi1d}, the luminosities in the quark-quark, quark-antiquark,
gluon-gluon, and gluon-quark channels are shown at $13$ TeV, as ratios of the
fits with HT and PCs to the corresponding NNLO, NNLO+MHOU, and
aN$^3$LO+MHOU~\cite{NNPDF:2024nan} standard fits. As expected from the PDF
plots, the effects of HT and PCs at both high and low invariant masses are
modest, well within one sigma, and largely disappear once MHOUs are added, and
in the \annnlo fit. In fact the only really significant change due to the
inclusion of PCs is in the gluon-gluon channel at invariant masses around the
Higgs mass. This is particularly noticeable in the standard NNLO fit: here there
is a one percent drop in the gluon-gluon luminosity for invariant masses
between $100$ and $250$ GeV, just over two sigma. Once MHOUs are included the
significance of this effect is reduced, and the \annnlo
corrections reduce the impact to well within one sigma. A related effect
is seen in the
gluon-quark channel, though here it is never more than one sigma.

\subsection{Predictions for LHC observables}

\begin{figure}[t!]
  \centering
  \includegraphics[width=0.46\textwidth]{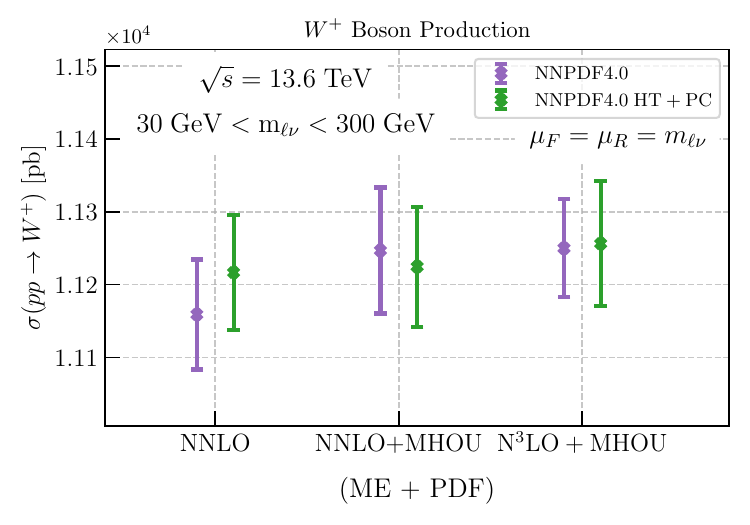}
  \includegraphics[width=0.46\textwidth]{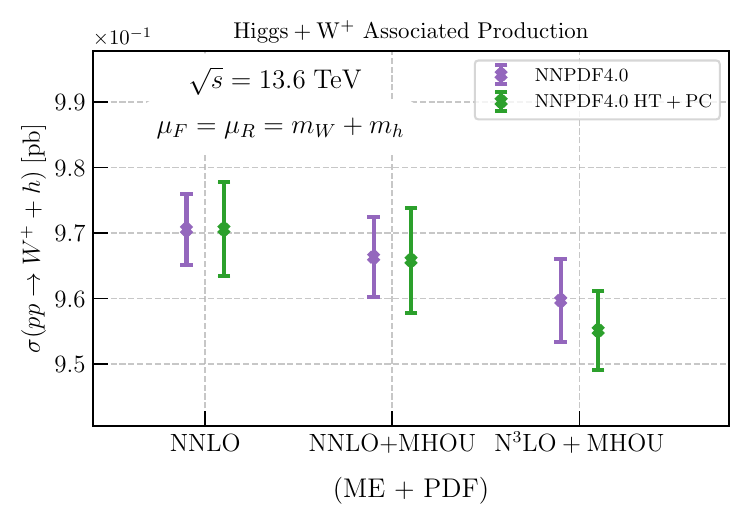}
  \includegraphics[width=0.46\textwidth]{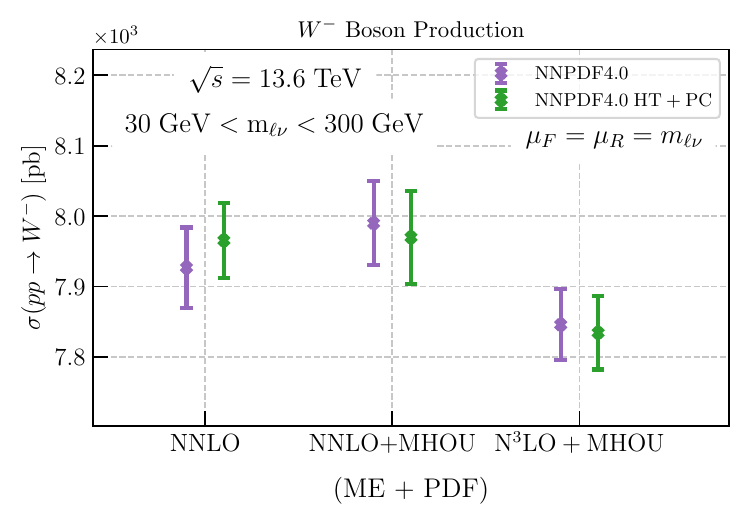}
  \includegraphics[width=0.46\textwidth]{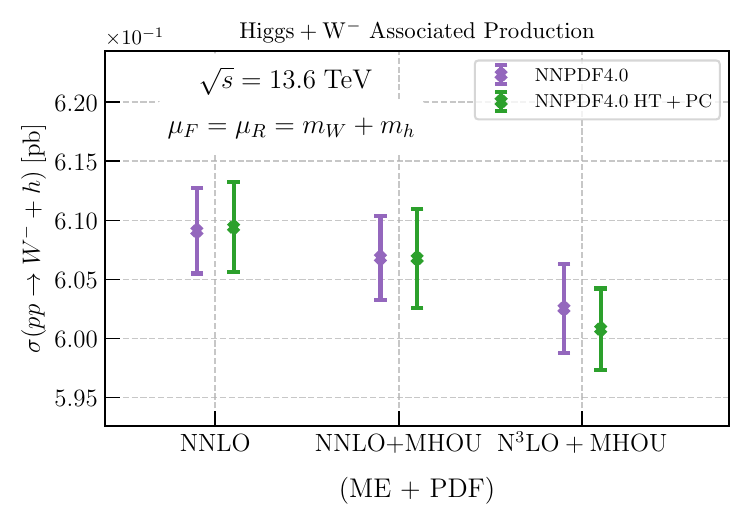}
  \includegraphics[width=0.46\textwidth]{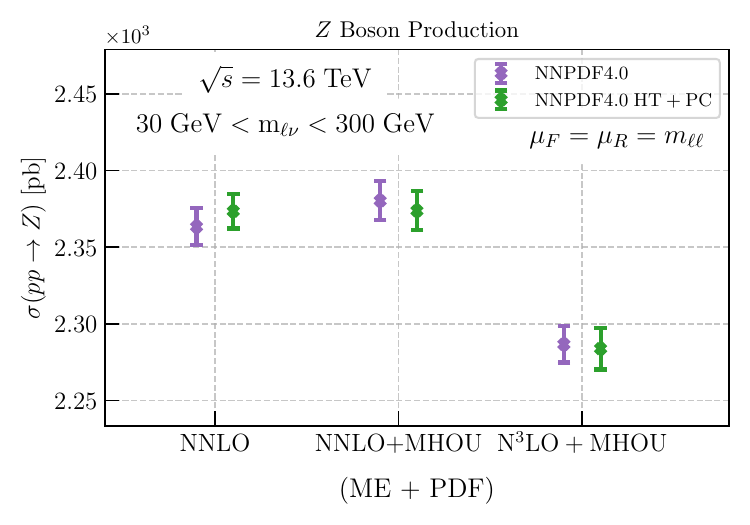}
  \includegraphics[width=0.46\textwidth]{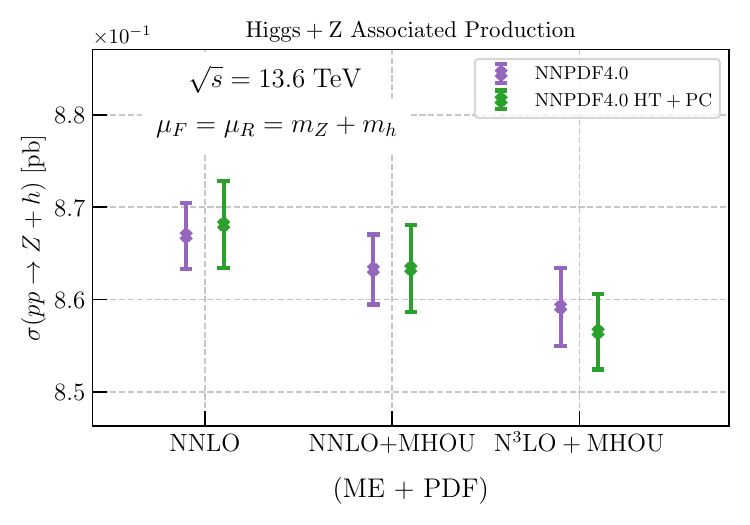}
   \includegraphics[width=0.46\textwidth]{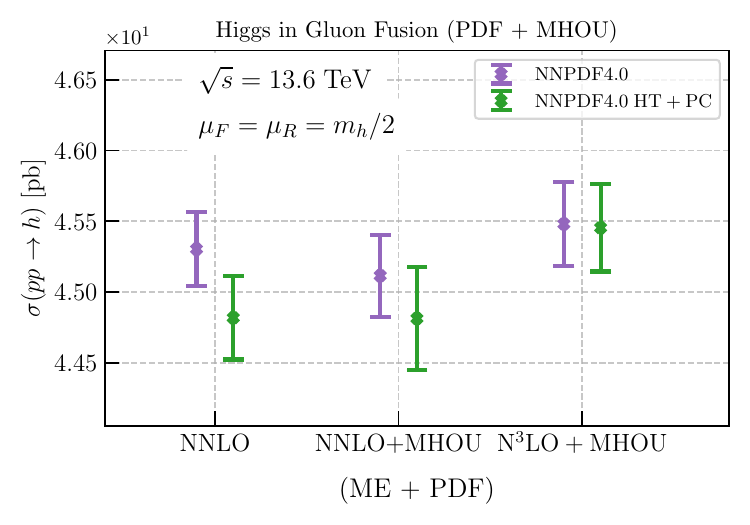}
  \includegraphics[width=0.46\textwidth]{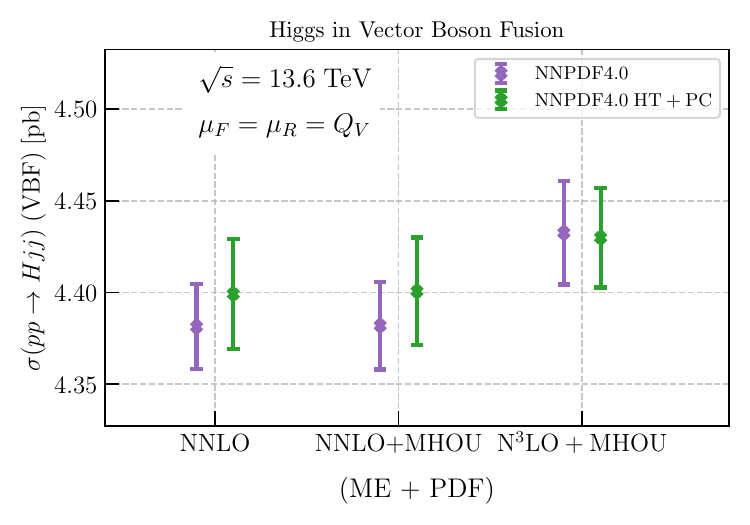}
  \caption{Total cross-sections at $13.6$ TeV for vector boson production (upper
    left), production of Higgs in association with a vector boson (upper right),
    Higgs in gluon fusion (bottom left) and Higgs in vector boson fusion (bottom
    right). Each is computed with matrix element and PDFs evaluated consistently
    at NNLO, NNLO+MHOU and N$^3$LO+MHOU, and with and without HT and PCs
    included in the determination of the PDFs. Error bars include only PDF
    uncertainties: in particular, the MHOU of the ME is not included.}
  \label{fig:pheno_dy}
\end{figure}
\begin{figure}[h!]
  \centering
  \includegraphics[width=0.6\textwidth]{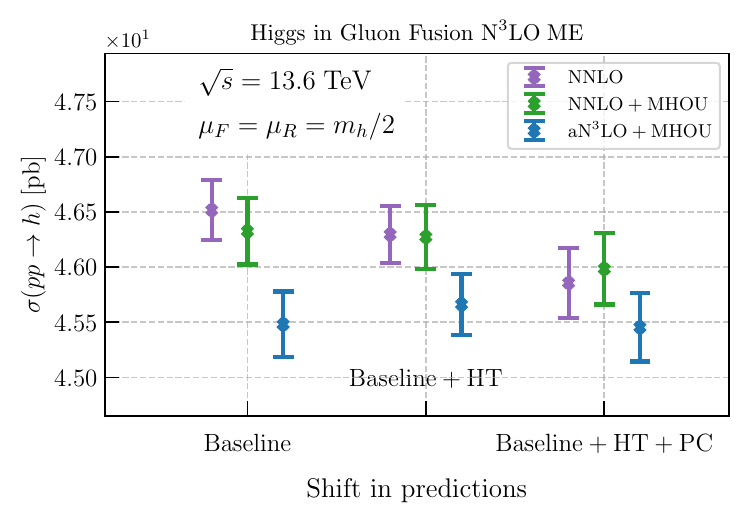}
  \caption{Inclusive Higgs production cross-section in gluon fusion with the
  matrix element evaluated at N$^3$LO, but with the PDFs at NNLO, NNLO+MHOU,
  N$^3$LO+MHOU PDFs without HT and PCs (left), with the PDFs including the HT
  corrections (centre), and finally with the PDFs including PCs for
  single-inclusive jets and dijets (right). Error bars include only PDF
  uncertainties: in particular, the MHOU of the ME is not included.}
  \label{fig:processes_higgs}
\end{figure}

We next analyse the impact of power corrections and higher twist on various
totally inclusive LHC observables related to DY and Higgs production,
specifically vector boson production, associated production of Higgs with a
vector boson, then Higgs production in gluon fusion, and in vector boson fusion
(VBF). The predictions use hard cross-sections computed up to N$^3$LO using
{\sc\small n3loxs}~\cite{Baglio:2022wzu} for associated production, {\sc\small
ggHiggs}~\cite{Bonvini:2014jma} for gluon fusion, and {\sc\small proVBFH}
code~\cite{Dreyer:2018qbw} for VBF.

In Fig.~\ref{fig:pheno_dy} we collect the results of these calculations at NNLO,
NNLO+MHOU, and \annnlo+MHOU, with the matrix element and PDFs computed at the
same order in each case. The results with and without HT and PCs in the PDFs are
then compared (no PCs are known for the matrix elements). In each case, the
effect of the HT+PCs in the PDFs is small, less than one sigma. For vector boson
production and Higgs production (in gluon fusion and VBF) the effect is less at
N$^3$LO than at NNLO, while for associated production the effect is a little
larger at N$^3$LO than at NNLO. In some cases there is a slight increase in
uncertainty once the HT and PCs are included.

The largest effect of the HT and PCs is seen in NNLO Higgs production by gluon
fusion, as expected from the discussion of the gluon-gluon luminosity in the
previous section. Here, the HT+PCs reduce the total cross-section by around 1\%
(though still under one sigma). However when MHOUs are included, this effect
drops to around 0.5\%, and at N$^3$LO it is almost negligible.

To explore this further, in Fig.~\ref{fig:processes_higgs} we show Higgs
production in gluon fusion with the matrix element held fixed (at N$^3$LO), but
with the PDFs computed at NNLO, NNLO+MHOU, aN$^3$LO+MHOU, then in each case with
only the HT added, and then finally with both the HT and PCs included (to single
inclusive jets and dijets). Without HT or PCs, going from NNLO to \annnlo PDFs
reduces the cross-section by 2.1\% (almost three sigma). When the HT is
included, the NNLO result drops by around 0.4\%, while the N$^3$LO goes up (by
0.4\%), so the difference is now only 1.3\%. When the PCs are added, the NNLO
result drops again, by 0.8\%, while the N$^3$LO only drops by 0.4\%, so the
difference between NNLO and N$^3$LO PDFs is now only 0.9\% (a little over one
sigma). Therefore including HT and PCs significantly reduces the difference between
using NNLO or aN$^3$LO PDFs in combination with the N$^3$LO matrix element.

\subsection{Effect on the Strong Coupling}

\begin{figure}[t!]
  \centering
  \includegraphics[width=0.6\textwidth]{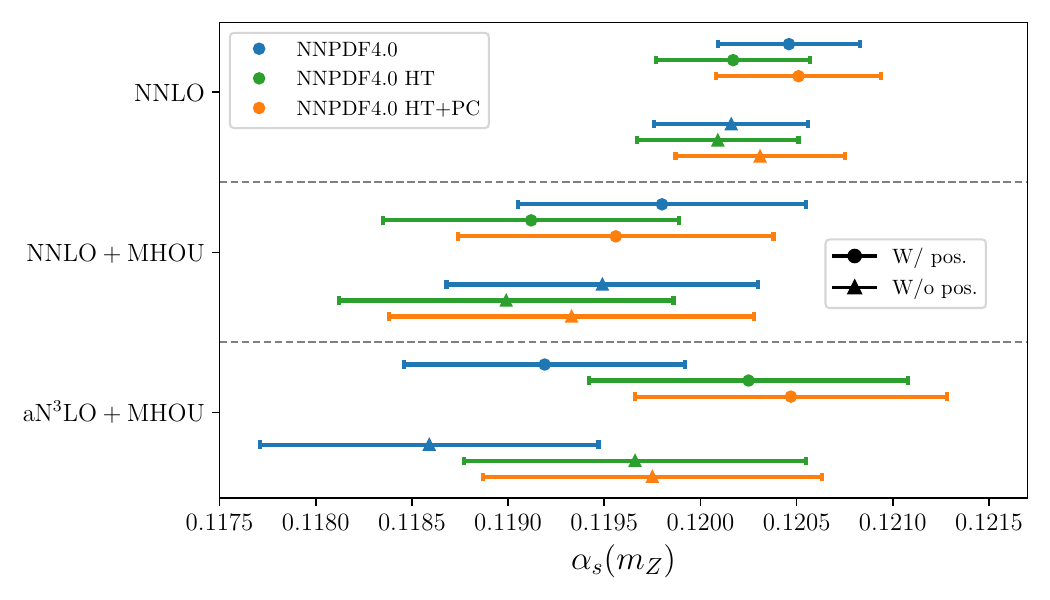}
  \caption{Results for the strong coupling $\alpha_s(m_Z)$ determined in the
    global fit, with no HT or PC (blue), with HT but no PC (green), and with
    both HT and PC (red). Results are given at NNLO, NNLO+MHOU and N$^3$LO+MHOU,
    in each case with and without constraints of PDF positivity.}
  \label{fig:alphas}
\end{figure}

We finally consider the effect of HT and PCs on the extraction of the strong
coupling at the mass of the $Z$-boson, $\alpha_s(m_Z)$, from the global dataset.
We again use the theory covariance methodology, precisely as in
Ref.~\cite{Ball:2025xgq}, but now including also the HT and PC contributions to
the theory covariance matrix. The results are shown in Fig.~\ref{fig:alphas}.

In the NNLO fit, without MHOU, the HT reduces $\alpha_s$ a little, while the jet
PCs increase it, in each case by around one standard deviation. When MHOU is
added, the best fit value of $\alpha_s$ drops significantly, and its uncertainty
increases substantially, as found in Ref.~\cite{Ball:2025xgq}. Again the HT
reduces $\alpha_s$, and the PCs increase it, by about one standard deviation.
However, at \annnlo the pattern changes: while the \annnlo corrections reduce
$\alpha_s$ a little, the HT correction now results in a substantial increase
(more than a standard deviation). We think that
this is because at \annnlo the HT is less contaminated by MHO perturbative
corrections to the leading twist, and its effect on $\alpha_s$ is thus less
easily hidden by the MHOU. 
The effect of the jet PCs is positive as before, but smaller (the
jet cross-sections are still evaluated at NNLO+MHOU in the \annnlo fits). The result is a value of $\alpha_s(m_Z)$ at
\annnlo which is very consistent with that obtained at NNLO, but with
substantially larger uncertainty.

As we noted in Ref.~\cite{Ball:2025xgq}, all these results are subject to a
further uncertainty due to the implementation of positivity constraints on the
PDF fit. Since any PDF positivity constraint necessarily introduces a
nongaussianity in the distribution of the PDFs, the closure testing of the
methodology only works when the positivity constraints are removed. As can be
seen in Fig.~\ref{fig:alphas}, releasing the positivity constraints results in
lower values of $\alpha_s$ consistently across all the various determinations.

%% file: tables/tab_total_chi2.tex
\begin{tabularx}{\textwidth}{Xcccccccccc}
  \toprule
  \multirow{2}{*}{Dataset}
  & \multirow{2}{*}{$N_{\rm data}$}
  & \multicolumn{3}{c}{NNLO}
  & \multicolumn{3}{c}{NNLO+MHOU}
  & \multicolumn{3}{c}{aN$^3$LO+MHOU} \\
  \cmidrule(lr){3-5}
  \cmidrule(lr){6-8}
  \cmidrule(lr){9-11}
  && \tiny baseline & \tiny only shift & \tiny shift+unc.
  &  \tiny baseline & \tiny only shift & \tiny shift+unc.
  &  \tiny baseline & \tiny only shift & \tiny shift+unc. \\
  \midrule
  DIS NC & 2100
    & 1.230 & 1.228 & 1.211   %
    & 1.204 & 1.200 & 1.195     %
    & 1.203 & 1.201 & 1.196 \\  %
  DIS CC & 989
    & 0.902 & 0.912 & 0.900
    & 0.901 & 0.898 & 0.897
    & 0.916 & 0.920 & 0.912 \\
  DY NC & 735
    & 1.201 & 1.184 & 1.184
    & 1.150 & 1.156 & 1.150
    & 1.159 & 1.140 & 1.143 \\
  DY CC & 157
    & 1.470 & 1.486 & 1.460
    & 1.366 & 1.347 & 1.317
    & 1.370 & 1.405 & 1.383 \\
  Top pairs & 64
    & 1.250 & 1.32 & 1.293
    & 1.440 & 1.459 & 1.381
    & 1.407 & 1.347 & 1.310 \\
  Single-inclusive jets & 356
    & 0.954 & 0.928 & 0.942
    & 0.809 & 0.787 & 0.791
    & 0.832 & 0.808 & 0.804 \\
  Dijets & 144
    & 2.018 & 1.928 & 1.929
    & 1.713 & 1.479 & 1.485
    & 1.674 & 1.480 & 1.477 \\
  Prompt photons & 53
    & 0.756 & 0.735 & 0.734
    & 0.670 & 0.686 & 0.688
    & 0.680 & 0.696 & 0.691 \\
  Single top & 17
    & 0.364 & 0.372 & 0.374
    & 0.376 & 0.384 & 0.380
    & 0.355 & 0.354 & 0.353 \\
  \midrule
  Total & 4615
    & 1.171 & 1.164 & 1.154
    & 1.135 & 1.123 & 1.116
    & 1.140 & 1.130 & 1.124 \\
\bottomrule
\end{tabularx}

%% file: sec-conclusions.tex
\section{Conclusions}
\label{sec:conclusions}
We have presented a determination of higher twist corrections in DIS, and linear
power corrections in single-inclusive jets and dijets in a joint fit with global
PDFs. We find good agreement with earlier results for higher twist. This is the
first time an empirical determination of power corrections in jets has
been performed in
the context of PDF fits, but we find agreement with theoretical expectations, and in particular with estimates of hadronization corrections made using final state Monte Carlos.

While the higher twist corrections are mainly taken care of
by cuts in $Q^2$ of a few GeV$^2$, and in $W^2$ at a few tens of GeV$^2$, we
find that the effect of the linear power corrections in jets can extend up to
hard scales of several hundred GeV. This means in particular that when
incorporating LHC data on single-inclusive jets in global PDF fits, it may be
necessary for PDFs accurate to within 1\% to either cut some of the low $p_T$
data, or include a linear power correction in the theoretical predictions (or
indeed both, similar to the strategy for DIS). The size of the power
correction  could be reduced by first correcting the data (from hadron to
parton level) using a correction factor estimate using final state Monte
Carlos, if such factors were determined reliably and consistently across all
available jet datasets. 

We have produced PDFs which include both higher twist for DIS and linear power
corrections for single-inclusive jets and dijets in the theoretical predictions
used in the fit, along with their corresponding uncertainties encoded in a
theory covariance matrix. We have assessed the impact of these corrections on a
series of representative LHC observables and found that while they are for the
most part small, they can be as large as one percent for processes sensitive to
the gluon-gluon luminosity in the intermediate mass region, such as Higgs in
gluon fusion. We have likewise examined the effect on the value of $\alpha_s$
determined from hadronic processes, and again found effects as large as one
percent.

Since the effects of higher twist and linear power corrections in jets can be
relevant at the current level of precision for various processes, they should be
accounted for in future global PDF determinations that hope to achieve a
precision of 1\% in LHC observables. The empirical determination of power
corrections in Drell-Yan and vector boson production processes, and the possible
linear power corrections in top production~\cite{Makarov:2023uet}, are left for
future studies.

The PDF sets with power corrections are made available in LHAPDF
format~\cite{Buckley:2014ana} on the NNPDF website
\begin{center}
\url{https://nnpdf.mi.infn.it/nnpdf4-0-higher-twist/}
\end{center}
under the following names:
\begin{flushleft}
  \ttfamily{ NNPDF40\_nnlo\_as\_01180\_ht \\
    NNPDF40\_nnlo\_as\_01180\_mhou\_ht \\
    NNPDF40\_an3lo\_as\_01180\_mhou\_ht . }
\end{flushleft}

\bigskip
\paragraph{Acknowledgements:}
We would like to thank Gavin Salam for discussions at Gargnano in September
2023, where he emphasised the possible significance of linear power corrections
to jet observables in global PDF fits. We would also like to thank the members
of the NNPDF collaboration for insightful discussions during the course of this
work. Finally, we are grateful to Felix
Hekhorn and Emanuele Nocera for providing valuable feedback
on the earlier drafts of this manuscript.
R.D.B. and R.S. are supported by the Science and Technology Facilities Council
(STFC) via grant awards ST/T000600/1 and ST/X000494/1.